\documentclass[sigconf,9pt]{acmart}

\usepackage[english]{babel}
\usepackage{blindtext}
\usepackage{subfigure}
\usepackage{xcolor}
\usepackage{array}
\usepackage{tikz}
\usepackage{ctable}
\usepackage{balance}
\usepackage{enumitem}
\usepackage{amsfonts}
\usepackage{xspace}
\usepackage{siunitx}
\usepackage[ruled,vlined,linesnumbered]{algorithm2e}
\usepackage{setspace}
\usepackage{bbding}
\usepackage{makecell}
\usepackage{multirow}
\usepackage{booktabs}
\usepackage{colortbl}
\usepackage{pifont}
\usepackage{diagbox}

\usepackage[most]{tcolorbox}
\usepackage{lipsum} 

\newcommand{\ie}{\textit{i}.\textit{e}.}
\newcommand{\eg}{\textit{e}.\textit{g}.}
\newcommand{\etc}{\textit{etc}}

\newcommand{\Rmnum}[1]{\uppercase\expandafter{\romannumeral#1}}

\newcommand{\projectName}{\textsc{RANPilot}\xspace}

\acmYear{2026}\copyrightyear{2026}
\setcopyright{cc}
\setcctype[4.0]{by}
\acmConference[SIGCOMM '26]{ACM SIGCOMM 2026 Conference}{August 17--21, 2026}{Denver, CO, USA}
\acmBooktitle{ACM SIGCOMM 2026 Conference (SIGCOMM '26), August 17--21, 2026, Denver, CO, USA}
\acmDOI{10.1145/3789240.3829146}
\acmISBN{979-8-4007-2467-1/26/08}

\begin{document}

\title[\textsc{RANPilot}: Making AI Functionalities Robust to Dynamic O-RAN Reconfigurations]{\textsc{RANPilot}: Making AI Functionalities Robust to\\ Dynamic O-RAN Reconfigurations}

\author[S. Yu, L. Shen, J. Zhang, X. Li, X. Xia, Y. Zheng, Y. Xie]{{Shiming Yu\textsuperscript{\dag}}, {Leming Shen\textsuperscript{\dag}}, {Jianing Zhang\textsuperscript{\dag}}, {Xin Li\textsuperscript{\dag}}, {Xianjin Xia\textsuperscript{\dag}},{Yuanqing Zheng\textsuperscript{\dag}}, {Yaxiong Xie\textsuperscript{\ddag}}
}

\affiliation{
  \institution{\textsuperscript{\dag} The Hong Kong Polytechnic University, 
  \textsuperscript{\ddag} University at Buffalo, SUNY}
  \country{}
}

\begin{abstract}
The Open Radio Access Network (O-RAN) promises unprecedented
flexibility through its reconfigurable architecture and AI-driven control.
However, this agility exposes a critical fragility: 
AI models trained on one network
configuration suffer significant performance degradation after an upgrade due to dramatic data drift.
The standard solution, reactive retraining, is unacceptably slow,
leaving the network in a suboptimal state for tens of
minutes and undermining the core benefits of O-RAN's dynamism.
This paper introduces \projectName, the first framework to address this challenge through proactive AI adaptation.
\projectName constructs a lightweight "virtual O-RAN" (a trace-driven emulator) to synthesize high-fidelity training
data representing the post-reconfiguration state before the physical change occurs,
allowing AI models to be adapted in advance.
Extensive experiments on a real-world 5G testbed demonstrate
that \projectName achieves near interruption-free AI services upon reconfiguration,
reducing AI downtime by 85\% to 94\% against reactive baselines.
By shifting the AI evolution paradigm from reactive redevelopment to proactive preparation,
\projectName explores a digital-leadoff approach to enable robust AI in reconfigurable O-RAN deployments.
\end{abstract}



\maketitle

\section{Introduction}

Cellular Radio Access Networks (RAN) are undergoing a paradigm shift, moving away from the rigid,
monolithic designs of the past toward the Open RAN (O-RAN) \cite{ORANAlliance,cannata2024towards,foukas2025ranbooster}.
This new paradigm is founded on the principles of disaggregation, virtualization, and open interfaces,
splitting traditional base station functions into distinct, interoperable software/hardware components. 
This modularity, coupled with the introduction of the RAN Intelligent Controller (RIC),
unlocks two transformative capabilities.
First, it enables unprecedented reconfigurability,
allowing network operators to dynamically add or upgrade radio
units to enhance coverage and edge processing capabilities \cite{ge2025iridescence,mahimkar2022aurora,ge2023chroma,mahimkar2021auric}.
Second, it fosters programmability, where AI-driven applications, known as xApps,
can be deployed on the RIC to intelligently manage the network, optimizing functions like resource allocation \cite{ko2024edgeric,balasingam2024application},
traffic steering \cite{lacava2023programmable,ntassah2023xapp}, and interference mitigation \cite{sun2024spotlight,kilinc2022jade,ghosh2024sparc}.
Together, these innovations promise a future of smarter, more adaptive,
and cost-efficient cellular networks \cite{khan2023ai,AIRANAlliance,heidari2025phasemo,permal2024towards,aslan2025fairric,schiavo2024cloudric}.

\begin{figure}[t]
    \setlength{\abovecaptionskip}{2pt}
    \subfigtopskip=-2pt
    \subfigcapskip=-2pt
    \centering
    \subfigure[Reactive redevelopment of existing paradigms]{
    \begin{minipage}[t]{0.45\textwidth}
    \centerline{\includegraphics[width = 0.95\textwidth]{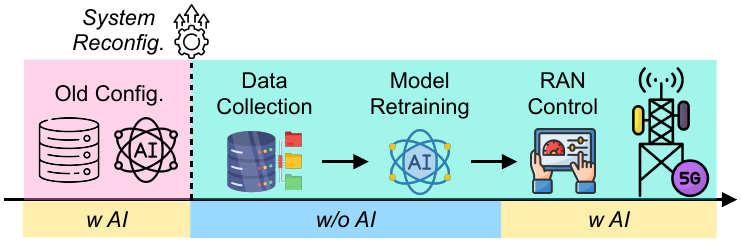}
    \label{fig:framework_a}}
    \end{minipage}
    }
    \centering
    \subfigure[Proactive adaptation of \projectName]{
    \begin{minipage}[t]{0.45\textwidth}
    \centerline{\includegraphics[width = 0.95\textwidth]{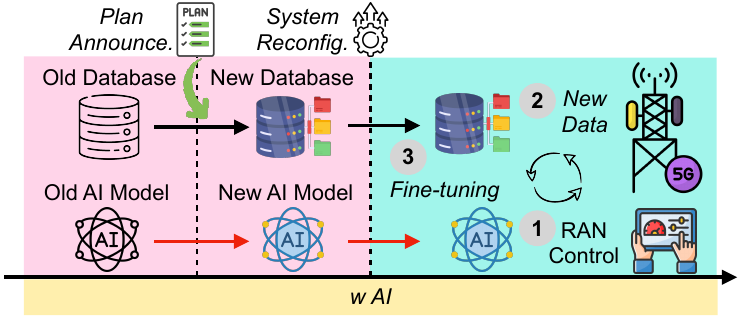}
    \label{fig:framework_b}}
    \end{minipage}
    }
    \caption{Illustration of AI development in O-RAN: (a) existing paradigms reactively retrain/adapt AI models after system reconfiguration; (b) \projectName proactively adapts AI models before system reconfiguration.}
    \label{fig:framework}
    \vspace{-10pt}
\end{figure}

The promising AI functionalities in O-RAN expose a brittle truth when confronted with flexible reconfigurability: \textbf{today’s AI models in O-RAN systems often break even under \textit{planned} network upgrades}.
For example, an AI model trained on last month's network data may struggle with this month's new configuration or tuned parameters,
often yielding dramatic performance degradations, following planned upgrades.
Such network reconfigurations or upgrades are necessary in routine operations and essential for AI innovations. For example, network operators need to regularly modify hardware and software components, upgrade radio units, change network policies, tune AI models, \etc.~
In practice, however, such necessary changes can invalidate previously well-trained AI models.
Our measurements on a live 5G O-RAN testbed underscore the severity of this problem (\S\ref{SubSec:Measurement}).
After a common cell-addition,
a pretrained Quality of Experience (QoE) predictor fails to anticipate sharp throughput fluctuations,
causing video streaming to suffer from a high frequency of playback stalls (Figure~\ref{fig:QoE_Prediction}).
Simultaneously, an anomaly detector is flooded with false alarms,
misclassifying benign inter-cell interference from the new cell as a malicious event (Figure~\ref{fig:AnomalyDetection_Motivation}).

The culprit is dramatic data drift caused by network reconfigurations and component upgrades: the reconfiguration introduces novel RAN interactions
and traffic dynamics that the AI model was never trained on.
We quantify this drift, finding that while most traffic patterns remain similar,
a critical 8\% of post-reconfiguration traffic shows a significant deviation (maximum cosine similarity < 0.7) from the original training data (Figure~\ref{fig:QoE_Drift}).
The model's performance collapses precisely on this drifted data,
with the QoE predictor's error spiking above 10 Mbps,
demonstrating that the AI model fails at the most critical moments
when it must understand the network's new operational logic.

Existing solutions (\eg, reactively collecting new data to perform model adaptation after deployment as illustrated in Figure~\ref{fig:framework_a})
are agonizingly slow and operationally untenable for cellular networks.
Our measurements show that adapting a QoE prediction model via
continuous learning takes 29 minutes to regain its original performance,
while an anomaly detection model requires 20 minutes (Figure~\ref{fig:AdaptTime}).
It took almost the same amount of time to train new models from scratch (31 and 21 minutes, respectively), indicating the data drift is too severe for simple fine-tuning. 
This prolonged period of suboptimal performance (effectively \textit{AI outage}) negates the core O-RAN promise of agility and flexibility: If every dynamic upgrade incurs a half-hour of degraded service and potential user complaints, the flexibility comes at too high a cost, 
rendering cellular operators reluctant to adopt AI models or deliver new AI functionalities via system upgrades.
Instead of reactively fixing AI after major network reconfigurations or upgrades,
\textit{can AI plan for such changes and adapt beforehand?}
We present \projectName,
a new framework for proactive O-RAN AI adaptation (Figure~\ref{fig:framework_b}).
Its design rests on three pillars (\S\ref{SystemDesignSection}): (1) At its core, \projectName builds a \textit{lightweight virtual O-RAN} that emulates the planned network changes
with sufficient fidelity to synthesize training data for the future scenarios ahead of time.
This differs from a heavyweight, ray-tracing-based physical-layer reconstruction \cite{iye2025open,polese2024colosseum}.
Instead, \projectName targets data- and control-level fidelity by focusing on the structural network behaviors and RAN unit interactions that drive post-reconfiguration data drift.
(2) \projectName uses a dedicated meta-learning strategy to intelligently augment the synthetic KPM data that helps the model generalize to
diverse operational conditions; 
(3) \projectName rapidly closes the remaining simulation-to-reality gap by prioritizing important real-world KPMs for incremental tuning.

Our extensive evaluation on a live 5G O-RAN testbed \cite{srsRAN}
shows that \projectName achieves near interruption-free AI services through reconfigurations,
\textit{reducing AI downtime by 85–94\% compared to reactive retraining baselines}.
This transforms the adaptation process from a lengthy recovery into a swift, final calibration.
For instance, \projectName prepares anomaly detection models that are ready in just 1.3–2.4
minutes and a resource allocation model that achieves near-Oracle performance immediately,
resolving minor discrepancies within four minutes of deployment.
From the moment of deployment,
\projectName-prepared models operate at full accuracy; 
our anomaly detector achieves 89.1–97.5\% accuracy from the start,
while the Vanilla model languishes at 30–50\%. 
This is made possible as our synthesized data faithfully reproduces critical network dynamics
of the new interaction patterns among reconfigured RAN units,
such as throughput fluctuations,
that are essential for robust AI but overlooked by benchmarks.
By enabling AI to evolve in lockstep with network updates,
\projectName closes a critical gap in today's O-RAN architecture.
In summary, this paper makes the following contributions:
\vspace{-2pt}
\begin{itemize}[leftmargin = 0.3cm]
\setlength{\itemsep}{2pt}
    \item For the first time, we draw attention to the problem of AI robustness in dynamic O-RAN reconfigurations, identifying its unique challenges and implications (\S\ref{BackgroundandMotivationSection}).
    \item We propose \projectName, a new framework that supports the evolution of AI functionalities prior to physical RAN reconfiguration. \projectName marks a paradigm shift from reactive remedy to proactive adaptation for AI functionalities in reconfigurable O-RAN (\S\ref{SystemDesignSection}).
    \item We design, implement, and evaluate \projectName on a realistic 5G RAN testbed across various use cases. 
    Results indicate that \projectName can deliver performant and robust AI functionalities and support diverse applications (\S\ref{ImplementationSection}-\S\ref{EvaluationSection}). 
\end{itemize}
\vspace{-2pt}
\noindent\textbf{Ethics}: This work does not raise any ethical issues.

\vspace{-3pt}
\section{Related Work}
\vspace{-2pt}

\noindent\textbf{AI functionalities in O-RAN.} AI for cellular networks has gained substantial attention in recent years \cite{mei2022realtime,ye2024dissecting,shafin2020artificial,mungari2025ran,aslan2024fair}. The rise of open and programmable control in O-RAN has further accelerated this trend, enabling developers to integrate AI/ML-driven mechanisms into the RAN for tasks such as network slicing \cite{chen2023channel,balasingam2024application,budhdev2021fsa}, traffic steering \cite{lacava2023programmable,ntassah2023xapp}, interference mitigation \cite{zumegen2024beamarmor,kilinc2022jade}, anomaly detection \cite{wen20245g,wen20246g,sun2024spotlight,nguyen2019gee}, and automated control \cite{abubakar2023energy,lozano2025kairos,foukas2021concordia,kalia2025towards,hidalgo2025aegisran}. 
The AI functionalities have been undeniably promising for smarter, more adaptive NextG cellular networks \cite{khan2023ai,AIRANAlliance}.

\noindent\textbf{Service degradation after RAN reconfigurations.} Early measurement studies of large operational networks \cite{mahimkar2010detecting,mahimkar2011rapid,mahimkar2013robust,li2021nationwide} revealed that configuration changes can induce subtle yet persistent performance degradations. To reduce user-visible impact, previous work focused on minimizing service disruption through carefully coordinated, network-wide upgrades \cite{xu2015magus,qureshi2017coordinating}. More recent efforts have further enhanced system robustness and achieve near–zero downtime during network updates \cite{ramanathan2022resiliency,khooi2025update,lazarev2023resilient,xing2023enabling}.

With the emergence of O-RAN, however, AI-driven control and optimization have become integral to RAN operation, introducing a new vulnerability: learned models themselves can degrade under frequent and flexible reconfigurations. Existing AI functionalities remain largely rigid and tightly coupled to static configurations, making them particularly sensitive to configuration shifts \cite{ananthanarayanan2025distributed}. 


\noindent\textbf{AI robustness against data drift.} Prior work \cite{amiri2023deep} and our measurements (\S\ref{SubSec:Measurement}) show that O-RAN reconfigurations can induce severe data drift that quickly invalidates deployed AI models. The standard remedy is to adapt models to new data distributions through continuous or transfer learning \cite{bhardwaj2022ekya,kemker2018measuring,weiss2016survey,wang2024comprehensive}, and recent studies have attempted to apply these techniques to cellular systems \cite{benzaid2024federated,mhatre2025transfer,nagib2023safe}. 
However, unlike conventional AI settings, cellular networks face a practical challenge: collecting real-world data is slow, resource‑intensive, and often requires lengthy over‑the‑air measurements \cite{ananthanarayanan2025distributed}. 
Motivated by recent advances in synthetic data generation \cite{gong2025data,chi2024rf,chen2023rf,nguyen2024digital,polese2024colosseum}, we explore whether AI models can learn ahead of time using synthetic traces tailored to emulate upcoming network reconfiguration, thereby improving robustness under dynamic O‑RAN reconfigurations.

\noindent
\textbf{AI-based network management and model adaptation.} Networking systems adopt AI for network management and model adaptation across diverse domains. In wide-area networks, AI-driven traffic engineering \cite{alqiam2024transferable, perry2023dote} aims to enhance resilience under dynamic traffic demands and topologies. In data center networks, xWeaver \cite{wang2018neural} leverages AI to adapt topology decisions to evolving traffic patterns. Similarly, cellular networks increasingly rely on AI models to execute complex operational tasks \cite{sun2024spotlight,ye2024dissecting}. However, AI in production networks suffers from severe performance degradation under planned or unplanned system changes, calling for efficient model adaptation mechanisms \cite{liu2021understanding}. To tackle this degradation, LEAF~\cite{liu2023leaf} characterizes concept drift in large-scale cellular infrastructure and mitigates accuracy drops through a combination of targeted retraining, data forgetting, and sample oversampling. Argus \cite{yan2012argus} continuously monitors service anomalies and adapts AI models on the fly to absorb data distribution variations. 
Unlike these works, \projectName specifically addresses data drift in dynamic O-RAN reconfigurations and proposes a proactive paradigm that adapts AI models using synthetic data.

\vspace{-0.2cm}
\section{Background and Motivation}
\label{BackgroundandMotivationSection}



\subsection{O-RAN and AI Functionalities}
\label{SubSec:Background}

\begin{figure}[t]
\setlength{\abovecaptionskip}{2pt}
        \subfigtopskip=-2pt
        \subfigcapskip=-2pt
\subfigure[5G architecture]{
\centering
\includegraphics[height = 3.3cm]{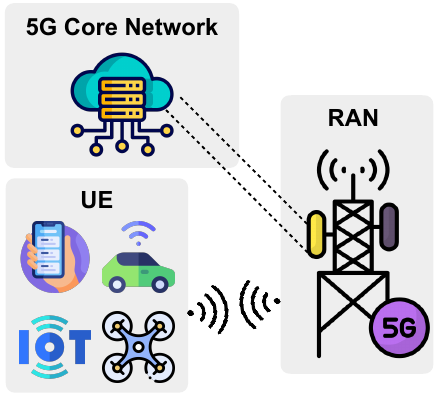}
\label{fig:background_a}
}
\hspace{0.1cm}
\subfigure[O-RAN architecture]{
\centering
\includegraphics[height = 3.3cm]{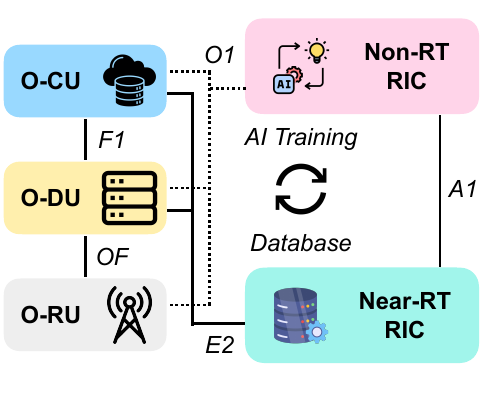}
\label{fig:background_b}
}
\caption{Background of 5G Open RAN: disaggregated RAN units integrated with AI/ML functionalities.}
\label{fig:background}
\vspace{-12pt}
\end{figure}

\textbf{5G Networks.} 
As illustrated in \figureautorefname{}~\ref{fig:background_a}, 5G
networks consist of three primary components.
\textit{User Equipment (UE)} includes devices such as smartphones, IoT devices,
and connected vehicles that access the network.
\textit{Radio Access Network (RAN)} provides wireless connectivity,
enabling communication between UEs and the network infrastructure. 
\textit{5G Core Network (5GC)} manages essential functions like user authentication and mobility management,
while also routing user data between the RAN and backhaul networks.

\noindent\textbf{O-RAN vs. Traditional RAN.} 
By decoupling traditional RAN hardware and standardizing interfaces, O-RAN fosters a multi-vendor ecosystem that promotes reconfigurability and innovation.
As illustrated in \figureautorefname{}~\ref{fig:background_b}, O-RAN disaggregates RAN functions into three logical components:  
O-Radio Unit (O-RU), O-Distributed Unit (O-DU), and O-Central Unit (O-CU).
\textit{O-RU} transmits and receives radio signals over the air interface, and implements lower PHY functions (\eg, FFT/iFFT).
\textit{O-DU} executes higher PHY functions (\eg, scrambling, modulation), MAC scheduling, and RLC operations.
\textit{O-CU} handles higher-layer functions (\eg, RRC signaling, mobility management, and PDCP processing). It orchestrates UE behavior, manages handovers,
and enforces QoS policies.
O-RAN further integrates RAN Intelligent Controller (RIC) to coordinate AI functionalities.
O-RAN collects and stores the Key Performance Metrics (KPMs) of RAN components.





\begin{figure}[t]
\setlength{\abovecaptionskip}{0pt}
        \subfigtopskip=-2pt
        \subfigcapskip=-2pt
\centering
\begin{minipage}[b]{0.235\textwidth}
\subfigure[Stall frequency]{
\centering
\includegraphics[height = 2.9cm]{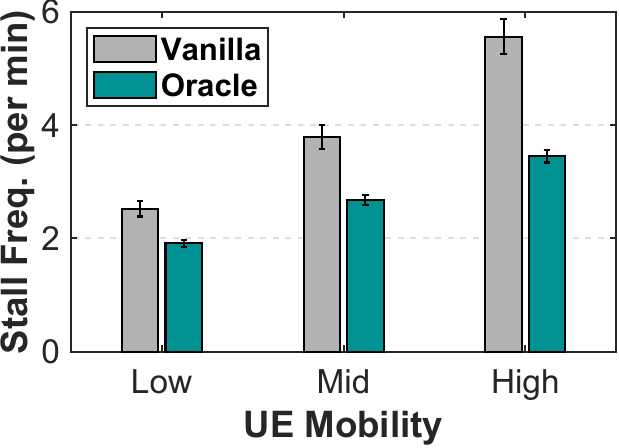}
\label{fig:QoE_Prediction_a}
}
\end{minipage}
\begin{minipage}[b]{0.235\textwidth}
\subfigure[Failure moment]{
\centering
\includegraphics[height = 2.9cm]{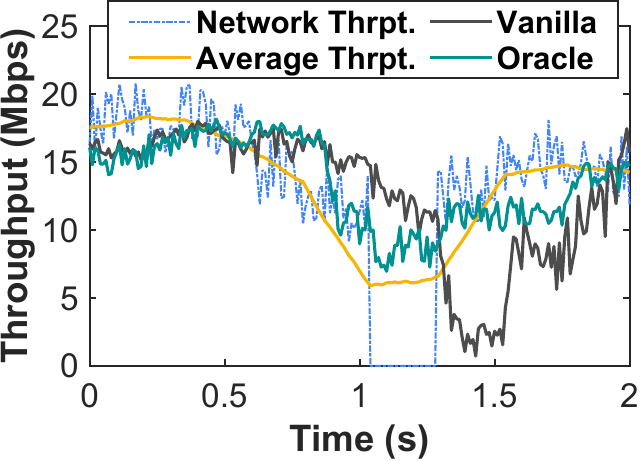}
\label{fig:QoE_Prediction_b}
}
\end{minipage}
\caption{QoE prediction after RAN reconfiguration: (a) Overall video streaming stalls; (b) Throughput fluctuation in one of the prediction failure moments.}
\label{fig:QoE_Prediction}
\vspace{-10pt}
\end{figure}

\begin{figure}
\setlength{\abovecaptionskip}{2pt}
    \centering
    \includegraphics[height = 2.9cm]{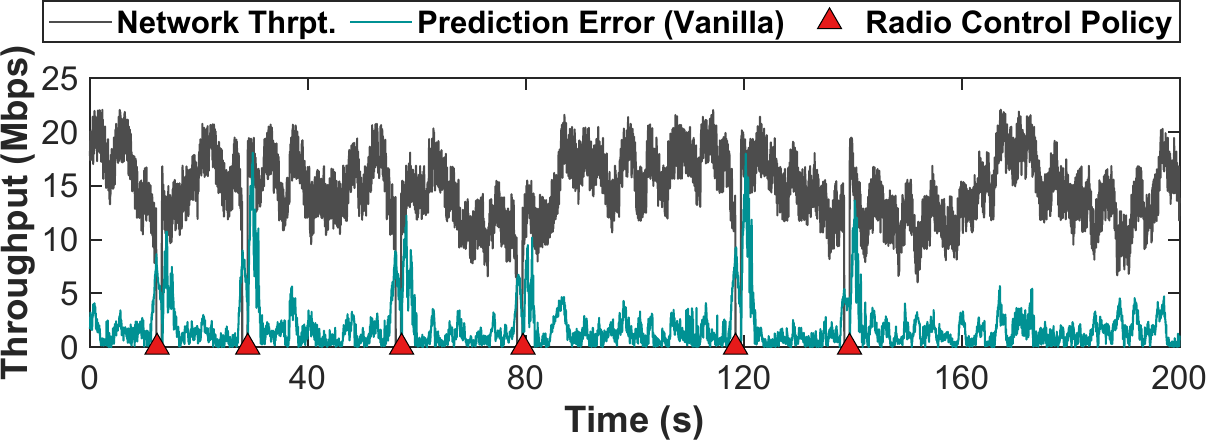}
    \caption{The failure of Vanilla QoE prediction model aligns with the execution of radio control policies.}
    \label{fig:QoE_Timeline}
    \vspace{-12pt}
\end{figure}

\noindent\textbf{Need for robust and interruption-free AI services.} AI services are fundamental to network control and optimization.
The flexible reconfigurability and stringent downtime requirements \cite{lazarev2023resilient,qureshi2017coordinating} pose an urgent demand for robust and interruption-free AI functionalities, particularly during the initial phase following RAN reconfiguration. 
In other words, AI functionalities are expected to maintain consistent performance with ideally zero service downtime after major reconfiguration as well as system upgrades.

\vspace{-0.2cm}
\subsection{RAN Reconfiguration and Data Drift}
\label{SubSec:Measurement}

To demonstrate the fragility of current AI approach and trained models even facing planned reconfiguration and system upgrades,
we conduct a series of preliminary studies (\S\ref{ImplementationSection}-\S\ref{EvaluationSection}). 
We measure the performance degradation of AI functionalities following routine reconfiguration operations and updates.
We test two standard AI functionalities with representative model architectures and training methods:




\vspace{-2pt}
\begin{itemize}[leftmargin = 0.3cm]
    \item \textbf{QoE prediction and serving} \cite{ye2024dissecting} predict QoE based on historical traffic patterns. 
    Accurate predictions enable real‑time, proactive adjustment of application and network parameters to maintain high user satisfaction, which is especially critical for latency‑sensitive and immersive applications like AR/VR and cloud gaming \cite{meng2022achieving,cheng2024grace,an2025tooth,meng2024hairpin,jia2025towards}.
    \item \textbf{Anomaly detection and localization} \cite{sun2024spotlight} adopt a distribution learning model to analyze RAN KPMs to detect system failures and irregularities. 
    Accurate and timely detections enable automated maintenance workflows and rapid recovery from network anomalies, thereby reducing operating expenses and enhancing reliability \cite{wen20245g,wen20246g,lakshmanan2021stealthy}.
\end{itemize}
\vspace{-2pt}

\noindent We tested various system reconfigurations including both hardware and software changes and upgrades.
To make our discussion concrete,
we start with one routine system upgrade, \ie, adding a cell tower for better coverage.


\begin{figure}
\setlength{\abovecaptionskip}{2pt}
    \centering
    \includegraphics[height = 2.9cm]{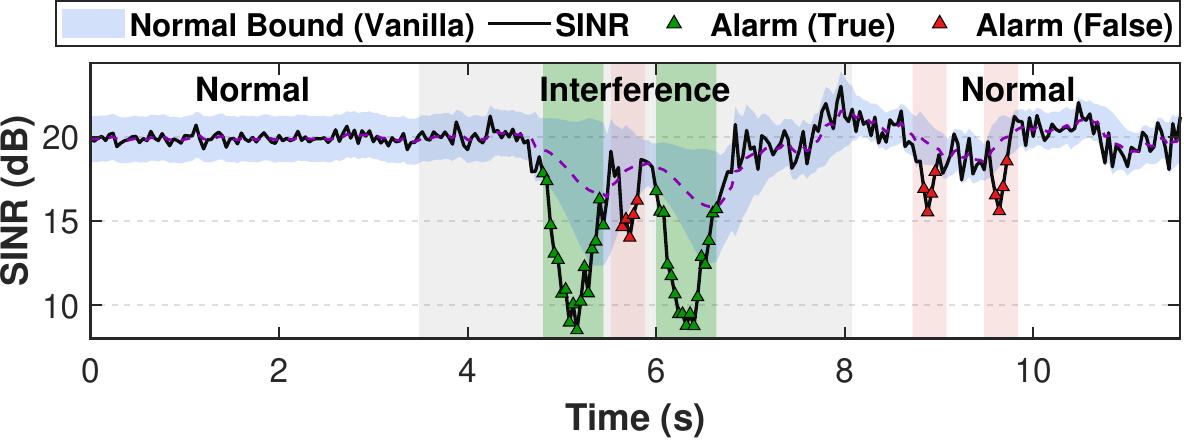}
    \caption{Anomaly detection after RAN reconfiguration: normal traffic frequently crosses the normal bound of Vanilla model, raising extensive false alarms.}
    \label{fig:AnomalyDetection_Motivation}
    \vspace{-15pt}
\end{figure}

\noindent\textbf{AI Service Interruption after Routine Updates.} 
After a reconfiguration,
video streaming quality degrades significantly when managed by the Vanilla QoE predictor (\ie, model before reconfiguration), exhibiting a high frequency of playback stalls that worsen with UE mobility, as shown in \figureautorefname~\ref{fig:QoE_Prediction_a}.
The root cause is that the current AI approach does not train the model to cope with the sharp data drift caused by major network reconfiguration.
As a result, the Vanilla model lags behind the actual throughput as shown in Figure~\ref{fig:QoE_Prediction_b}, leading to performance degradation and affect user experience.


By cross-checking these failures with RAN operation logs,
we discovered a notable pattern: the most failures for the
Vanilla model coincide with the updates to radio control policies, as shown in Figure~\ref{fig:QoE_Timeline}.
These newly integrated policies (\eg, possible handover to/from the new RAN unit) 
cause network dynamics that were unseen by the Vanilla model, rendering the model ineffective. 

Our measurements on anomaly detection unveil similar limitations of current AI approach and trained models. 
\figureautorefname~\ref{fig:AnomalyDetection_Motivation} illustrates 
that following a system reconfiguration, the Vanilla anomaly detection model 
frequently misclassifies benign network behavior (\eg, new patterns of inter-cell interference) as malicious anomalies. 
These subtle fluctuations are typically caused by the reconfigured
topology but rarely seen in the collected data.


In both cases, we observe that the AI models cannot stay effective and robust throughout major reconfigurations and system upgrades.
\noindent\textbf{Impact of RAN Reconfiguration on Data Distribution.} 
We further analyze the data distribution variation of network traffic induced by RAN reconfiguration. 
To quantify data distribution variations, we adopt a unique metric, \textit{maximum cosine similarity}, which assesses the affinity of each new KPM sample to the training set used by the Vanilla model.
Specifically, for each new KPM sample, we compute its cosine similarity with all training samples and select the maximum value as a proxy, representing the closest distance between the KPM sample and the training set \cite{shen2024outlier}. 
This approach can eliminate the influence of the inherently high variance of cellular traffic. To benchmark the impact of reconfiguration, we measure two 20-minute traffic segments: one before and one after the reconfiguration. We segment each trace into \SI{2}{s} intervals and compute their similarities to the training set.

\begin{figure}[t]
\setlength{\abovecaptionskip}{0pt}
        \subfigtopskip=-2pt
        \subfigcapskip=-2pt
\centering
\begin{minipage}[b]{0.235\textwidth}
\subfigure[Data similarity distribution]{
\centering
\includegraphics[height = 2.9cm]{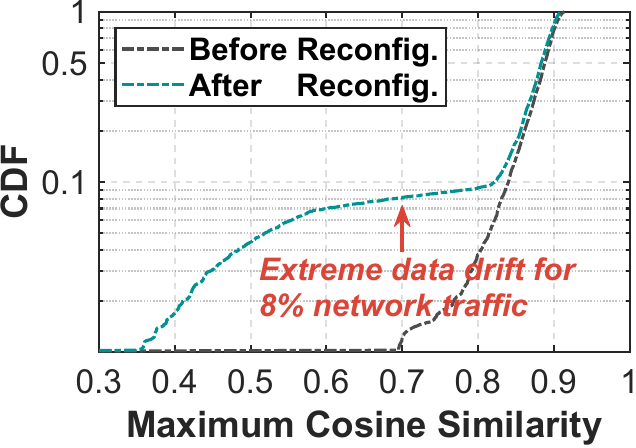}
\label{fig:QoE_Drift_a}
}
\end{minipage}
\begin{minipage}[b]{0.235\textwidth}
\subfigure[Accuracy vs. data drift]{
\centering
\includegraphics[height = 2.9cm]{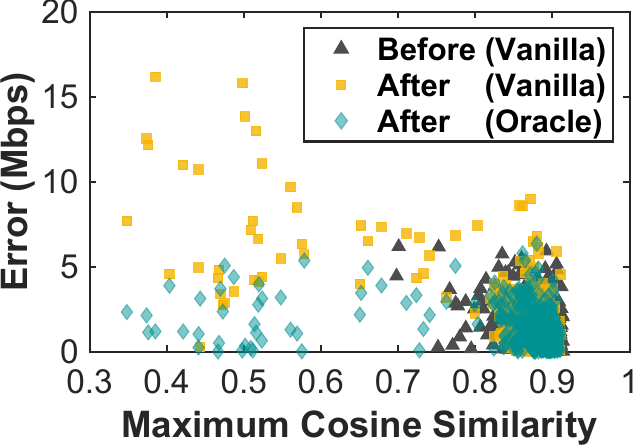}
\label{fig:QoE_Drift_b}
}
\end{minipage}
\caption{Data drift from RAN reconfiguration: (a) network traffic experiences large data distribution drifts; (b) Vanilla model degrades sharply on the drifted data.}
\label{fig:QoE_Drift}
\vspace{-8pt}
\end{figure}

Our empirical analysis shows that, although overall data similarity remains broadly stable after reconfiguration, a notable portion of network traffic diverges significantly from the training distribution (see \figureautorefname~\ref{fig:QoE_Drift_a}). These low‑similarity segments are unrepresented in the original training set and invalidate the Vanilla model. As illustrated in \figureautorefname~\ref{fig:QoE_Drift_b}, throughput predictions become highly unreliable in these low‑affinity regions. These results indicate that RAN reconfigurations introduce substantial data drift, undermining the robustness of pretrained, static O‑RAN AI models.

\noindent\textbf{Limitations of Existing Solutions.}
Data drift is a well-known challenge in AI.
Moderate data drift can be addressed  
by continuous learning \cite{bhardwaj2022ekya,kemker2018measuring,weiss2016survey,wang2024comprehensive}. However, this reactive approach is untenable in O-RAN due to two unique challenges:
1) \textit{Lack of post-reconfiguration data.} 
Adapting AI models requires collecting new training data in target domains (\ie,
after reconfiguration),
which is unavailable until the reconfiguration is enforced.
2) \textit{Stringent downtime requirement.}
Cellular networks impose a stringent service downtime requirement \cite{lazarev2023resilient,
qureshi2017coordinating}. Continuous learning \cite{wang2024comprehensive} generally takes a long time to collect sufficient 
post-reconfiguration data for fine-tuning models.
During this process, the AI models inevitably suffer from drastic performance degradation, rendering cellular operators reluctant to adopt AI models or deliver new AI functionalities via system upgrades.

\begin{figure}[t]
\setlength{\abovecaptionskip}{4pt}
        \subfigtopskip=-2pt
        \subfigcapskip=-2pt
\centering
\begin{minipage}[b]{0.24\textwidth}
\subfigure[QoE prediction]{
\centering
\includegraphics[height = 2.9cm]{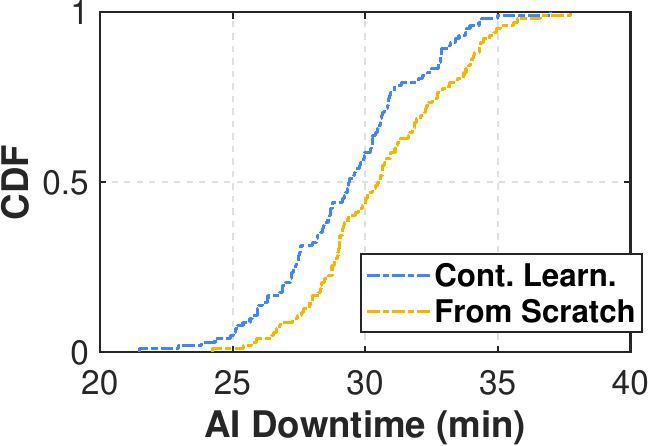}
\label{fig:AdaptTime_a}
}
\end{minipage}
\begin{minipage}[b]{0.23\textwidth}
\subfigure[Anomaly detection]{
\centering
\includegraphics[height = 2.9cm]{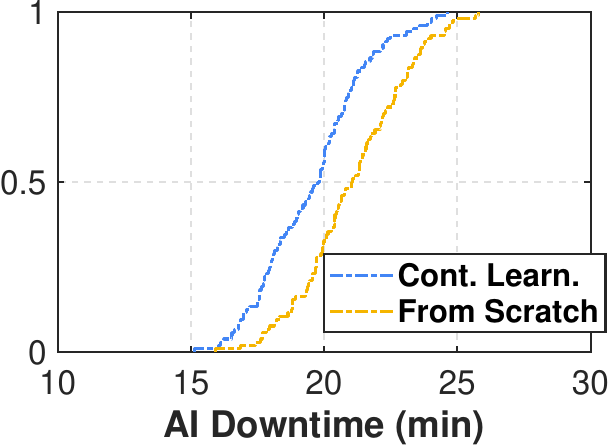}
\label{fig:AdaptTime_b}
}
\end{minipage}
\vspace{-5pt}
\caption{End-to-end adaptation time of different AI functionalities, during which the reconfigured RAN suffers from suboptimal performance.}
\label{fig:AdaptTime}
\vspace{-15pt}
\end{figure}

To quantify the \textit{reconfiguration tax}, we measure the time required to adapt the models.  
Specifically, we consider two adaptation approaches:
1) updating the Vanilla models via continuous learning \cite{tong2025continual}, and 2) collecting new KPM traces after reconfiguration and training new models from scratch. 
We measure the \textit{AI downtime}, which is defined as the duration from the activation of RAN reconfiguration till the adapted AI model achieves 95\% of its best performance (denoted as Oracle).
The two models require 29 and 20 minutes for continuous learning, respectively (\figureautorefname~\ref{fig:AdaptTime}).
Notably, due to the time-consuming data collection and continuous learning required to handle extreme data drift, the adaptation efforts are almost the same as developing a new model from scratch, which takes about 31 and 21 minutes, respectively. These adaptation methods lead to extended AI service downtime and prolonged periods of suboptimal network performance, negating O-RAN's flexible reconfigurability.

In summary, we have two key observations: 

\vspace{-2pt}
\begin{itemize}[leftmargin = 0.3cm]
    \item Major reconfiguration and system upgrades introduce extreme data drift, obsoleting AI functionalities developed on legacy data.
    \item Existing continuous learning or fine-tuning solutions lead to unacceptable service downtime with performance degradation.
\end{itemize}
\vspace{-2pt}

\vspace{-0.2cm}
\section{\projectName Design}
\label{SystemDesignSection}

\begin{figure}
\setlength{\abovecaptionskip}{4pt}
    \centering
    \includegraphics[width=0.45\textwidth]{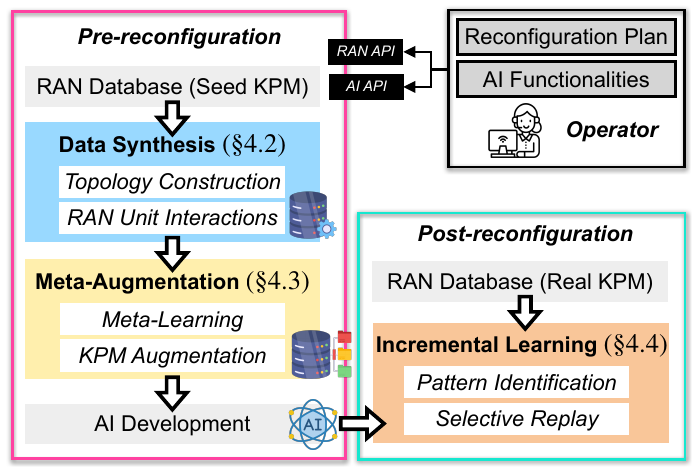}
    \vspace{-5pt}
    \caption{System overview of \projectName.}
    \label{fig:systemArchitecture}
    \vspace{-12pt}
\end{figure}

\subsection{System Overview}

\textit{Is it possible to develop robust AI models that can look ahead and accommodate RAN system reconfiguration without service interruption?} 
A fundamental obstacle is that AI adaptation requires training data, yet real post‑reconfiguration data is unavailable prior to deployment, and collecting it afterward is prohibitively slow.
To address this gap, we propose \projectName, a novel framework that proactively adapts AI models ahead of RAN reconfigurations using synthetic data. 
By enabling AI adaptation before reconfiguration, \projectName mitigates the data drift that could break the deployed models during major system upgrades.

However, effective data synthesis and AI adaptation for dynamic O-RAN reconfigurations face three key challenges: 
\textit{(i) Capturing System Changes.} O-RAN introduces disaggregated units and programmable control policies with highly flexible and dynamic reconfigurations. These changes alter traffic patterns and even system topology, making it difficult to synthesize data that faithfully reflects underlying network behaviors. 
\textit{(ii) Ensuring Data Generalization for AI Robustness.} 
To adapt the AI models with sufficient robustness, synthetic data must comprehensively cover diverse operational scenarios and network dynamics. 
\textit{(iii) Bridging the Reality Gap.} Even the best synthetic data cannot fully replicate real-world nuances such as environmental dynamics. Thus, post-reconfiguration, we must efficiently use real data to close this gap as quickly as possible.

\figureautorefname~\ref{fig:systemArchitecture} illustrates the overview of \projectName. To faithfully emulate O‑RAN reconfigurations, \projectName builds a virtual O-RAN that abstracts disaggregated RAN units, their interactions, and control policies (\S\ref{SubSec:DataSynthesis}). Driven by real network traces and expert knowledge, this virtual O‑RAN emulates diverse reconfiguration changes while preserving essential system behaviors.
To improve the AI robustness, \projectName augments the synthesized data from virtual O-RAN with rich operational scenarios and network dynamics (\S\ref{SubSec:DataAugmentation}). A modified meta‑learning module \cite{finn2017model} learns how to augment O‑RAN KPM traces and continuously incorporates synthesized data from unseen configurations into a local database, enabling progressive refinement and improved generalization.
Finally, to rapidly close the simulation-to-reality gap after deployment, \projectName detects biased traffic patterns and assigns priority weights for selective replay, enabling efficient incremental training using real KPM data (\S\ref{SubSec:IncrementalLearning}).
When operators plan an O-RAN reconfiguration, they notify \projectName of the target RAN configuration and desired AI functionalities via RAN and AI APIs; \projectName then synthesizes and augments KPM data to adapt the AI models, which are subsequently refined online after deployment to ensure accuracy under live network conditions.

\vspace{-0.2cm}
\subsection{Building Virtual O-RAN}
\label{SubSec:DataSynthesis}

\subsubsection{Insight \& Solution.} 
\begin{figure}
\setlength{\abovecaptionskip}{4pt}
    \centering
    \includegraphics[width=0.45\textwidth]{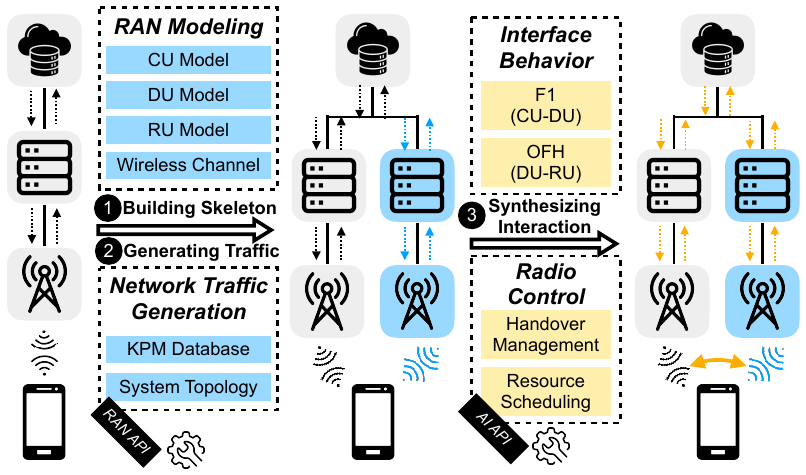}
    \caption{Virtual O-RAN construction based on real traces and abstracted RAN units.}
    \label{fig:dataSynthesis}
    \vspace{-15pt}
\end{figure}

Proactive AI adaptation requires future state data before a physical reconfiguration occurs. To synthesize effective data for AI adaptation, our design is grounded on three core principles: (1) \textit{Data and control fidelity over PHY exactness.} Instead of simulating every waveform, we abstract radio units and model their interactions at the control and scheduling layers.
This captures the system-level variations that dominate KPM drift and AI performance.
(2) \textit{Interface‑ and policy‑awareness.} We explicitly emulate open interfaces and their contention dynamics, along with operator-defined radio control policies. (3) \textit{Trace‑driven realism.} We bootstrap the synthesis with real KPM traces from the current RAN, then perturb them to reflect new topology, mobility, and interference. As illustrated in \figureautorefname{}~\ref{fig:dataSynthesis}, our design follows three stages: building a static skeleton of the target O-RAN configuration, generating O-RAN traffic, and synthesizing O-RAN unit interactions. 
The result is a \textbf{lightweight virtual O-RAN} -- a trace-driven emulator that generates post-reconfiguration KPM data for downstream AI adaptation.  
It departs from PHY‑exact digital twins \cite{iye2025open,polese2024colosseum}, and instead captures system‑level variations that ultimately reshape KPM traces after reconfiguration.


\vspace{-2pt}
\subsubsection{Workflow \& Input/Output Schema.}
The virtual O-RAN is a trace‑driven emulator built from softwarized RAN units. It takes seed KPM traces as input and allows planned reconfigurations to be applied to the software units. In addition, we explicitly model external factors that affect KPM distribution (\eg, RU placement, wireless channel, O-RAN interface contention) to align the emulator with real‑world physical variations. The following control knobs support diverse RAN reconfigurations (detailed in Table~\ref{tab:control_knobs}): (1) \textit{Topology}, which represents the target O-RAN as a directed graph. The nodes consist of abstracted CU, DU, and RU, while the edges represent the F1 (CU–DU) and Open Fronthaul (DU–RU) interfaces. The adjacency encodes hierarchy and enables per-interface latency/bandwidth effects to propagate into synthesized KPMs; (2) \textit{Cell configuration}, which defines cell identities and adjacent cell topologies used for tracking handovers; (3) \textit{RAN unit configuration}, which modifies softwarized RAN (RU, DU, CU) variables; (4) \textit{xApp}, which executes programmable control logic. For example, cell addition can be emulated by changing the topology, cell configurations, and RU variables; PRB scheduling or handover policy upgrades can be emulated by modifying specific control logic in the xApp and DU/CU variables.

\vspace{-2pt}
\subsubsection{Building O-RAN Skeleton.} 
To emulate a reconfigured O-RAN system, we first build its structural skeleton by initializing software RAN units at each layer within the input RAN topology. Rather than modeling the full protocol stack, we utilize functional abstractions that capture the essential RAN unit behaviors that drive KPM drift. In our design, RU represents a lightweight PHY-layer abstraction, handling signal propagation by executing commands from the upper layer and reporting PHY KPMs; DU serves as a control and buffering entity, and its functions include aggregating metrics from lower layers and performing basic scheduling; CU operates as a policy-driven controller, tasked with managing UE state and enforcing radio resource control. This layered skeleton (\ding{182} in Figure~\ref{fig:dataSynthesis}) allows flexible construction of a target O-RAN system with various radio unit abstractions forming different topologies.



\vspace{-2pt}
\subsubsection{Generating O-RAN Traffic.} 
We then generate synthetic KPM traces that activate the static O-RAN skeleton into a live virtual system.
\textbf{(1) KPM-driven bootstrapping.} 
We seed the virtual O-RAN with a KPM database that spans radio resources, UE channel quality, and QoS metrics (detailed in \tableautorefname{}~\ref{tab:KPM_database}). 
The KPMs are grounded on the current \textit{real} RAN operational states and provide the statistical priors for our synthesis. 
\textbf{(2) Traffic reconstruction and generation.}
We generate O-RAN traffic with a bifurcated strategy. For unchanged RAN units, we directly reconstruct traffic by replaying real-world KPM traces to maintain absolute baseline fidelity.
For newly added or reconfigured units, we reuse behavioral priors from real traces but adjust them based on the new topology and network dynamics. 
This ensures realistic physical coupling (\eg, inter-cell interference) and produces plausible KPMs that reflect the target topology.
\textbf{(3) PHY consistency via wireless modeling.}
We integrate a lightweight wireless channel model as an important external factor that links the software emulator to real-world PHY impacts of changed RUs, ensuring the generated KPMs remain physically consistent.
Instead of complex ray-tracing \cite{iye2025open}, we account for both mobility-induced and stochastic channel dynamics \cite{hlawatsch2011wireless}:
\begin{equation}
\setlength{\abovedisplayskip}{2pt}
\setlength{\belowdisplayskip}{2pt}
    \text{SINR}(t) = P_{\text{tx}} - PL(d(t)) + X_\sigma + F(t) - N_0
\end{equation}
with path loss $PL(d(t))$, shadowing $X_\sigma$, and fading $F(t)$; mobility and environment enter via $d(t)$ and $F(t)$, respectively (detailed in \S\ref{subsec:wirelessChannelDetails}).
To map input traces and configurations to the wireless channel model, RANPilot executes a trace-driven parameter tuning procedure:
Specifically, we tune SINR parameters to match seed KPM traces, and retain SINR environment parameters for the reconfigured RU while adjusting distance/mobility parameters to simulate PHY changes.
Overall, this real-KPM-driven generation (\ding{183} in Figure~\ref{fig:dataSynthesis}) outputs basic synthetic KPM traces that the AI models and RAN controllers can operate on.

\vspace{-2pt}
\subsubsection{Synthesizing O-RAN Unit Interactions.}
We next synthesize the essential RAN unit interactions that reshape the network behaviors and drive data drift.
\textbf{(1) Open-interface behaviors.} 
While software RAN units can be directly updated within the emulator, the open-interface behaviors that affect KPM distribution must be considered.
Specifically, we integrate an Open Fronthaul (OFH) interface contention model as an important external factor that replays O-RAN interface behavior. When multiple RUs share a constrained OFH, concurrent packet arrivals can lead to collisions \cite{lin20255g,sun2024spotlight}.
To map input traces and configurations to the OFH contention model, we calculate the packet loss probability $P_{loss}$ as a function of the aggregate fronthaul throughput $S$, derived by the average throughput per RU times the number of contending RUs.
As $S$ approaches the interface capacity $C$, the probability of collision increases: $P_{loss}(S) = 1 - e^{-\lambda(S/C)}$. We use $\lambda=2$, calibrated from empirical measurements.
These collisions trigger higher-layer retransmissions, which we derive into observable KPM impacts: increased latency due to backoff timers and a proportional degradation in effective UE throughput. 
\textbf{(2) Radio control policies.}
O‑RAN decouples control policies through programmable xApps \cite{polese2023understanding,ORANAlliance}. 
These customizable policies \cite{handoverHeterogeneity} are not mere configuration details. They act as first-order drivers of data drift during reconfigurations and reshape traffic patterns. 
This AI-in-the-loop emulation approach enables the model to approximate the empirical mapping between policy updates and resulting network behaviors. 
For AI models that do not directly affect network policies or control loops, the new models are reconfigured in virtual O-RAN for emulation and model adaptation.
The interaction synthesis (\ding{184} in Figure~\ref{fig:dataSynthesis}) produces essential but underrepresented network behaviors beyond the Vanilla KPM database, mitigating data drift in the post-reconfiguration state.

\vspace{-0.2cm}
\subsection{Harnessing Network Dynamics}
\label{SubSec:DataAugmentation}

\subsubsection{Insight \& Solution.} 
\begin{figure}
\setlength{\abovecaptionskip}{4pt}
    \centering
    \includegraphics[width=0.45\textwidth]{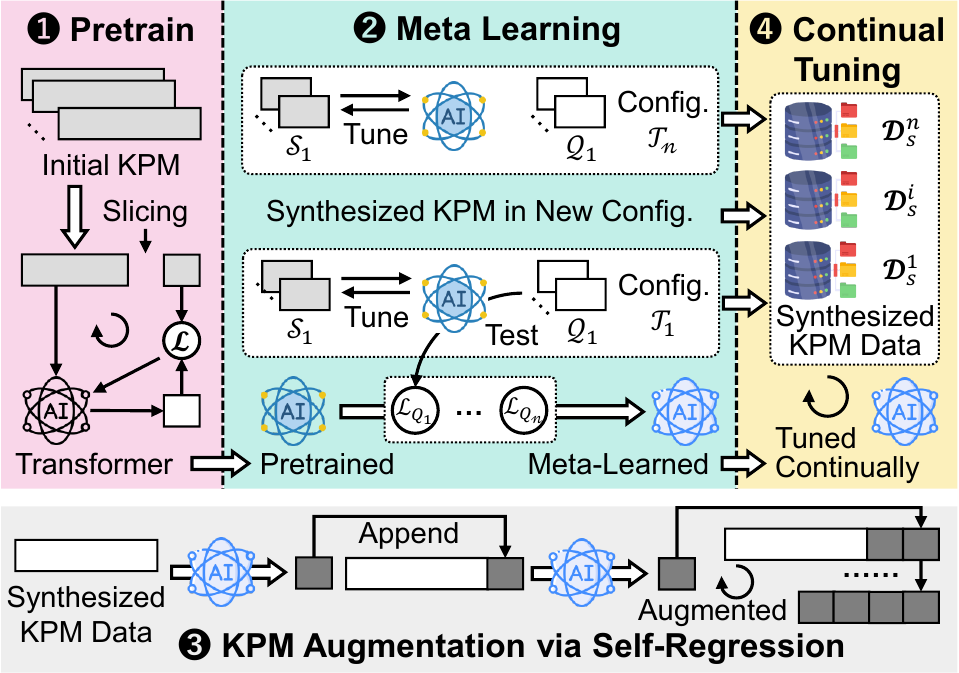}
    \caption{Learning-to-augment paradigm to harness network dynamics under evolving configurations.}
    \label{fig:continuous_meta_augmentation}
    \vspace{-15pt}
\end{figure}

The virtual O-RAN can effectively generate target‑state KPM traces for new RAN configurations, but network dynamics and cross‑scenario variability can be under-represented. Moreover, the limited quantity of synthetic KPM traces is insufficient to train robust AI models and can lead to overfitting. 
To address these issues, augmenting the synthetic KPM data is a promising approach to improving both quantity and quality.
Unfortunately, hand‑crafted augmentation (\eg, random jittering or resampling) distorts temporal dependencies and cross‑metric correlations that downstream AI relies on. 
To tackle this problem, we propose a \textit{continuous meta-augmentation} paradigm that first teaches AI models to learn how to augment KPM data and continuously evolve by dynamically maintaining a configuration database as operators plan additional reconfigurations.
\figureautorefname{}~\ref{fig:continuous_meta_augmentation} illustrates the workflow of our \textit{continuous meta-augmentation} paradigm. Given the synthetic KPM data (\S~\ref{SubSec:DataSynthesis}), the paradigm generates an augmented KPM dataset with enhanced quantity and quality through four steps: \textbf{pretraining} that captures network dynamics, \textbf{meta-learning} that learns how to augment, \textbf{KPM augmentation}, and \textbf{continual tuning} in response to emerging O-RAN reconfigurations.
This approach captures the intricate spatio-temporal dependencies inherent in multi-dimensional KPM data.

\vspace{-2pt}
\subsubsection{Capturing Network Dynamics.}
\label{sec:pretraining}
We pretrain a lightweight, six-layer Transformer \cite{vaswani2017attention} on synthetic KPM data to serve as a foundation model. 
Given a KPM sample, this model can predict subsequent KPM segments, which capture diverse network dynamics that are often deeply buried within KPM data.
Specifically, we first segment each KPM sample $\mathcal{K}$ into a set of slices $\mathcal{K}_s=\{k_1,k_2,\cdots\}$ by truncating the data at varying indices. The KPM data following each slicing point serves as the ground-truth label. We then feed $\mathcal{K}_s$ into the model to predict the next KPM trace and train the model by minimizing the loss between the predicted KPM and the ground truth. This method has two main advantages: (1) the entire training process is autoregressive and self-supervised, eliminating the need for manual labeling; (2) by generating multiple slices from varying truncation points, the model can effectively learn diverse network dynamics even from a limited quantity of synthetic KPM data.

\vspace{-2pt}
\subsubsection{Learning to Augment.}
We redirect the pretrained model $\mathcal{M}$ to learn how to augment KPM traces across diverse RAN configurations via meta-learning \cite{finn2017model}. As new RAN configurations emerge, the synthesized KPM data is stored in a progressively updated database $\mathcal{D}$. For each configuration $\mathcal{T}_i$ in $\mathcal{D}$, we split the KPM traces into a support set $\mathcal{S}_i$ and a query set $\mathcal{Q}_i$. We then perform a meta-update: a copy of $\mathcal{M}$ is trained on $\mathcal{S}_i$ and evaluated on $\mathcal{Q}_i$. The evaluation losses across all configurations $\{\mathcal{T}_1,\mathcal{T}_2, \cdots\}$ are summed to update the original $\mathcal{M}$. This alternating optimization process is repeated for several epochs until $\mathcal{M}$ converges. Such a meta-trained model $\hat{\mathcal{M}}$ can effectively generalize its KPM augmentation capabilities across various configurations.

\vspace{-2pt}
\subsubsection{KPM Augmentation.}
To perform augmentation, we segment KPM traces from each configuration $\mathcal{T}_i$ into multiple slices and leverage the meta-trained model $\hat{\mathcal{M}}$ to generate subsequent KPM traces in an autoregressive manner (\S~\ref{sec:pretraining}). The augmented KPM data $\mathcal{D}^a=\{\mathcal{D}_1, \mathcal{D}_2, \cdots\}$, along with the sliced synthetic data $\mathcal{K}_s$, are used to retrain the O-RAN AI models prior to physical RAN reconfigurations.

\vspace{-2pt}
\subsubsection{Continual Tuning.}
Unlike conventional meta-learning methods that rely on static datasets, we implement a dynamic, continuous meta-learning based on the progressively maintained database $\mathcal{D}$.
An intuitive approach would be to continuously fine-tune the augmentation model $\hat{\mathcal{M}}$ using the entire $\mathcal{D}$ as new RAN configurations arrive. Nonetheless, we observe that this strategy leads to \textit{intransigence}. As the database expands, early-arriving configurations are repeatedly sampled during training, causing the model to overfit to historical data and overshadow recent samples. Consequently, the model exhibits suboptimal adaptability to new RAN configurations. To mitigate this, we adopt a linear decay weighting scheme during continual tuning. Specifically, during loss calculation, we assign a decay weight of $1/n$ to each RAN configuration, where $n$ represents the cumulative number of times that configuration has been used for tuning. 
We select a $1/n$ decay rate, as faster-decaying functions penalize historical data too aggressively and can cause catastrophic forgetting on early RAN states. The harmonic $1/n$ decay satisfies standard stochastic approximation conditions and allows the model to continuously adapt to emerging configurations while retaining historical network behaviors.
Algorithm~\ref{alg:continuous_meta_augmentation} shows the detailed process.

\vspace{-0.2cm}
\subsection{Closing Simulation-to-Reality Gap}
\label{SubSec:IncrementalLearning}

\subsubsection{Insight \& Solution}
Admittedly, the virtual O-RAN and meta-augmentation may not be able to cover all possible dynamics in real diverse scenarios. 
The real‑world nuances could arise from local implementation details, user behavior, and environmental factors that cannot be fully captured. 
An intuitive solution is to adopt incremental learning \cite{masana2022class} that replays useful data to reinforce AI models. Unfortunately, traditional incremental learning cannot be applied in O‑RAN because useful KPM patterns (capturing data drift) are sparse and naive replay is dominated by typical KPM patterns, which could lead to catastrophic forgetting.
To remedy this, we apply a new incremental learning strategy with selective replay that detects useful, drift‑carrying patterns on the fly and assigns them higher priority during incremental tuning to prevent forgetting. This allows \projectName to adapt to real-world data and close the simulation-to-reality gap. 

\vspace{-2pt}
\subsubsection{Module Breakdown.}
\label{sec:hyperparameters}
\textbf{(1) Useful pattern identification.} We first define the \textit{dominant pattern} $K_d$ as the KPM traces that produce normal inference results from O-RAN AI models (\eg, KPM that does not trigger alarms in anomaly detection). Then, we adopt a sliding window of length $l$ to sequentially slice the incoming KPM traces. For each KPM slice,
we quantify its data distribution affinity to the \textit{dominant pattern} by calculating the Kullback-Leibler divergence (KLD) \cite{lin2002divergence} between the two KPM traces.
A larger KLD value indicates a greater disparity in data distributions between the \textit{dominant pattern} and the current KPM slice. The \textit{useful pattern} is then identified as the KPM slice whose KLD value exceeds a predefined threshold $\tau$. 
\textbf{(2) Selective replay based on weighted priority} assigns different priority weights to distinct KPM slices for tuning. The underlying rationale is that, as \textit{useful patterns} in incoming KPM traces typically yield higher KLD values, we assign them higher priority for incremental training, while assigning lower priority for \textit{dominant patterns}. Specifically, at regular intervals, we collect $t$ incoming KPM slices $K_1, K_2, \cdots, K_t$. The priority weight $p_i$ of $K_i$ is then determined to be equal to its computed KLD value. Next, the top-$b$ KPM slices with the highest priority are selected and stacked into a KPM buffer of size $b$. Once the buffer is full, we incrementally train the O-RAN AI model using the priority weight of each KPM slice during loss calculation.
Note that we set $t\gg b$ to ensure that the collected KPM slices encompass a sufficient variety of traffic patterns, with the buffer containing more \textit{useful patterns} than \textit{dominant patterns}. As such, both types of KPM patterns will be replayed for incremental training.

\vspace{-0.2cm}
\section{Implementation}
\label{ImplementationSection}

\begin{figure}[t]
\vspace{-2pt}
\setlength{\abovecaptionskip}{4pt}
\centerline{\includegraphics[height = 4cm]{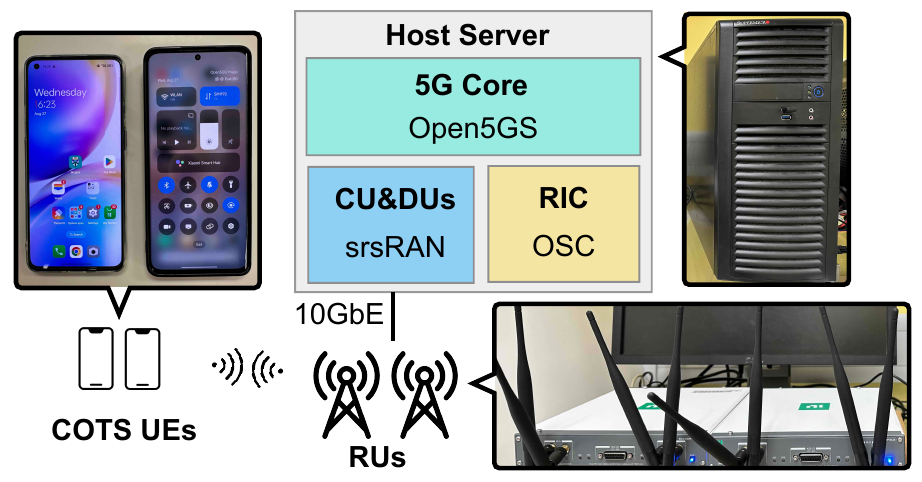}}
\vspace{-5pt}
\caption{5G testbed with COTS UEs and SDRs.}
\label{fig:implementation}
\vspace{-10pt}
\end{figure}

\noindent\textbf{Testbed Setting.} We implement \projectName on the open-source \texttt{srsRAN} stack~\cite{srsRAN}, utilizing software-defined radios (SDRs) and commercial UEs, as illustrated in~\figureautorefname~\ref{fig:implementation}. The UEs include OnePlus 8T and Xiaomi 13 Pro smartphones equipped with programmable SIM/USIM cards (sysmoISIM-SJA2~\cite{sysmoISIM}). The RAN stack is hosted on a workstation running Ubuntu 22.04.1 LTS, equipped with an Intel Xeon(R) E5-2620 v4 CPU (32 GB RAM) and an NVIDIA RTX 4090 GPU (24 GB VRAM). Two USRP X310 SDRs serve as the RUs. We deploy Open5GS~\cite{Open5GS} on the same workstation to provide core network functionality. For control-plane support and AI integration, we utilize the OSC RIC framework~\cite{OSCRIC}.
The KPM augmentation model contains six Transformer blocks with a batch size of four.
Pre-training the KPM augmentation model on a 1-hour KPM database
takes approximately 18 minutes, and continually tuning it on a 10-minute KPM database after each reconfiguration takes around 3 minutes. The GPU VRAN consumption is roughly \SI{1.5}{GB}. 

\noindent
\textbf{Integration with O-RAN Architecture.} 
\projectName enforces strict compute isolation to protect time-sensitive control loops in O-RAN operation.
Only the lightweight KPM collection module is implemented as an xApp within the Near-RT RIC, while the heavy data synthesis and augmentation modules are deployed entirely in the Non-RT RIC.
In addition to the basic KPMs provided by \texttt{srsRAN}, we integrate custom hooks \cite{foukas2023taking} into the stack to capture more granular and informative KPMs every \SI{40}{ms}.  
Operators configure \projectName via open APIs, enabling it to leverage the realistic database to emulate reconfigured RAN and develop diverse AI functionalities. Once the new configuration is online, the updated AI models are immediately activated to provide robust network control and optimization.

 
\vspace{-0.2cm}
\section{Evaluation}
\label{EvaluationSection}

The key takeaways of our evaluations are:

\vspace{-1pt}
\begin{itemize}[leftmargin = 0.3cm]
\setlength{\itemsep}{2pt}
    \item \projectName synthesizes KPM traces that faithfully capture system dynamics and data drifts across various reconfiguration types and deployment environments (\S\ref{subsec:evaluation_fidelity}).
    \item Across three downstream AI use cases, \projectName substantially improves AI robustness under diverse O‑RAN reconfigurations, enabling superior plug‑and‑play performance and dramatically reducing AI downtime against baselines (\S\ref{subsec:evaluation_generalization}).
    \item \projectName incurs modest and practical pre‑deployment overhead across a wide range of reconfiguration scenarios and system topologies (\S\ref{subsec:evaluation_overhead}).
\end{itemize}
\vspace{-1pt}

\vspace{-0.2cm}
\subsection{Methodology}

\begin{figure*}[t]
\subfigure[Indoor - Cell Add.]{
\begin{minipage}[t]{0.15\textwidth}
\centerline{\includegraphics[height = 2.7cm]{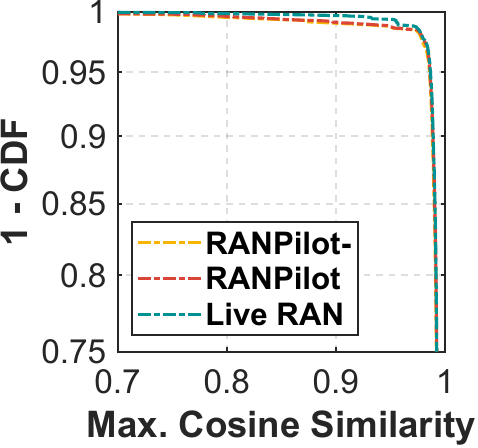}}
\end{minipage}
}
\subfigure[Indoor - HO Policy]{
\begin{minipage}[t]{0.15\textwidth}
\centerline{\includegraphics[height = 2.7cm]{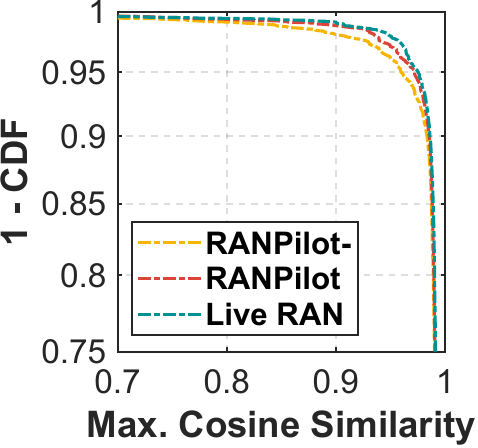}}
\end{minipage}
}
\subfigure[Outdoor - Cell Add.]{
\begin{minipage}[t]{0.15\textwidth}
\centerline{\includegraphics[height = 2.7cm]{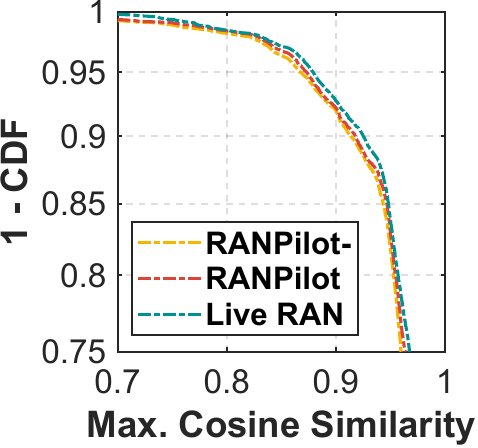}}
\end{minipage}
}
\subfigure[Outdoor - PHY Change]{
\begin{minipage}[t]{0.15\textwidth}
\centerline{\includegraphics[height = 2.7cm]{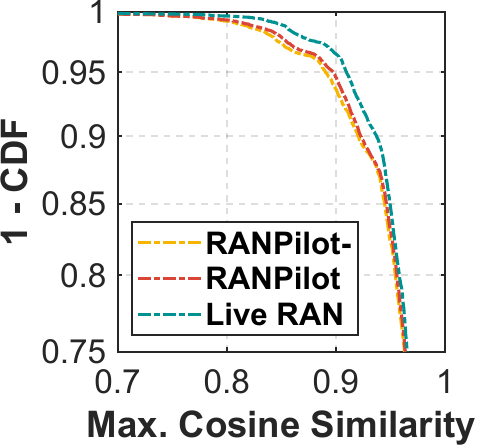}}
\end{minipage}
}
\subfigure[Outdoor - PHY + Sched.]{
\begin{minipage}[t]{0.15\textwidth}
\centerline{\includegraphics[height = 2.7cm]{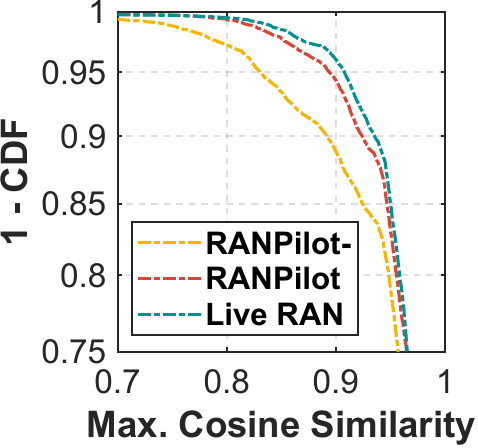}}
\end{minipage}
}
\subfigure[Outdoor - New Site]{
\begin{minipage}[t]{0.15\textwidth}
\centerline{\includegraphics[height = 2.7cm]{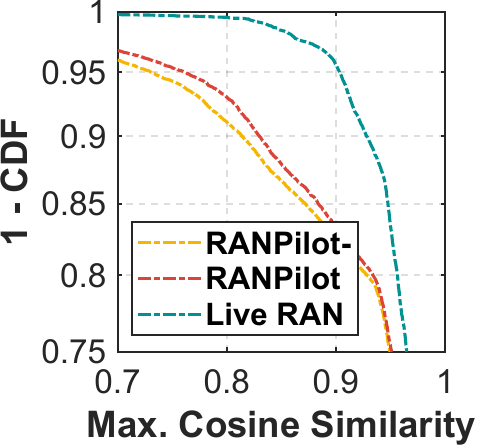}}
\end{minipage}
}
\vspace{-12pt}
\caption{Quantitative fidelity validation across reconfiguration types and deployment environments.}
\label{fig:quantitative_fidelity}
\vspace{-10pt}
\end{figure*}

\noindent\textbf{Setting.} 
We collect over 30 million KPM measurement points in our testbed, each point includes $>$60 metrics with details shown in \tableautorefname{}~\ref{tab:KPM_database}. The collected database covers various channel conditions (\eg, indoor and outdoor, low and high mobilities).
All results presented in this section are based on real-world KPM traces and over-the-air experiments.

\noindent
\textbf{Baselines.} 
We compare \projectName against various representative baselines along two distinct dimensions:
\vspace{-2pt}
\begin{itemize}[leftmargin = 0.3cm]
\setlength{\itemsep}{2pt}
    \item \textbf{Data synthesis:} We compare with (1) \textit{OWDT} \cite{iye2025open}, a SOTA O‑RAN digital twin that employs ray tracing to model physical-layer signal propagation and UE mobility; and (2) \textit{Calibrated Noise Augmentation}, which synthesizes data by injecting random Gaussian noise and temporal shifting, with the variance calibrated from the seed KPM trace.
    \item \textbf{AI adaptation:} We compare with (1) \textit{TenaxDoS} \cite{benzaid2024federated}, an online learning framework that incrementally fine-tunes models using new KPM data after reconfiguration, serving as a standard continual learning benchmark; and (2) \textit{CORAL} \cite{sun2016return}, a domain adaptation method using a correlation alignment layer to minimize domain shift. 
\end{itemize}
\vspace{-2pt}

\noindent
\textbf{Variants of \projectName.} 
To assess the contribution of each module, we construct distinct ablation variants by disabling specific modules within \projectName, including Virtual O-RAN (V.), Meta-Augmentation (M.), and Incremental Learning (I.). We also implement \projectName-Lite, which replaces the transformer with a standard LSTM to test the performance of simpler sequence generators.


\begin{figure}[t]
\setlength{\abovecaptionskip}{0pt}
        \subfigtopskip=-2pt
        \subfigcapskip=-2pt
\centering
\begin{minipage}[b]{0.235\textwidth}
\subfigure[SINR]{
\centering
\includegraphics[height = 3cm]{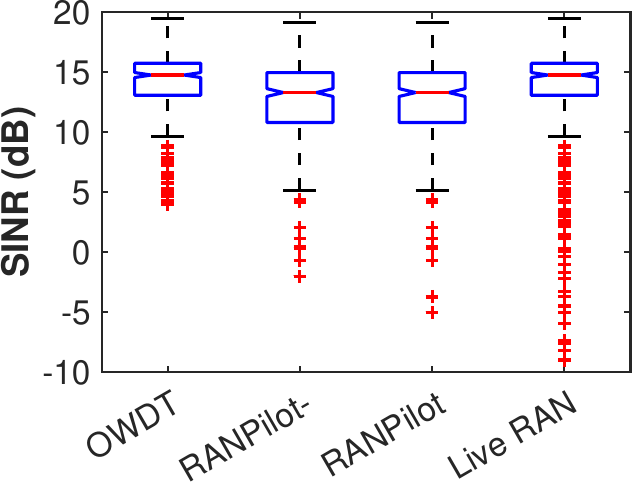}
\label{fig:fidelity_evaluation_a}
}
\end{minipage}
\begin{minipage}[b]{0.235\textwidth}
\subfigure[Throughput]{
\centering
\includegraphics[height = 3cm]{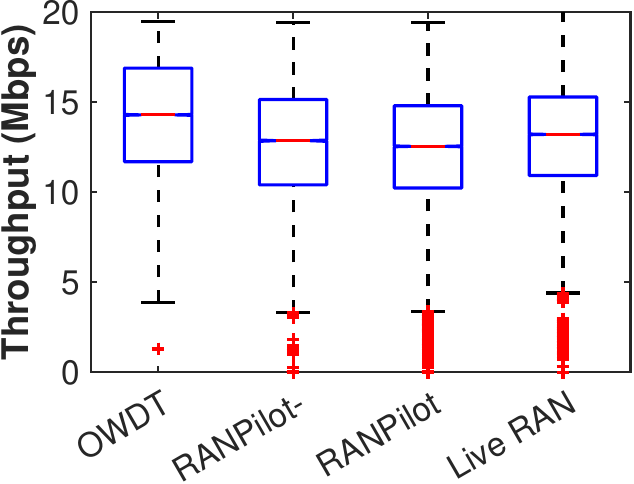}
\label{fig:fidelity_evaluation_b}
}
\end{minipage}
\vspace{-5pt}
\caption{Statistics of \projectName-synthesized network traffic: (a) SINR and (b) Throughput.}
\label{fig:fidelity_evaluation}
\vspace{-12pt}
\end{figure}

\vspace{-0.2cm}
\subsection{Synthetic Data Fidelity}
\label{subsec:evaluation_fidelity}

\vspace{-2pt}

To quantitatively validate the fidelity of the synthesized traces, we evaluate the data similarity across diverse reconfiguration types and physical deployment environments. Following the metric defined in \S\ref{SubSec:Measurement}, we quantify data fidelity by measuring the maximum cosine similarity distribution of 2~s intervals across 10 minutes of synthesized KPM traces, utilizing a post-reconfiguration trace from the operational \textit{Live RAN} as the ground-truth reference. In these experiments, the UEs randomly alternate traffic loads between video streaming, file downloads, and \texttt{iperf} UDP tests to generate realistic, dynamic workloads. 
Since the Incremental Learning module does not contribute to the data synthesis, we use \projectName- to represent the ablated variant without the Meta-Augmentation module in this subsection for simplicity. 
Additionally, we plot the similarity distribution of a temporally independent trace from the post-reconfiguration \textit{Live RAN} to provide a statistical benchmark for natural temporal variations. \projectName is evaluated across two physical environments spanning six reconfiguration scenarios: an indoor deployment testing (1) cell addition and (2) handover policy change, followed by an outdoor deployment testing (3) cell addition, (4) PHY changes via Tx/Rx gain adjustments, (5) a combined upgrade modifying both the PHY gains and the PRB scheduling rules simultaneously, and (6) a complete system migration to a geographically unserved site. To assess the impact of historical configurations on Meta-Augmentation, we initiate the augmentation model at the start of both indoor and outdoor environments (\ie, scenario (1) and (3)) and incrementally train it on the databases from the subsequent configurations.

\noindent
\textbf{Results.} \figureautorefname{}~\ref{fig:quantitative_fidelity} compares synthetic and live RAN KPM distributions across six scenarios. In the indoor setting, \projectName closely matches the live distribution under cell addition (\figureautorefname{}~\ref{fig:quantitative_fidelity}(a)). Under handover policy changes (\figureautorefname{}~\ref{fig:quantitative_fidelity}(b)), the synthetic KPMs largely align with the real network, with minor discrepancies. This slight variation stems from the mismatch between the idealized state updates and the real timing jitter.
This gap is partially mitigated by the Meta-Augmentation module, which augments operational dynamics into the synthetic KPMs. In the challenging outdoor environment, \projectName remains robust despite higher traffic and channel variability, accurately reproducing post-reconfiguration distributions for both cell addition (\figureautorefname{}~\ref{fig:quantitative_fidelity}(c)) and PHY changes (\figureautorefname{}~\ref{fig:quantitative_fidelity}(d)). This robustness arises because site-specific propagation and traffic characteristics are embedded in the seed KPMs, providing a strong statistical prior for synthesis. For the combined upgrade involving both PHY and PRB scheduling policy changes (\figureautorefname{}~\ref{fig:quantitative_fidelity}(e)), the fidelity degrades slightly, as simultaneous reconfigurations introduce multiple interdependent variables into the emulation loop. The Meta-Augmentation module can mitigate this gap by leveraging historical reconfiguration data to augment the synthetic KPMs. 
However, a clear limitation emerges when transferring to an entirely unserved site (\figureautorefname{}~\ref{fig:quantitative_fidelity}(f)), where fidelity drops noticeably. The lack of representative seed KPMs for the new environment constrains the synthesis pipeline, necessitating post-deployment online calibration. 

\begin{figure}[t]
\setlength{\abovecaptionskip}{0pt}
        \subfigtopskip=-2pt
        \subfigcapskip=-2pt
\centering
\begin{minipage}[b]{0.235\textwidth}
\subfigure[Indoor]{
\centering
\includegraphics[height = 3cm]{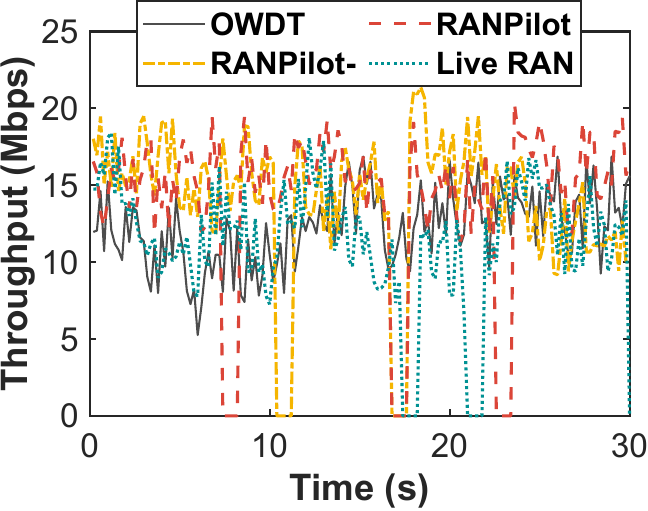}
\label{fig:fidelity_timeline_a}
}
\end{minipage}
\begin{minipage}[b]{0.235\textwidth}
\subfigure[Outdoor]{
\centering
\includegraphics[height = 3cm]{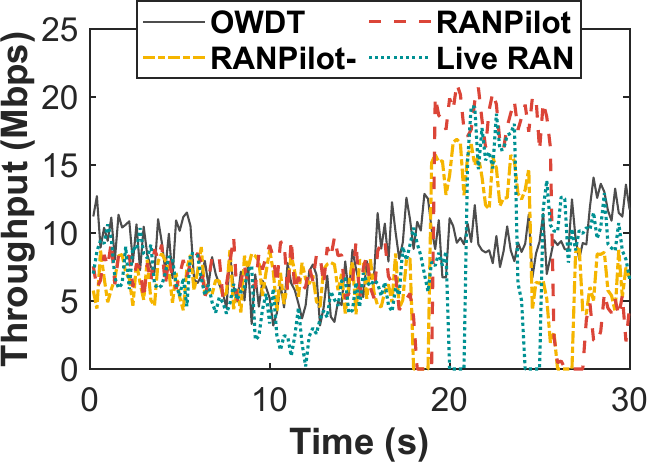}
\label{fig:fidelity_timeline_b}
}
\end{minipage}
\vspace{-5pt}
\caption{Time-aligned comparison of \projectName-synthesized downlink throughput traces.}
\label{fig:fidelity_timeline}
\vspace{-12pt}
\end{figure}


Following prior work \cite{lai2023starrynet}, we further illustrate the statistical characteristics of two representative metrics, SINR and throughput.
\figureautorefname{}~\ref{fig:fidelity_evaluation} shows that OWDT approximates UE mobility and produces data distributions broadly similar to the live RAN. However, it lacks the pronounced outliers that reflect real-world data drift. This gap arises because OWDT focuses primarily on PHY effects and does not account for system-level interactions such as inter-cell interference, handovers, or other uncertainties. In contrast, \projectName achieves higher fidelity by jointly modeling RU behavior and interaction patterns, enabling it to capture the full spectrum of network fluctuations. Compared with its ablated variant, \projectName-, the full model better reproduces transient variations and drift events, \ie, key characteristics needed to train robust AI models for O-RAN. 
\figureautorefname{}~\ref{fig:fidelity_timeline} plots the time-aligned downlink throughput after reconfiguration in indoor and outdoor environments. Although OWDT matches the general trend of the live system, it reflects an overly simplified emulation that omits the complex interactions of dynamic RAN changes. The \projectName-synthesized KPMs track the live RAN more closely, capturing the critical fluctuation periods introduced by updated configurations and control policies.


\vspace{-3pt}
\subsection{\projectName Generalization}
\vspace{-2pt}
\label{subsec:evaluation_generalization}

We next evaluate how \projectName generalizes across diverse RAN reconfigurations and use cases. \textit{Specifically, we test various reconfigurations in daily RAN operations, including hardware additions, handover policy changes, AI model updates, and topology modifications, and integrate them into three representative AI-driven use cases}: \textit{QoE prediction} \cite{ye2024dissecting}, \textit{resource allocation} \cite{ko2024edgeric}, and \textit{anomaly detection} \cite{sun2024spotlight}. 
These applications span diverse AI characteristics, including different model architectures, training paradigms, execution timescales, and input data modalities.
We use two metrics to benchmark the AI adaptation performance: (1) \textit{plug-and-play performance}: the performance of AI models immediately after system reconfiguration, which evaluates the effectiveness of synthetic data on downstream AI robustness; (2) \textit{AI downtime}: the duration from the activation of RAN reconfiguration till the adapted AI model achieves 95\% of its best performance (denoted as Oracle).
For each reconfiguration, we use a 10-minute seed KPM trace to bootstrap the virtual O-RAN, generate a 10-minute synthetic KPM trace, and then extend it into a 1-hour KPM trace via Meta-Augmentation.
For fair comparison, all methods use identical model architectures and differ only in the training data and adaptation mechanisms.



\begin{figure}[t]
\setlength{\abovecaptionskip}{0pt}
        \subfigtopskip=-2pt
        \subfigcapskip=-2pt
\centering
\begin{minipage}[b]{0.235\textwidth}
\subfigure[Cell reconfiguration]{
\centering
\includegraphics[height = 3.2cm]{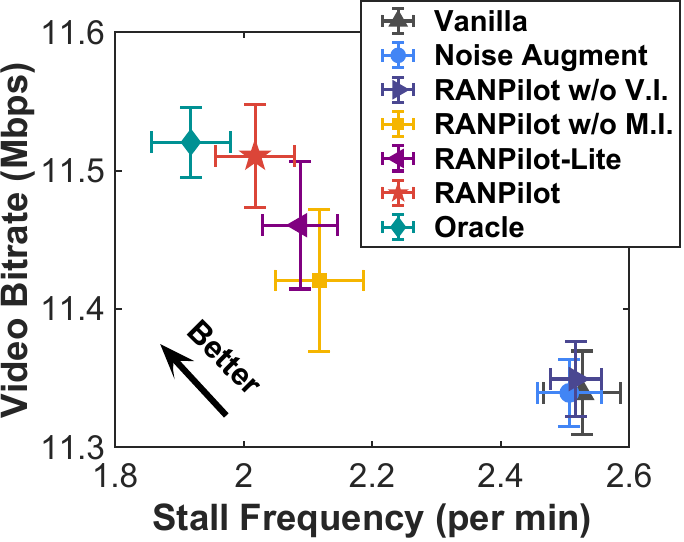}
\label{fig:QoE_Reconfig_a}
}
\end{minipage}
\begin{minipage}[b]{0.235\textwidth}
\subfigure[Handover policy change]{
\centering
\includegraphics[height = 3.2cm]{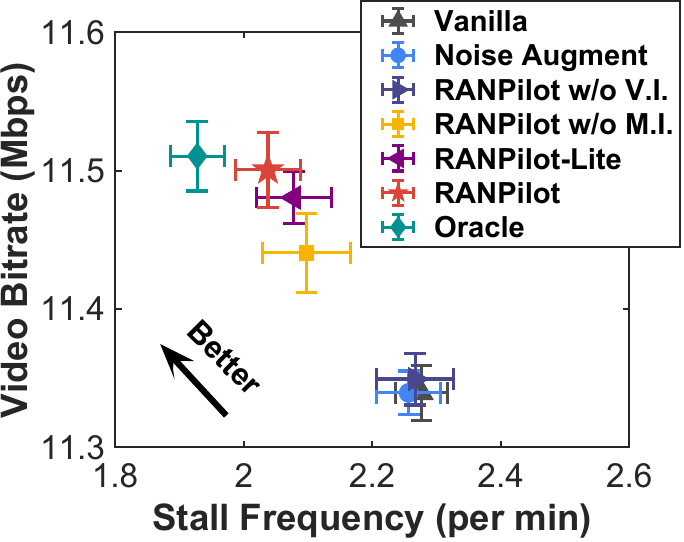}
\label{fig:QoE_Reconfig_b}
}
\end{minipage}
\vspace{-5pt}
\caption{Performance of \projectName-developed QoE predictor: Interaction QoEs immediately after (a) cell reconfiguration; and (b) handover policy change.}
\label{fig:QoE_Reconfig}
\vspace{-15pt}
\end{figure}

\vspace{-2pt}
\subsubsection{QoE prediction for mobile video streaming} 

We evaluate \projectName in an adaptive video streaming application, where the QoE predictor forecasts the next two seconds’ average throughput to adjust video bitrates and prevent playback stalls. 
Following the SOTA QoE predictor \cite{ye2024dissecting}, we apply the same RNN model but remove the input CA parameters for throughput prediction. For calibrated noise augmentation, we synthesize the same amount of data by randomly injecting calibrated noise and temporal shifts derived from seed KPM. For the \projectName w/o V.I., we skip the virtual O-RAN and directly apply the Meta-Augmentation module to synthesize the same amount of data based on the seed KPM.
We introduce two successive RAN reconfigurations to assess generalization.
The first adds a new cell to the deployment with a default SINR-based handover control policy, altering coverage and UE associations. 
After the model adapts to the new deployment, the second reconfiguration upgrades the handover policy to incorporate load balancing, incorporating both signal strength and UE workload distribution across available cells.
Together, these reconfigurations stress the QoE predictor under hardware and control policy changes.

\noindent\textbf{Results.}
We measure application-level QoE immediately after each reconfiguration using stall frequency and achieved bitrate.
After the cell reconfiguration (\figureautorefname~\ref{fig:QoE_Reconfig_a}). \projectName achieves the lowest stall rate ($\sim$2 stalls/min), substantially outperforming Vanilla and closely approaching Oracle performance.
After the handover policy reconfiguration (\figureautorefname~\ref{fig:QoE_Reconfig_b}), Vanilla shows partial improvement as it adapts to the presence of the new cell, but remains highly sensitive to the changed handover logic, resulting in significantly more playback stalls than Oracle. In contrast, \projectName consistently delivers near‑Oracle QoE across both reconfigurations.
This robustness is enabled by Virtual O-RAN, which incorporates new cell deployments and handover policies into the training data for proactive adaptation, and Meta‑Augmentation, which improves tolerance to unseen topologies and policy changes by increasing volume and diversity of synthetic data. These mechanisms enable \projectName to quickly align its predictions with evolving RAN conditions, translating network reconfiguration storms into stable end‑user streaming experiences.
Notably, simple synthetic pretraining approaches, such as omitting virtual O-RAN or applying calibrated noise augmentation, perform nearly as poorly as Vanilla. The reason is that the underlying data drift is driven by RAN unit behaviors and new control policies that cannot be captured through simple data augmentation. Furthermore, \projectName-Lite exhibits a performance degradation compared to the full \projectName, as the LSTM sequence generator possesses lower modeling capacity than the Transformer. Nevertheless, \projectName-Lite still delivers performance gains over the variant without Meta-Augmentation.

\begin{figure}
\setlength{\abovecaptionskip}{4pt}
    \centering
    \includegraphics[height = 2.9cm]{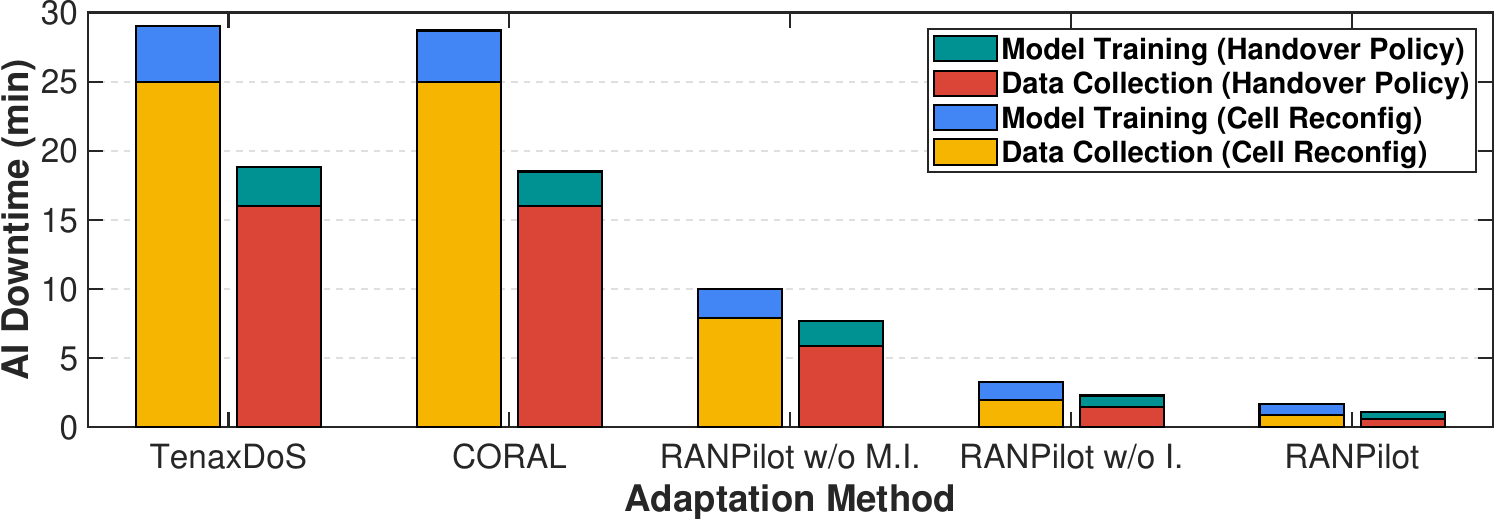}
    \vspace{-3pt}
    \caption{AI downtime of QoE prediction model after different RAN reconfigurations.}
    \label{fig:QoE_AdaptTime}
    \vspace{-15pt}
\end{figure}

To compare the reconfiguration tax, we measure the downtime of different AI adaptation strategies after each RAN reconfiguration.
As shown in \figureautorefname~\ref{fig:QoE_AdaptTime}, during the first reconfiguration (cell addition), the standard continual learning baseline TenaxDoS \cite{benzaid2024federated} requires 29 minutes to converge. Most of this delay (25 minutes) is spent collecting sufficient high quality data from the reconfigured network before meaningful learning can begin. Adopting the domain adaptation baseline CORAL \cite{sun2016return} offers negligible performance benefits. This failure stems from a fundamental mismatch between general domain shift and RAN reconfiguration data drift: the data drift between the pre- and post-reconfiguration KPM traces (\ie, the source and target domains) is not a global distribution shift, but rather a distortion of a small but essential portion of data, localized to critical execution of updated radio control policies.
Coarse data-level feature alignment is difficult to mitigate changes in underlying system behavior.
In contrast, even without Meta‑Augmentation and Incremental Learning, \projectName’s synthesized data sharply reduces this data‑collection phase to 7.9 minutes by exposing the model to the new configuration in advance. Adding Meta‑Augmentation further reduces the overhead to 2 minutes by improving data diversity, while Incremental Learning cuts it to just 0.9 minutes by prioritizing and reusing the most informative live samples. As data quality improves, training itself also becomes faster. End‑to‑end, \projectName completes adaptation in 1.7 minutes, achieving a 94\% reduction in downtime compared to the SOTA baseline.
A similar trend appears under the second reconfiguration (handover policy change). The time required to collect useful data drops from 16 minutes to 6 minutes with Virtual O-RAN, 1.5 minutes with Meta‑Augmentation, and 0.6 minutes with Incremental Learning, yielding a 95\% reduction in total adaptation time. These results demonstrate that \projectName remains effective across different reconfiguration types, from physical changes (adding cells) to control‑plane changes (handover policies), with each module contributing a distinct and critical performance benefit.

\begin{figure}[t]
\setlength{\abovecaptionskip}{0pt}
        \subfigtopskip=-2pt
        \subfigcapskip=-2pt
\centering
\begin{minipage}[b]{0.235\textwidth}
\subfigure[Cell reconfiguration]{
\centering
\includegraphics[height = 3cm]{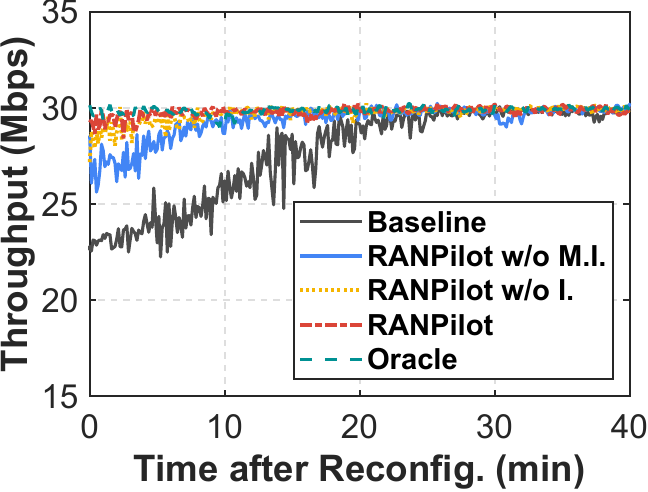}
\label{fig:ResourceAllocation_Evaluation_a}
}
\end{minipage}
\begin{minipage}[b]{0.235\textwidth}
\subfigure[RL policy change]{
\centering
\includegraphics[height = 3cm]{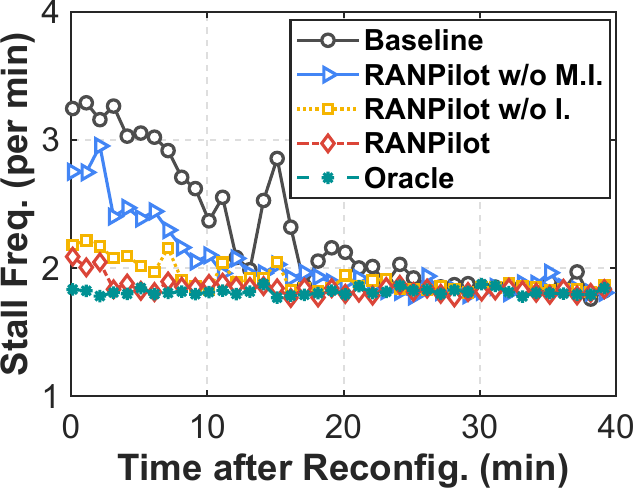}
\label{fig:ResourceAllocation_Evaluation_b}
}
\end{minipage}
\vspace{-5pt}
\caption{Performance of \projectName-developed resource allocator: (a) aggregate throughput after cell reconfiguration; and (b) stalls after RL policy change.}
\label{fig:ResourceAllocation_Evaluation}
\vspace{-15pt}
\end{figure}

\vspace{-2pt}
\subsubsection{Resource allocation for multi-user streaming.}

We evaluate \projectName in a multi‑user streaming scenario where an RL‑based resource allocation agent manages handovers and PRB assignments across multiple UEs. 
Following the SOTA resource allocator \cite{ko2024edgeric}, we apply the same RL agent and training method but enlarge the action space to support handover control.
We assess generalization under two successive RAN reconfigurations. The agent is initially trained for file-transfer workloads with a pure throughput objective. The first reconfiguration expands the deployment by adding a new cell, changing the handover topology and interference conditions. After the model stabilizes, we replace file transfer with AR streaming, introducing QoE constraints. To support this, the second reconfiguration updates the RL agent’s state with per‑UE buffer size and penalizes low‑buffer conditions that may cause playback stalls. We use \textit{baseline} to denote the \textit{Vanilla+TenaxDoS} approach, which applies the Vanilla model after reconfiguration and adapts through continuous learning with real data.
\noindent\textbf{Results.} After the cell reconfiguration (\figureautorefname~\ref{fig:ResourceAllocation_Evaluation_a}), the baseline remains suboptimal for an extended period, requiring 26 minutes to reach Oracle‑level throughput. In contrast, pre‑adapting the agent using Virtual O-RAN enables immediate high throughput, reducing AI downtime to 12 minutes. Adding Meta‑Augmentation further improves robustness to unseen RAN layouts by increasing the diversity and generality of synthetic data, lowering downtime to 8 minutes. After deployment, Incremental Learning accelerates adaptation using real traces, cutting downtime to only 4 minutes. After the RL policy change (\figureautorefname~\ref{fig:ResourceAllocation_Evaluation_b}), we see that the baseline requires 21 minutes to minimize playback stalls. Even when initialized with parameters from the Vanilla model, significant downtime remains because the RL policy has changed. With \projectName, AI downtime drops from 21 minutes to 15 minutes with Virtual O-RAN, 9 minutes with Meta‑Augmentation, and 3 minutes with Incremental Learning. Overall, \projectName maintains AI robustness across both reconfigurations (\ie, cell addition and AI model update).

\vspace{-2pt}
\subsubsection{Anomaly detection for system diagnosis}

We evaluate \projectName in an anomaly detection application, where the detector identifies anomalous behaviors during RAN operation and reports them to operators. 
Following the SOTA anomaly detector \cite{sun2024spotlight}, we apply the same model architecture for distribution learning.
We study model generalization under two representative cell reconfiguration scenarios that alter RAN topology: (1) shared-cell deployment, and (2) dual-cell deployment. For each scenario, we measure accuracy, recall, and precision to assess the plug-and-play performance. 

\begin{table}[t]
\centering
\footnotesize
\setlength{\tabcolsep}{2.2pt}
\begin{tabular}{lcccc}
\toprule
\textbf{Method} & \textbf{Accuracy} & \textbf{Recall} & \textbf{Precision} & \textbf{Downtime (min)} \\
\midrule
Vanilla   & 30.3\% & 0 & 0 & 19.6 \\
\projectName w/o M.I.    & 80.4\% & 92.8\% & 87.6\% & 4.7 \\
\projectName w/o I.     & 89.1\% & 95.2\% & 90.7\% & 2.9 \\
\projectName     & 89.1\% & 95.2\% & 90.7\% & 1.3 \\
Oracle & 99.9\% & 100\% & 99.9\% & N/A \\
\bottomrule
\end{tabular}
\caption{Performance of \projectName-developed anomaly detection model on RU contention.}
\label{tab:anomaly_RUcontention}
\vspace{-23pt}
\end{table}

\begin{table}[t]
\centering
\footnotesize
\setlength{\tabcolsep}{2.2pt}
\begin{tabular}{lcccc}
\toprule
\textbf{Method} & \textbf{Accuracy} & \textbf{Recall} & \textbf{Precision} & \textbf{Downtime (min)} \\
\midrule
Vanilla   & 50.1\% & 86.5\% & 46.16\% & 23.7\\
\projectName w/o M.I.    & 88.8\% & 91.3\% & 68.08\% & 6.5\\
\projectName w/o I.    & 97.5\% & 99.1\% & 95.37\% & 4.1\\
\projectName     & 97.5\% & 99.1\% & 95.37\% & 2.4\\
Oracle & 99.8\% & 99.9\% & 99.7\% & N/A \\
\bottomrule
\end{tabular}
\caption{Performance of \projectName-developed anomaly detection model on radio interference recognition.}
\label{tab:anomaly_radio}
\vspace{-23pt}
\end{table}

\noindent\textbf{Results.}
In the shared-cell deployment, two RUs are connected to a single DU to extend coverage. This configuration introduces contention among traffic streams from concurrent RUs, resulting in packet collisions and retransmissions. As shown in \tableautorefname~\ref{tab:anomaly_RUcontention}, the Vanilla model completely fails to detect RU contention anomalies, yielding zero precision. Even when continuously training after deployment \cite{benzaid2024federated}, Vanilla requires an average of 19.6 minutes to adapt to the new configuration. In contrast, \projectName reduces the reconfiguration tax by 93\%, recovering Oracle-level performance within only 1.3 minutes. 
In the dual-cell deployment, two RUs operate as independent cells, where over-the-air signal overlap leads to inter-cell interference. \tableautorefname~\ref{tab:anomaly_radio} summarizes the anomaly detection performance after this reconfiguration. The Vanilla model exhibits extremely low precision and a high false-positive rate, as it frequently misclassifies benign inter-cell interference as illegal radio interference. Correspondingly, Vanilla incurs an average downtime of 23.7 minutes before stabilizing. By comparison, \projectName reduces downtime by 90\%, adapting in just 2.4 minutes. These gains stem from the complementary roles of \projectName’s components. The Virtual O-RAN and Meta-Augmentation modules proactively adapt the model using synthetic and augmented traces that anticipate post-reconfiguration behaviors, while the Incremental Learning further minimizes residual downtime by enabling rapid online refinement. Together, these modules enable \projectName to achieve robust plug-and-play performance under reconfigurations.



\vspace{-0.2cm}
\subsection{Scalability and Overhead}
\label{subsec:evaluation_overhead}
\vspace{-2pt}

We then evaluate the scalability and overhead of \projectName in supporting different RAN reconfigurations. Specifically, we test four consecutive reconfiguration plans with growing topological complexity: (1) QoE predictor update; (2) cell addition with integrated handover policy; (3) PHY parameter modification with handover policy upgrade; and (4) extension of P2 to a four-cell deployment. For each plan, we first extract 10‑minute representative KPM traces from real-world datasets and apply the Virtual O-RAN module to generate KPM data consistent with the target reconfigured RAN. We then invoke the meta‑augmentation module to expand the dataset in both volume and network dynamics, producing one hour of KPM traces. We record the time overhead incurred at each stage. To evaluate sensitivity to UE scale, we conduct experiments using both real UEs and ZMQ‑emulated UEs with varying UE counts.

\begin{figure}[t]
\setlength{\abovecaptionskip}{4pt}
        \subfigtopskip=0pt
        \subfigcapskip=0pt
\centering
\begin{minipage}[b]{0.235\textwidth}
\subfigure[Impact of reconfiguration plan]{
\centering
\includegraphics[height = 3cm]{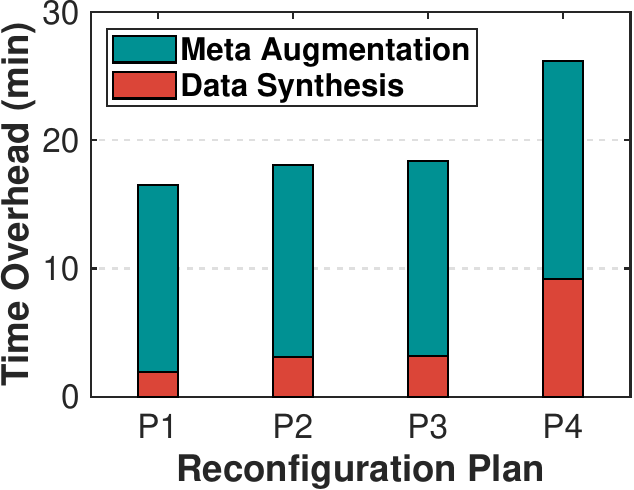}
\label{fig:Overhead_Evaluation_a}
}
\end{minipage}
\begin{minipage}[b]{0.235\textwidth}
\subfigure[Impact of UEs]{
\centering
\includegraphics[height = 3cm]{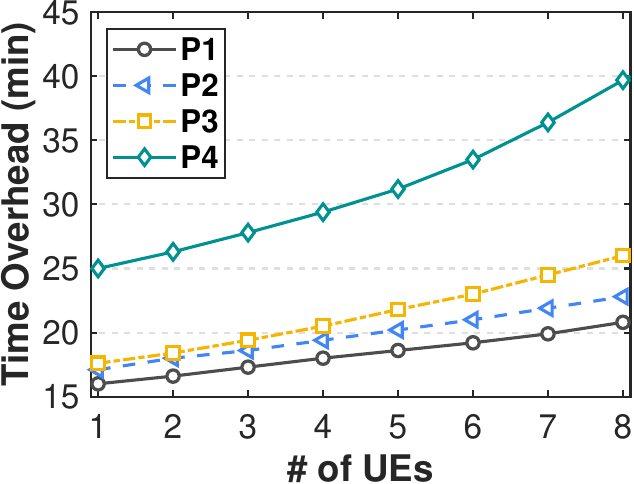}
\label{fig:Overhead_Evaluation_b}
}
\end{minipage}
\vspace{-5pt}
\caption{Overhead of \projectName under different (a) reconfiguration plan and (b) number of UEs.}
\label{fig:Overhead_Evaluation}
\vspace{-18pt}
\end{figure}

\noindent\textbf{Results.}
\figureautorefname~\ref{fig:Overhead_Evaluation_a} shows that the time overhead of data synthesis grows with the complexity of the RAN topology. In contrast, meta augmentation incurs a higher overhead (14$\sim$15 minutes) but increases only marginally as the topology becomes more complex.
The synthesis overhead is driven primarily by topological complexity, whereas the augmentation overhead is dominated by the generative model’s inference time.
Overall, compared to manually collecting real-world network traffic, \projectName reduces the time required to obtain an equivalent data volume by up to 50\%. More importantly, \projectName shifts the majority of the data‑collection costs from post-reconfiguration to pre-reconfiguration, enabling proactive AI adaptation using synthetic KPMs. Given that reconfiguring even small-scale private cellular network \cite{ericsson_private,johnson2024dauth} can take hours or days after a plan is finalized, the additional pre‑deployment overhead introduced by \projectName is modest and practically acceptable.
\figureautorefname{}~\ref{fig:Overhead_Evaluation_b} illustrates the time overhead as the number of UEs increases. We observe a upward trend in overhead, driven by the growing complexity of UE behavior emulation. More complex configurations (\eg, P4) exhibit both a higher baseline overhead and a steeper growth rate, owing to the expanded dimensionality of the network topology and its interaction with UE dynamics. These findings suggest that for more intricate RAN topologies and larger UE numbers, \projectName requires a longer but acceptable buffering period prior to physical reconfiguration in order to generate high-quality data and facilitate robust AI functionality development.





\vspace{-2pt}
\section{Discussion and Future Work}
\label{Sec:Limitation}
\vspace{-2pt}

\noindent
\textbf{Design scope and assumptions.} 
\projectName is designed to facilitate proactive AI adaptation for 
operator-initiated reconfigurations within individual base stations, rather than citywide, synchronized cellular rollouts spanning massive cell sites. 
Given that typical base stations operate with 1$\sim$4 cells, our system evaluation is grounded on a typical O-RAN base station mirroring this cell settings. 
The \projectName's capability of synthesizing KPM traces and adapting AI models is based on two assumptions: (1) the seed KPMs harvested during the pre-reconfiguration stage should capture the site-specific complexities of cellular networks (\eg, outdoor nature, number of users, mobility and traffic patterns, channel conditions, cell densities and capacities); and (2) the planned reconfiguration only alters RAN components, whereas external environmental features and user behavior profiles remain largely consistent.

\noindent
\textbf{Limitations and failure boundary.} 
(1) \textit{Seed KPM Consistency}: \projectName's data synthesis relies on the empirical consistency of seed KPMs. A failure boundary emerges when this consistency is disrupted, such as deploying new RUs in unserved areas or experiencing extreme spatial variations (\figureautorefname{}~\ref{fig:quantitative_fidelity}(f)). In these scenarios, the lack of site-specific historical traffic patterns and channel dynamics can compromise the fidelity of synthetic data. \projectName can resolve these inconsistencies after system changes by incremental learning. (2) \textit{Advanced PHY Techniques}: modern 5G/6G networks adopt advanced PHY technologies (\eg, massive MIMO, beamforming, mmWave), which expand the network's operational and reconfiguration space. Fully mapping these PHY changes to system-level performance impacts remains an open challenge. \projectName abstracts PHY changes into signal strength variations using a simplified wireless channel model. Our future work will explore integrating PHY-foreseeing emulation 
to support proactive adaptation for complex PHY policy changes (\eg, beamforming).

\noindent
\textbf{Robustness for unplanned changes.} \projectName targets data drift introduced by planned, operator-initiated RAN reconfigurations, where network changes are known ahead of deployment.
In real-world operations, unplanned changes may also introduce sudden or gradual data drift that invalidates deployed AI models.
Examples include sudden CBRS license changes due to higher-priority incumbents, bursty UE surges triggered by mass gatherings, and environmental dynamics such as new transmitters/reflectors that alter radio propagation. 
Mitigating such unplanned changes requires orthogonal mechanisms. For example, scaling the training dataset to encompass diverse unplanned dynamics, thereby enhancing the AI model's inherent robustness, or deploying reactive, online calibration strategies to fix performance gaps after the drift is observed \cite{yan2012argus}.
Fully addressing AI robustness under unplanned network changes requires \projectName to work together with these online calibration mechanisms and remains a promising direction for future work.

\noindent\textbf{AI‑native data synthesis.}
While \projectName enables proactive AI adaptation, it still relies on human intervention to bootstrap the data‑synthesis pipeline: operators must explicitly provide reconfiguration plans via RAN and AI APIs. Looking ahead, the broader AI‑RAN vision \cite{AIRANAlliance,ananthanarayanan2025distributed} for NextG cellular systems aims for a fully \textit{AI‑native} architecture, in which the RAN autonomously determines when and how to reconfigure. Realizing such a paradigm would require the control plane to automatically infer performance bottlenecks, synthesize candidate reconfiguration plans, and seamlessly trigger \projectName to generate synthetic KPM data and update deployed AI models, entirely without operator guidance. Our future work will explore AI‑native data synthesis for the evolution of intelligent, self‑optimizing RANs.

\vspace{-2pt}
\section{Conclusion}
\vspace{-2pt}

This paper introduces \projectName, a novel AI evolution framework that bridges the gap between O-RAN's architectural flexibility and rigid AI development paradigm. Unlike conventional methods, \projectName facilitates proactive adaptation of AI models prior to planned system upgrades, based on novel \textit{virtual O-RAN powered data synthesis}. This digital-leadoff approach enables robust AI functionalities with sustained performance and seamless transition to new network configurations. 
Extensive evaluations demonstrate the advantages of \projectName in reducing AI service downtime during RAN reconfiguration. 
We believe \projectName can potentially empower more adaptive AI-driven control and optimization in dynamic O-RAN reconfigurations.

\begin{acks}
We sincerely thank the anonymous shepherd and reviewers for their comments and feedback. 
This work is supported in part by HK RGC GRF (Grant No. 15204926, 15216325, and 15211924), HK RGC CRF (Grant No. C5085-25G and C6086-25G), and HK PolyU RIAIoT P0051269.
Xianjin Xia and Yuanqing Zheng are the corresponding authors.
\end{acks}

\bibliographystyle{ACM-Reference-Format}
\bibliography{ref}


\begin{thebibliography}{94}


\ifx \showCODEN    \undefined \def \showCODEN     #1{\unskip}     \fi
\ifx \showISBNx    \undefined \def \showISBNx     #1{\unskip}     \fi
\ifx \showISBNxiii \undefined \def \showISBNxiii  #1{\unskip}     \fi
\ifx \showISSN     \undefined \def \showISSN      #1{\unskip}     \fi
\ifx \showLCCN     \undefined \def \showLCCN      #1{\unskip}     \fi
\ifx \shownote     \undefined \def \shownote      #1{#1}          \fi
\ifx \showarticletitle \undefined \def \showarticletitle #1{#1}   \fi
\ifx \showURL      \undefined \def \showURL       {\relax}        \fi
\providecommand\bibfield[2]{#2}
\providecommand\bibinfo[2]{#2}
\providecommand\natexlab[1]{#1}
\providecommand\showeprint[2][]{arXiv:#2}

\bibitem[38.331(2018)]%
        {handoverHeterogeneity}
\bibfield{author}{\bibinfo{person}{3GPP Standard~(TS) 38.331}.} \bibinfo{year}{2018}\natexlab{}.
\newblock \bibinfo{booktitle}{\emph{NR; Radio Resource Control (RRC); Protocol Specification, Version 15.0.0}}.
\newblock
\newblock
\shownote{\url{"http://www.3gpp.org/DynaReport/38331.htm"}}.


\bibitem[3GPP(2024)]%
        {3GPP}
\bibfield{author}{\bibinfo{person}{3GPP}.} \bibinfo{year}{2024}\natexlab{}.
\newblock \bibinfo{booktitle}{\emph{3rd Generation Partnership Project}}.
\newblock
\newblock
\shownote{\url{"https://www.3gpp.org/"}}.


\bibitem[Abubakar et~al\mbox{.}(2023)]%
        {abubakar2023energy}
\bibfield{author}{\bibinfo{person}{Attai~Ibrahim Abubakar}, \bibinfo{person}{Oluwakayode Onireti}, \bibinfo{person}{Yusuf Sambo}, \bibinfo{person}{Lei Zhang}, \bibinfo{person}{GK Ragesh}, {and} \bibinfo{person}{Muhammad~Ali Imran}.} \bibinfo{year}{2023}\natexlab{}.
\newblock \showarticletitle{Energy efficiency of open radio access network: A survey}. In \bibinfo{booktitle}{\emph{2023 IEEE 97th Vehicular Technology Conference (VTC2023-Spring)}}. IEEE, \bibinfo{pages}{1--7}.
\newblock


\bibitem[Alliance(2024a)]%
        {AIRANAlliance}
\bibfield{author}{\bibinfo{person}{AI-RAN Alliance}.} \bibinfo{year}{2024}\natexlab{a}.
\newblock \bibinfo{booktitle}{\emph{AI-RAN}}.
\newblock
\newblock
\shownote{\url{"https://ai-ran.org/"}}.


\bibitem[Alliance(2024b)]%
        {ORANAlliance}
\bibfield{author}{\bibinfo{person}{O-RAN Alliance}.} \bibinfo{year}{2024}\natexlab{b}.
\newblock \bibinfo{booktitle}{\emph{O-RAN}}.
\newblock
\newblock
\shownote{\url{"https://www.o-ran.org"}}.


\bibitem[AlQiam et~al\mbox{.}(2024)]%
        {alqiam2024transferable}
\bibfield{author}{\bibinfo{person}{Abd~AlRhman AlQiam}, \bibinfo{person}{Yuanjun Yao}, \bibinfo{person}{Zhaodong Wang}, \bibinfo{person}{Satyajeet~Singh Ahuja}, \bibinfo{person}{Ying Zhang}, \bibinfo{person}{Sanjay~G Rao}, \bibinfo{person}{Bruno Ribeiro}, {and} \bibinfo{person}{Mohit Tawarmalani}.} \bibinfo{year}{2024}\natexlab{}.
\newblock \showarticletitle{Transferable neural wan te for changing topologies}. In \bibinfo{booktitle}{\emph{Proceedings of the ACM SIGCOMM 2024 Conference}}. \bibinfo{pages}{86--102}.
\newblock


\bibitem[Amiri et~al\mbox{.}(2023)]%
        {amiri2023deep}
\bibfield{author}{\bibinfo{person}{Esmaeil Amiri}, \bibinfo{person}{Ning Wang}, \bibinfo{person}{Mohammad Shojafar}, \bibinfo{person}{Mutasem~Q Hamdan}, \bibinfo{person}{Chuan~Heng Foh}, {and} \bibinfo{person}{Rahim Tafazolli}.} \bibinfo{year}{2023}\natexlab{}.
\newblock \showarticletitle{Deep reinforcement learning for robust vnf reconfigurations in o-ran}.
\newblock \bibinfo{journal}{\emph{IEEE Transactions on Network and Service Management}} \bibinfo{volume}{21}, \bibinfo{number}{1} (\bibinfo{year}{2023}), \bibinfo{pages}{1115--1128}.
\newblock


\bibitem[An et~al\mbox{.}(2025)]%
        {an2025tooth}
\bibfield{author}{\bibinfo{person}{Congkai An}, \bibinfo{person}{Huanhuan Zhang}, \bibinfo{person}{Shibo Wang}, \bibinfo{person}{Jingyang Kang}, \bibinfo{person}{Anfu Zhou}, \bibinfo{person}{Liang Liu}, \bibinfo{person}{Huadong Ma}, \bibinfo{person}{Zili Meng}, \bibinfo{person}{Delei Ma}, \bibinfo{person}{Yusheng Dong}, {et~al\mbox{.}}} \bibinfo{year}{2025}\natexlab{}.
\newblock \showarticletitle{Tooth: Toward Optimal Balance of Video $\{$QoE$\}$ and Redundancy Cost by $\{$Fine-Grained$\}$$\{$FEC$\}$ in Cloud Gaming Streaming}. In \bibinfo{booktitle}{\emph{22nd USENIX Symposium on Networked Systems Design and Implementation (NSDI 25)}}. \bibinfo{pages}{635--651}.
\newblock


\bibitem[Ananthanarayanan et~al\mbox{.}(2025)]%
        {ananthanarayanan2025distributed}
\bibfield{author}{\bibinfo{person}{Ganesh Ananthanarayanan}, \bibinfo{person}{Matthew Balkwill}, \bibinfo{person}{Xenofon Foukas}, \bibinfo{person}{Zhihua Lai}, \bibinfo{person}{Bozidar Radunovic}, \bibinfo{person}{Connor Settle}, {and} \bibinfo{person}{Yongguang Zhang}.} \bibinfo{year}{2025}\natexlab{}.
\newblock \showarticletitle{Distributed AI Platform for the 6G RAN}. In \bibinfo{booktitle}{\emph{Proceedings of the 2nd ACM Workshop on Open and AI RAN}}. \bibinfo{pages}{15--21}.
\newblock


\bibitem[Aslan et~al\mbox{.}(2025)]%
        {aslan2025fairric}
\bibfield{author}{\bibinfo{person}{Fatih Aslan}, \bibinfo{person}{Jose~A Ayala-Romero}, \bibinfo{person}{Andres Garcia-Saavedra}, \bibinfo{person}{Xavier Costa-Perez}, {and} \bibinfo{person}{George Iosifidis}.} \bibinfo{year}{2025}\natexlab{}.
\newblock \showarticletitle{FairRIC: Real-Time Fair Allocation in O-RAN with Shared Computing}. In \bibinfo{booktitle}{\emph{IEEE INFOCOM 2025-IEEE Conference on Computer Communications}}. IEEE, \bibinfo{pages}{1--10}.
\newblock


\bibitem[Aslan et~al\mbox{.}(2024)]%
        {aslan2024fair}
\bibfield{author}{\bibinfo{person}{Fatih Aslan}, \bibinfo{person}{George Iosifidis}, \bibinfo{person}{Jose~A Ayala-Romero}, \bibinfo{person}{Andres Garcia-Saavedra}, {and} \bibinfo{person}{Xavier Costa-Perez}.} \bibinfo{year}{2024}\natexlab{}.
\newblock \showarticletitle{Fair resource allocation in virtualized o-ran platforms}.
\newblock \bibinfo{journal}{\emph{Proceedings of the ACM on Measurement and Analysis of Computing Systems}} \bibinfo{volume}{8}, \bibinfo{number}{1} (\bibinfo{year}{2024}), \bibinfo{pages}{1--34}.
\newblock


\bibitem[Balasingam et~al\mbox{.}(2024)]%
        {balasingam2024application}
\bibfield{author}{\bibinfo{person}{Arjun Balasingam}, \bibinfo{person}{Manikanta Kotaru}, {and} \bibinfo{person}{Paramvir Bahl}.} \bibinfo{year}{2024}\natexlab{}.
\newblock \showarticletitle{$\{$Application-Level$\}$ Service Assurance with 5G $\{$RAN$\}$ Slicing}. In \bibinfo{booktitle}{\emph{21st USENIX Symposium on Networked Systems Design and Implementation (NSDI 24)}}. \bibinfo{pages}{841--857}.
\newblock


\bibitem[Benza{\"\i}d et~al\mbox{.}(2024)]%
        {benzaid2024federated}
\bibfield{author}{\bibinfo{person}{Chafika Benza{\"\i}d}, \bibinfo{person}{Fahim~Muhtasim Hossain}, \bibinfo{person}{Tarik Taleb}, \bibinfo{person}{Pedro~Merino G{\'o}mez}, {and} \bibinfo{person}{Michael Dieudonne}.} \bibinfo{year}{2024}\natexlab{}.
\newblock \showarticletitle{A federated continual learning framework for sustainable network anomaly detection in o-ran}. In \bibinfo{booktitle}{\emph{2024 IEEE Wireless Communications and Networking Conference (WCNC)}}. IEEE, \bibinfo{pages}{1--6}.
\newblock


\bibitem[Bhardwaj et~al\mbox{.}(2022)]%
        {bhardwaj2022ekya}
\bibfield{author}{\bibinfo{person}{Romil Bhardwaj}, \bibinfo{person}{Zhengxu Xia}, \bibinfo{person}{Ganesh Ananthanarayanan}, \bibinfo{person}{Junchen Jiang}, \bibinfo{person}{Yuanchao Shu}, \bibinfo{person}{Nikolaos Karianakis}, \bibinfo{person}{Kevin Hsieh}, \bibinfo{person}{Paramvir Bahl}, {and} \bibinfo{person}{Ion Stoica}.} \bibinfo{year}{2022}\natexlab{}.
\newblock \showarticletitle{Ekya: Continuous learning of video analytics models on edge compute servers}. In \bibinfo{booktitle}{\emph{19th USENIX Symposium on Networked Systems Design and Implementation (NSDI 22)}}. \bibinfo{pages}{119--135}.
\newblock


\bibitem[Budhdev et~al\mbox{.}(2021)]%
        {budhdev2021fsa}
\bibfield{author}{\bibinfo{person}{Nishant Budhdev}, \bibinfo{person}{Raj Joshi}, \bibinfo{person}{Pravein~Govindan Kannan}, \bibinfo{person}{Mun~Choon Chan}, {and} \bibinfo{person}{Tulika Mitra}.} \bibinfo{year}{2021}\natexlab{}.
\newblock \showarticletitle{FSA: Fronthaul slicing architecture for 5G using dataplane programmable switches}. In \bibinfo{booktitle}{\emph{Proceedings of the 27th Annual International Conference on Mobile Computing and Networking}}. \bibinfo{pages}{723--735}.
\newblock


\bibitem[Cannat{\`a} et~al\mbox{.}(2024)]%
        {cannata2024towards}
\bibfield{author}{\bibinfo{person}{Raphael Cannat{\`a}}, \bibinfo{person}{Haoxin Sun}, \bibinfo{person}{Dan~Mihai Dumitriu}, {and} \bibinfo{person}{Haitham Hassanieh}.} \bibinfo{year}{2024}\natexlab{}.
\newblock \showarticletitle{Towards Seamless 5G Open-RAN Integration with WebAssembly}. In \bibinfo{booktitle}{\emph{Proceedings of the 23rd ACM Workshop on Hot Topics in Networks}}. \bibinfo{pages}{121--131}.
\newblock


\bibitem[Chen and Zhang(2023)]%
        {chen2023rf}
\bibfield{author}{\bibinfo{person}{Xingyu Chen} {and} \bibinfo{person}{Xinyu Zhang}.} \bibinfo{year}{2023}\natexlab{}.
\newblock \showarticletitle{Rf genesis: Zero-shot generalization of mmwave sensing through simulation-based data synthesis and generative diffusion models}. In \bibinfo{booktitle}{\emph{Proceedings of the 21st ACM Conference on Embedded Networked Sensor Systems}}. \bibinfo{pages}{28--42}.
\newblock


\bibitem[Chen et~al\mbox{.}(2023)]%
        {chen2023channel}
\bibfield{author}{\bibinfo{person}{Yongzhou Chen}, \bibinfo{person}{Ruihao Yao}, \bibinfo{person}{Haitham Hassanieh}, {and} \bibinfo{person}{Radhika Mittal}.} \bibinfo{year}{2023}\natexlab{}.
\newblock \showarticletitle{Channel-Aware 5g RAN slicing with customizable schedulers}. In \bibinfo{booktitle}{\emph{20th USENIX Symposium on Networked Systems Design and Implementation (NSDI 23)}}. \bibinfo{pages}{1767--1782}.
\newblock


\bibitem[Cheng et~al\mbox{.}(2024)]%
        {cheng2024grace}
\bibfield{author}{\bibinfo{person}{Yihua Cheng}, \bibinfo{person}{Ziyi Zhang}, \bibinfo{person}{Hanchen Li}, \bibinfo{person}{Anton Arapin}, \bibinfo{person}{Yue Zhang}, \bibinfo{person}{Qizheng Zhang}, \bibinfo{person}{Yuhan Liu}, \bibinfo{person}{Kuntai Du}, \bibinfo{person}{Xu Zhang}, \bibinfo{person}{Francis~Y Yan}, {et~al\mbox{.}}} \bibinfo{year}{2024}\natexlab{}.
\newblock \showarticletitle{$\{$GRACE$\}$:$\{$Loss-Resilient$\}$$\{$Real-Time$\}$ video through neural codecs}. In \bibinfo{booktitle}{\emph{21st USENIX Symposium on Networked Systems Design and Implementation (NSDI 24)}}. \bibinfo{pages}{509--531}.
\newblock


\bibitem[Chi et~al\mbox{.}(2024)]%
        {chi2024rf}
\bibfield{author}{\bibinfo{person}{Guoxuan Chi}, \bibinfo{person}{Zheng Yang}, \bibinfo{person}{Chenshu Wu}, \bibinfo{person}{Jingao Xu}, \bibinfo{person}{Yuchong Gao}, \bibinfo{person}{Yunhao Liu}, {and} \bibinfo{person}{Tony~Xiao Han}.} \bibinfo{year}{2024}\natexlab{}.
\newblock \showarticletitle{RF-diffusion: Radio signal generation via time-frequency diffusion}. In \bibinfo{booktitle}{\emph{Proceedings of the 30th Annual International Conference on Mobile Computing and Networking}}. \bibinfo{pages}{77--92}.
\newblock


\bibitem[Ericsson(2025)]%
        {ericsson_private}
\bibfield{author}{\bibinfo{person}{Ericsson}.} \bibinfo{year}{2025}\natexlab{}.
\newblock \bibinfo{booktitle}{\emph{Private 5G, a network as dedicated as you are}}.
\newblock
\newblock
\shownote{\url{"https://www.ericsson.com/en/private-networks"}}.


\bibitem[Finn et~al\mbox{.}(2017)]%
        {finn2017model}
\bibfield{author}{\bibinfo{person}{Chelsea Finn}, \bibinfo{person}{Pieter Abbeel}, {and} \bibinfo{person}{Sergey Levine}.} \bibinfo{year}{2017}\natexlab{}.
\newblock \showarticletitle{Model-agnostic meta-learning for fast adaptation of deep networks}. In \bibinfo{booktitle}{\emph{International conference on machine learning}}. PMLR, \bibinfo{pages}{1126--1135}.
\newblock


\bibitem[Foukas and Radunovic(2021)]%
        {foukas2021concordia}
\bibfield{author}{\bibinfo{person}{Xenofon Foukas} {and} \bibinfo{person}{Bozidar Radunovic}.} \bibinfo{year}{2021}\natexlab{}.
\newblock \showarticletitle{Concordia: Teaching the 5G vRAN to share compute}. In \bibinfo{booktitle}{\emph{Proceedings of the 2021 ACM SIGCOMM 2021 Conference}}. \bibinfo{pages}{580--596}.
\newblock


\bibitem[Foukas et~al\mbox{.}(2023)]%
        {foukas2023taking}
\bibfield{author}{\bibinfo{person}{Xenofon Foukas}, \bibinfo{person}{Bozidar Radunovic}, \bibinfo{person}{Matthew Balkwill}, {and} \bibinfo{person}{Zhihua Lai}.} \bibinfo{year}{2023}\natexlab{}.
\newblock \showarticletitle{Taking 5G RAN analytics and control to a new level}. In \bibinfo{booktitle}{\emph{Proceedings of the 29th Annual International Conference on Mobile Computing and Networking}}. \bibinfo{pages}{1--16}.
\newblock


\bibitem[Foukas et~al\mbox{.}(2025)]%
        {foukas2025ranbooster}
\bibfield{author}{\bibinfo{person}{Xenofon Foukas}, \bibinfo{person}{Tenzin~Samten Ukyab}, \bibinfo{person}{Bozidar Radunovic}, \bibinfo{person}{Sylvia Ratnasamy}, {and} \bibinfo{person}{Scott Shenker}.} \bibinfo{year}{2025}\natexlab{}.
\newblock \showarticletitle{RANBooster: Democratizing advanced cellular connectivity through fronthaul middleboxes}. In \bibinfo{booktitle}{\emph{Proceedings of the ACM SIGCOMM 2025 Conference}}. \bibinfo{pages}{742--757}.
\newblock


\bibitem[Ge et~al\mbox{.}(2023)]%
        {ge2023chroma}
\bibfield{author}{\bibinfo{person}{Changhan Ge}, \bibinfo{person}{Zihui Ge}, \bibinfo{person}{Xuan Liu}, \bibinfo{person}{Ajay Mahimkar}, \bibinfo{person}{Yusef Shaqalle}, \bibinfo{person}{Yu Xiang}, {and} \bibinfo{person}{Shomik Pathak}.} \bibinfo{year}{2023}\natexlab{}.
\newblock \showarticletitle{Chroma: Learning and using network contexts to reinforce performance improving configurations}. In \bibinfo{booktitle}{\emph{Proceedings of the 29th Annual International Conference on Mobile Computing and Networking}}. \bibinfo{pages}{1--16}.
\newblock


\bibitem[Ge et~al\mbox{.}(2025)]%
        {ge2025iridescence}
\bibfield{author}{\bibinfo{person}{Changhan Ge}, \bibinfo{person}{Ajay Mahimkar}, \bibinfo{person}{Zihui Ge}, \bibinfo{person}{Romeo Fernandez}, \bibinfo{person}{Joseph Maniaci}, \bibinfo{person}{Shomik Pathak}, {and} \bibinfo{person}{Maulik Shah}.} \bibinfo{year}{2025}\natexlab{}.
\newblock \showarticletitle{Iridescence: Improving Configuration Tuning in the Presence of Confounders for 5G NSA Networks}.
\newblock \bibinfo{journal}{\emph{Proceedings of the ACM on Networking}} \bibinfo{volume}{3}, \bibinfo{number}{CoNEXT1} (\bibinfo{year}{2025}), \bibinfo{pages}{1--22}.
\newblock


\bibitem[Ghosh et~al\mbox{.}(2024)]%
        {ghosh2024sparc}
\bibfield{author}{\bibinfo{person}{Ushasi Ghosh}, \bibinfo{person}{Azuka Chiejina}, \bibinfo{person}{Nathan Stephenson}, \bibinfo{person}{Vijay~K Shah}, \bibinfo{person}{Srinivas Shakkottai}, {and} \bibinfo{person}{Dinesh Bharadia}.} \bibinfo{year}{2024}\natexlab{}.
\newblock \showarticletitle{SPARC: Spatio-Temporal Adaptive Resource Control for Multi-site Spectrum Management in NextG Cellular Networks}.
\newblock \bibinfo{journal}{\emph{Proceedings of the ACM on Networking}} \bibinfo{volume}{2}, \bibinfo{number}{CoNEXT4} (\bibinfo{year}{2024}), \bibinfo{pages}{1--18}.
\newblock


\bibitem[Gong et~al\mbox{.}(2025)]%
        {gong2025data}
\bibfield{author}{\bibinfo{person}{Chen Gong}, \bibinfo{person}{Bo Liang}, \bibinfo{person}{Wei Gao}, {and} \bibinfo{person}{Chenren Xu}.} \bibinfo{year}{2025}\natexlab{}.
\newblock \showarticletitle{Data Can Speak for Itself: Quality-guided Utilization of Wireless Synthetic Data}. In \bibinfo{booktitle}{\emph{Proceedings of the 23rd Annual International Conference on Mobile Systems, Applications and Services}}. \bibinfo{pages}{209--222}.
\newblock


\bibitem[Heidari et~al\mbox{.}(2025)]%
        {heidari2025phasemo}
\bibfield{author}{\bibinfo{person}{Adel Heidari}, \bibinfo{person}{Agrim Gupta}, \bibinfo{person}{Ish~Kumar Jain}, {and} \bibinfo{person}{Dinesh Bharadia}.} \bibinfo{year}{2025}\natexlab{}.
\newblock \showarticletitle{PhaseMO: A Universal Massive MIMO Architecture for Sustainable NextG}. In \bibinfo{booktitle}{\emph{IEEE INFOCOM 2025-IEEE Conference on Computer Communications}}. IEEE, \bibinfo{pages}{1--10}.
\newblock


\bibitem[Hidalgo et~al\mbox{.}(2025)]%
        {hidalgo2025aegisran}
\bibfield{author}{\bibinfo{person}{Ethan~Sanchez Hidalgo}, \bibinfo{person}{Jose~A Ayala-Romero}, \bibinfo{person}{J~Xavier~Salvat Lozano}, \bibinfo{person}{Andres Garcia-Saavedra}, {and} \bibinfo{person}{Xavier~Costa Perez}.} \bibinfo{year}{2025}\natexlab{}.
\newblock \showarticletitle{AegisRAN: A Fair and Energy-Efficient Computing Resource Allocation Framework for vRANs}.
\newblock \bibinfo{journal}{\emph{IEEE Transactions on Mobile Computing}} (\bibinfo{year}{2025}).
\newblock


\bibitem[Hlawatsch and Matz(2011)]%
        {hlawatsch2011wireless}
\bibfield{author}{\bibinfo{person}{Franz Hlawatsch} {and} \bibinfo{person}{Gerald Matz}.} \bibinfo{year}{2011}\natexlab{}.
\newblock \bibinfo{booktitle}{\emph{Wireless communications over rapidly time-varying channels}}.
\newblock \bibinfo{publisher}{Academic press}.
\newblock


\bibitem[Iye et~al\mbox{.}(2025)]%
        {iye2025open}
\bibfield{author}{\bibinfo{person}{Tetsuya Iye}, \bibinfo{person}{Masaya Sakamoto}, \bibinfo{person}{Shohei Takaya}, \bibinfo{person}{Eisaku Sato}, \bibinfo{person}{Yuki Susukida}, \bibinfo{person}{Yu Nagaoka}, \bibinfo{person}{Kazuki Maruta}, {and} \bibinfo{person}{Jin Nakazato}.} \bibinfo{year}{2025}\natexlab{}.
\newblock \showarticletitle{Open Wireless Digital Twin: End-to-End 5G Mobility Emulation in O-RAN Framework}.
\newblock \bibinfo{journal}{\emph{arXiv preprint arXiv:2503.12177}} (\bibinfo{year}{2025}).
\newblock


\bibitem[Jia et~al\mbox{.}(2025)]%
        {jia2025towards}
\bibfield{author}{\bibinfo{person}{Lianchen Jia}, \bibinfo{person}{Chao Zhou}, \bibinfo{person}{Chaoyang Li}, \bibinfo{person}{Jiangchuan Liu}, {and} \bibinfo{person}{Lifeng Sun}.} \bibinfo{year}{2025}\natexlab{}.
\newblock \showarticletitle{Towards User-level QoE: Large-scale Practice in Personalized Optimization of Adaptive Video Streaming}. In \bibinfo{booktitle}{\emph{Proceedings of the ACM SIGCOMM 2025 Conference}}. \bibinfo{pages}{1154--1166}.
\newblock


\bibitem[Johnson et~al\mbox{.}(2024)]%
        {johnson2024dauth}
\bibfield{author}{\bibinfo{person}{Matthew Johnson}, \bibinfo{person}{Sudheesh Singanamalla}, \bibinfo{person}{Nick Durand}, \bibinfo{person}{Esther Han~Boel Jang}, \bibinfo{person}{Spencer Sevilla}, {and} \bibinfo{person}{Kurtis Heimerl}.} \bibinfo{year}{2024}\natexlab{}.
\newblock \showarticletitle{dAuth: A Resilient Authentication Architecture for Federated Private Cellular Networks}. In \bibinfo{booktitle}{\emph{Proceedings of the ACM SIGCOMM 2024 Conference}}. \bibinfo{pages}{373--391}.
\newblock


\bibitem[Kalia et~al\mbox{.}(2025)]%
        {kalia2025towards}
\bibfield{author}{\bibinfo{person}{Anuj Kalia}, \bibinfo{person}{Nikita Lazarev}, \bibinfo{person}{Leyang Xue}, \bibinfo{person}{Xenofon Foukas}, \bibinfo{person}{Bozidar Radunovic}, {and} \bibinfo{person}{Francis~Y Yan}.} \bibinfo{year}{2025}\natexlab{}.
\newblock \showarticletitle{Towards Energy Efficient 5G vRAN Servers}. In \bibinfo{booktitle}{\emph{USENIX Symposium on Networked Systems Design and Implementation (NSDI)}}.
\newblock


\bibitem[Kemker et~al\mbox{.}(2018)]%
        {kemker2018measuring}
\bibfield{author}{\bibinfo{person}{Ronald Kemker}, \bibinfo{person}{Marc McClure}, \bibinfo{person}{Angelina Abitino}, \bibinfo{person}{Tyler Hayes}, {and} \bibinfo{person}{Christopher Kanan}.} \bibinfo{year}{2018}\natexlab{}.
\newblock \showarticletitle{Measuring catastrophic forgetting in neural networks}. In \bibinfo{booktitle}{\emph{Proceedings of the AAAI conference on artificial intelligence}}, Vol.~\bibinfo{volume}{32}.
\newblock


\bibitem[Khan and Schmid(2023)]%
        {khan2023ai}
\bibfield{author}{\bibinfo{person}{Naveed~Ali Khan} {and} \bibinfo{person}{Stefan Schmid}.} \bibinfo{year}{2023}\natexlab{}.
\newblock \showarticletitle{AI-RAN in 6G Networks: State-of-the-Art and Challenges}.
\newblock \bibinfo{journal}{\emph{IEEE Open Journal of the Communications Society}}  \bibinfo{volume}{5} (\bibinfo{year}{2023}), \bibinfo{pages}{294--311}.
\newblock


\bibitem[Khooi et~al\mbox{.}(2025)]%
        {khooi2025update}
\bibfield{author}{\bibinfo{person}{Xin~Zhe Khooi}, \bibinfo{person}{Anuj Kalia}, {and} \bibinfo{person}{Mun~Choon Chan}.} \bibinfo{year}{2025}\natexlab{}.
\newblock \showarticletitle{How to Update Your 5G vRAN}. In \bibinfo{booktitle}{\emph{Proceedings of the 31st Annual International Conference on Mobile Computing and Networking}}. \bibinfo{pages}{123--138}.
\newblock


\bibitem[Kilinc et~al\mbox{.}(2022)]%
        {kilinc2022jade}
\bibfield{author}{\bibinfo{person}{Caner Kilinc}, \bibinfo{person}{Mahesh~K Marina}, \bibinfo{person}{Muhammad Usama}, \bibinfo{person}{Salih Ergut}, \bibinfo{person}{Jon Crowcroft}, \bibinfo{person}{Tugrul Gundogdu}, {and} \bibinfo{person}{Ilhan Akinci}.} \bibinfo{year}{2022}\natexlab{}.
\newblock \showarticletitle{JADE: Data-driven automated jammer detection framework for operational mobile networks}. In \bibinfo{booktitle}{\emph{IEEE INFOCOM 2022-IEEE Conference on Computer Communications}}. IEEE, \bibinfo{pages}{1139--1148}.
\newblock


\bibitem[Ko et~al\mbox{.}(2024)]%
        {ko2024edgeric}
\bibfield{author}{\bibinfo{person}{Woo-Hyun Ko}, \bibinfo{person}{Ushasi Ghosh}, \bibinfo{person}{Ujwal Dinesha}, \bibinfo{person}{Raini Wu}, \bibinfo{person}{Srinivas Shakkottai}, {and} \bibinfo{person}{Dinesh Bharadia}.} \bibinfo{year}{2024}\natexlab{}.
\newblock \showarticletitle{EdgeRIC: Empowering real-time intelligent optimization and control in NextG cellular networks}. In \bibinfo{booktitle}{\emph{21st USENIX Symposium on Networked Systems Design and Implementation (NSDI 24)}}. \bibinfo{pages}{1315--1330}.
\newblock


\bibitem[Lacava et~al\mbox{.}(2023)]%
        {lacava2023programmable}
\bibfield{author}{\bibinfo{person}{Andrea Lacava}, \bibinfo{person}{Michele Polese}, \bibinfo{person}{Rajarajan Sivaraj}, \bibinfo{person}{Rahul Soundrarajan}, \bibinfo{person}{Bhawani~Shanker Bhati}, \bibinfo{person}{Tarunjeet Singh}, \bibinfo{person}{Tommaso Zugno}, \bibinfo{person}{Francesca Cuomo}, {and} \bibinfo{person}{Tommaso Melodia}.} \bibinfo{year}{2023}\natexlab{}.
\newblock \showarticletitle{Programmable and customized intelligence for traffic steering in 5G networks using open RAN architectures}.
\newblock \bibinfo{journal}{\emph{IEEE Transactions on Mobile Computing}} \bibinfo{volume}{23}, \bibinfo{number}{4} (\bibinfo{year}{2023}), \bibinfo{pages}{2882--2897}.
\newblock


\bibitem[Lai et~al\mbox{.}(2023)]%
        {lai2023starrynet}
\bibfield{author}{\bibinfo{person}{Zeqi Lai}, \bibinfo{person}{Hewu Li}, \bibinfo{person}{Yangtao Deng}, \bibinfo{person}{Qian Wu}, \bibinfo{person}{Jun Liu}, \bibinfo{person}{Yuanjie Li}, \bibinfo{person}{Jihao Li}, \bibinfo{person}{Lixin Liu}, \bibinfo{person}{Weisen Liu}, {and} \bibinfo{person}{Jianping Wu}.} \bibinfo{year}{2023}\natexlab{}.
\newblock \showarticletitle{$\{$StarryNet$\}$: empowering researchers to evaluate futuristic integrated space and terrestrial networks}. In \bibinfo{booktitle}{\emph{20th USENIX Symposium on Networked Systems Design and Implementation (NSDI 23)}}. \bibinfo{pages}{1309--1324}.
\newblock


\bibitem[Lakshmanan et~al\mbox{.}(2021)]%
        {lakshmanan2021stealthy}
\bibfield{author}{\bibinfo{person}{Nitya Lakshmanan}, \bibinfo{person}{Nishant Budhdev}, \bibinfo{person}{Min~Suk Kang}, \bibinfo{person}{Mun~Choon Chan}, {and} \bibinfo{person}{Jun Han}.} \bibinfo{year}{2021}\natexlab{}.
\newblock \showarticletitle{A stealthy location identification attack exploiting carrier aggregation in cellular networks}. In \bibinfo{booktitle}{\emph{30th usenix security symposium (usenix security 21)}}. \bibinfo{pages}{3899--3916}.
\newblock


\bibitem[Lazarev et~al\mbox{.}(2023)]%
        {lazarev2023resilient}
\bibfield{author}{\bibinfo{person}{Nikita Lazarev}, \bibinfo{person}{Tao Ji}, \bibinfo{person}{Anuj Kalia}, \bibinfo{person}{Daehyeok Kim}, \bibinfo{person}{Ilias Marinos}, \bibinfo{person}{Francis~Y Yan}, \bibinfo{person}{Christina Delimitrou}, \bibinfo{person}{Zhiru Zhang}, {and} \bibinfo{person}{Aditya Akella}.} \bibinfo{year}{2023}\natexlab{}.
\newblock \showarticletitle{Resilient baseband processing in virtualized rans with slingshot}. In \bibinfo{booktitle}{\emph{Proceedings of the ACM SIGCOMM 2023 Conference}}. \bibinfo{pages}{654--667}.
\newblock


\bibitem[Li et~al\mbox{.}(2021)]%
        {li2021nationwide}
\bibfield{author}{\bibinfo{person}{Yang Li}, \bibinfo{person}{Hao Lin}, \bibinfo{person}{Zhenhua Li}, \bibinfo{person}{Yunhao Liu}, \bibinfo{person}{Feng Qian}, \bibinfo{person}{Liangyi Gong}, \bibinfo{person}{Xianlong Xin}, {and} \bibinfo{person}{Tianyin Xu}.} \bibinfo{year}{2021}\natexlab{}.
\newblock \showarticletitle{A nationwide study on cellular reliability: Measurement, analysis, and enhancements}. In \bibinfo{booktitle}{\emph{Proceedings of the 2021 ACM SIGCOMM 2021 Conference}}. \bibinfo{pages}{597--609}.
\newblock


\bibitem[Lin(2002)]%
        {lin2002divergence}
\bibfield{author}{\bibinfo{person}{Jianhua Lin}.} \bibinfo{year}{2002}\natexlab{}.
\newblock \showarticletitle{Divergence measures based on the Shannon entropy}.
\newblock \bibinfo{journal}{\emph{IEEE Transactions on Information theory}} \bibinfo{volume}{37}, \bibinfo{number}{1} (\bibinfo{year}{2002}), \bibinfo{pages}{145--151}.
\newblock


\bibitem[Lin et~al\mbox{.}(2025)]%
        {lin20255g}
\bibfield{author}{\bibinfo{person}{Wei Lin}, \bibinfo{person}{Zongxiao Li}, \bibinfo{person}{Binbin Chen}, \bibinfo{person}{Jianwei Liu}, \bibinfo{person}{Ray-Guang Cheng}, {and} \bibinfo{person}{Fan Zhang}.} \bibinfo{year}{2025}\natexlab{}.
\newblock \showarticletitle{5G-Muffler: Covert DoS Attacks over Open Fronthaul Interface of O-RAN 5G Network}. In \bibinfo{booktitle}{\emph{IEEE INFOCOM 2025-IEEE Conference on Computer Communications}}. IEEE, \bibinfo{pages}{1--10}.
\newblock


\bibitem[Liu et~al\mbox{.}(2023)]%
        {liu2023leaf}
\bibfield{author}{\bibinfo{person}{Shinan Liu}, \bibinfo{person}{Francesco Bronzino}, \bibinfo{person}{Paul Schmitt}, \bibinfo{person}{Arjun~Nitin Bhagoji}, \bibinfo{person}{Nick Feamster}, \bibinfo{person}{Hector~Garcia Crespo}, \bibinfo{person}{Timothy Coyle}, {and} \bibinfo{person}{Brian Ward}.} \bibinfo{year}{2023}\natexlab{}.
\newblock \showarticletitle{Leaf: Navigating concept drift in cellular networks}.
\newblock \bibinfo{journal}{\emph{Proceedings of the ACM on Networking}} \bibinfo{volume}{1}, \bibinfo{number}{CoNEXT2} (\bibinfo{year}{2023}), \bibinfo{pages}{1--24}.
\newblock


\bibitem[Liu et~al\mbox{.}(2021)]%
        {liu2021understanding}
\bibfield{author}{\bibinfo{person}{Shinan Liu}, \bibinfo{person}{Francesco Bronzino}, \bibinfo{person}{Paul Schmitt}, \bibinfo{person}{Nick Feamster}, \bibinfo{person}{Ricardo Borges}, \bibinfo{person}{Hector~Garcia Crespo}, {and} \bibinfo{person}{Brian Ward}.} \bibinfo{year}{2021}\natexlab{}.
\newblock \showarticletitle{Understanding model drift in a large cellular network}.
\newblock \bibinfo{journal}{\emph{CoRR}} (\bibinfo{year}{2021}).
\newblock


\bibitem[Lozano et~al\mbox{.}(2025)]%
        {lozano2025kairos}
\bibfield{author}{\bibinfo{person}{J~Xavier~Salvat Lozano}, \bibinfo{person}{Jose~A Ayala-Romero}, \bibinfo{person}{Andres Garcia-Saavedra}, {and} \bibinfo{person}{Xavier Costa-Perez}.} \bibinfo{year}{2025}\natexlab{}.
\newblock \showarticletitle{Kairos: Energy-efficient radio unit control for O-RAN via advanced sleep modes}. In \bibinfo{booktitle}{\emph{IEEE INFOCOM 2025-IEEE Conference on Computer Communications}}. IEEE, \bibinfo{pages}{1--10}.
\newblock


\bibitem[Mahimkar et~al\mbox{.}(2022)]%
        {mahimkar2022aurora}
\bibfield{author}{\bibinfo{person}{Ajay Mahimkar}, \bibinfo{person}{Zihui Ge}, \bibinfo{person}{Xuan Liu}, \bibinfo{person}{Yusef Shaqalle}, \bibinfo{person}{Yu Xiang}, \bibinfo{person}{Jennifer Yates}, \bibinfo{person}{Shomik Pathak}, {and} \bibinfo{person}{Rick Reichel}.} \bibinfo{year}{2022}\natexlab{}.
\newblock \showarticletitle{Aurora: conformity-based configuration recommendation to improve LTE/5G service}. In \bibinfo{booktitle}{\emph{Proceedings of the 22nd ACM Internet Measurement Conference}}. \bibinfo{pages}{83--97}.
\newblock


\bibitem[Mahimkar et~al\mbox{.}(2011)]%
        {mahimkar2011rapid}
\bibfield{author}{\bibinfo{person}{Ajay Mahimkar}, \bibinfo{person}{Zihui Ge}, \bibinfo{person}{Jia Wang}, \bibinfo{person}{Jennifer Yates}, \bibinfo{person}{Yin Zhang}, \bibinfo{person}{Joanne Emmons}, \bibinfo{person}{Brian Huntley}, {and} \bibinfo{person}{Mark Stockert}.} \bibinfo{year}{2011}\natexlab{}.
\newblock \showarticletitle{Rapid detection of maintenance induced changes in service performance}. In \bibinfo{booktitle}{\emph{Proceedings of the Seventh COnference on Emerging Networking EXperiments and Technologies}}. \bibinfo{pages}{1--12}.
\newblock


\bibitem[Mahimkar et~al\mbox{.}(2013)]%
        {mahimkar2013robust}
\bibfield{author}{\bibinfo{person}{Ajay Mahimkar}, \bibinfo{person}{Zihui Ge}, \bibinfo{person}{Jennifer Yates}, \bibinfo{person}{Chris Hristov}, \bibinfo{person}{Vincent Cordaro}, \bibinfo{person}{Shane Smith}, \bibinfo{person}{Jing Xu}, {and} \bibinfo{person}{Mark Stockert}.} \bibinfo{year}{2013}\natexlab{}.
\newblock \showarticletitle{Robust assessment of changes in cellular networks}. In \bibinfo{booktitle}{\emph{Proceedings of the ninth ACM conference on Emerging networking experiments and technologies}}. \bibinfo{pages}{175--186}.
\newblock


\bibitem[Mahimkar et~al\mbox{.}(2021)]%
        {mahimkar2021auric}
\bibfield{author}{\bibinfo{person}{Ajay Mahimkar}, \bibinfo{person}{Ashiwan Sivakumar}, \bibinfo{person}{Zihui Ge}, \bibinfo{person}{Shomik Pathak}, {and} \bibinfo{person}{Karunasish Biswas}.} \bibinfo{year}{2021}\natexlab{}.
\newblock \showarticletitle{Auric: Using data-driven recommendation to automatically generate cellular configuration}. In \bibinfo{booktitle}{\emph{Proceedings of the 2021 ACM SIGCOMM 2021 Conference}}. \bibinfo{pages}{807--820}.
\newblock


\bibitem[Mahimkar et~al\mbox{.}(2010)]%
        {mahimkar2010detecting}
\bibfield{author}{\bibinfo{person}{Ajay~Anil Mahimkar}, \bibinfo{person}{Han~Hee Song}, \bibinfo{person}{Zihui Ge}, \bibinfo{person}{Aman Shaikh}, \bibinfo{person}{Jia Wang}, \bibinfo{person}{Jennifer Yates}, \bibinfo{person}{Yin Zhang}, {and} \bibinfo{person}{Joanne Emmons}.} \bibinfo{year}{2010}\natexlab{}.
\newblock \showarticletitle{Detecting the performance impact of upgrades in large operational networks}. In \bibinfo{booktitle}{\emph{Proceedings of the ACM SIGCOMM 2010 Conference}}. \bibinfo{pages}{303--314}.
\newblock


\bibitem[Masana et~al\mbox{.}(2022)]%
        {masana2022class}
\bibfield{author}{\bibinfo{person}{Marc Masana}, \bibinfo{person}{Xialei Liu}, \bibinfo{person}{Bart{\l}omiej Twardowski}, \bibinfo{person}{Mikel Menta}, \bibinfo{person}{Andrew~D Bagdanov}, {and} \bibinfo{person}{Joost Van De~Weijer}.} \bibinfo{year}{2022}\natexlab{}.
\newblock \showarticletitle{Class-incremental learning: survey and performance evaluation on image classification}.
\newblock \bibinfo{journal}{\emph{IEEE Transactions on Pattern Analysis and Machine Intelligence}} \bibinfo{volume}{45}, \bibinfo{number}{5} (\bibinfo{year}{2022}), \bibinfo{pages}{5513--5533}.
\newblock


\bibitem[Mei et~al\mbox{.}(2022)]%
        {mei2022realtime}
\bibfield{author}{\bibinfo{person}{Lifan Mei}, \bibinfo{person}{Jinrui Gou}, \bibinfo{person}{Yujin Cai}, \bibinfo{person}{Houwei Cao}, {and} \bibinfo{person}{Yong Liu}.} \bibinfo{year}{2022}\natexlab{}.
\newblock \showarticletitle{Realtime mobile bandwidth and handoff predictions in 4G/5G networks}.
\newblock \bibinfo{journal}{\emph{Computer Networks}}  \bibinfo{volume}{204} (\bibinfo{year}{2022}), \bibinfo{pages}{108736}.
\newblock


\bibitem[Meng et~al\mbox{.}(2022)]%
        {meng2022achieving}
\bibfield{author}{\bibinfo{person}{Zili Meng}, \bibinfo{person}{Yaning Guo}, \bibinfo{person}{Chen Sun}, \bibinfo{person}{Bo Wang}, \bibinfo{person}{Justine Sherry}, \bibinfo{person}{Hongqiang~Harry Liu}, {and} \bibinfo{person}{Mingwei Xu}.} \bibinfo{year}{2022}\natexlab{}.
\newblock \showarticletitle{Achieving consistent low latency for wireless real-time communications with the shortest control loop}. In \bibinfo{booktitle}{\emph{Proceedings of the ACM SIGCOMM 2022 Conference}}. \bibinfo{pages}{193--206}.
\newblock


\bibitem[Meng et~al\mbox{.}(2024)]%
        {meng2024hairpin}
\bibfield{author}{\bibinfo{person}{Zili Meng}, \bibinfo{person}{Xiao Kong}, \bibinfo{person}{Jing Chen}, \bibinfo{person}{Bo Wang}, \bibinfo{person}{Mingwei Xu}, \bibinfo{person}{Rui Han}, \bibinfo{person}{Honghao Liu}, \bibinfo{person}{Venkat Arun}, \bibinfo{person}{Hongxin Hu}, {and} \bibinfo{person}{Xue Wei}.} \bibinfo{year}{2024}\natexlab{}.
\newblock \showarticletitle{Hairpin: Rethinking packet loss recovery in edge-based interactive video streaming}. In \bibinfo{booktitle}{\emph{21st USENIX Symposium on Networked Systems Design and Implementation (NSDI 24)}}. \bibinfo{pages}{907--926}.
\newblock


\bibitem[Mhatre et~al\mbox{.}(2025)]%
        {mhatre2025transfer}
\bibfield{author}{\bibinfo{person}{Suvidha Mhatre}, \bibinfo{person}{Ferran Adelantado}, \bibinfo{person}{Kostas Ramantas}, {and} \bibinfo{person}{Christos Verikoukis}.} \bibinfo{year}{2025}\natexlab{}.
\newblock \showarticletitle{Transfer learning applied to deep reinforcement learning for 6G resource management in intra-and inter-slice RAN-edge domains}.
\newblock \bibinfo{journal}{\emph{IEEE Transactions on Consumer Electronics}} (\bibinfo{year}{2025}).
\newblock


\bibitem[Mungari et~al\mbox{.}(2025)]%
        {mungari2025ran}
\bibfield{author}{\bibinfo{person}{Federico Mungari}, \bibinfo{person}{Corrado Puligheddu}, \bibinfo{person}{Andres Garcia-Saavedra}, {and} \bibinfo{person}{Carla~Fabiana Chiasserini}.} \bibinfo{year}{2025}\natexlab{}.
\newblock \showarticletitle{O-RAN intelligence orchestration framework for quality-driven Xapp deployment and sharing}.
\newblock \bibinfo{journal}{\emph{IEEE Transactions on Mobile Computing}} (\bibinfo{year}{2025}).
\newblock


\bibitem[Nagib et~al\mbox{.}(2023)]%
        {nagib2023safe}
\bibfield{author}{\bibinfo{person}{Ahmad~M Nagib}, \bibinfo{person}{Hatem Abou-Zeid}, {and} \bibinfo{person}{Hossam~S Hassanein}.} \bibinfo{year}{2023}\natexlab{}.
\newblock \showarticletitle{Safe and accelerated deep reinforcement learning-based o-ran slicing: A hybrid transfer learning approach}.
\newblock \bibinfo{journal}{\emph{IEEE Journal on Selected Areas in Communications}} \bibinfo{volume}{42}, \bibinfo{number}{2} (\bibinfo{year}{2023}), \bibinfo{pages}{310--325}.
\newblock


\bibitem[Nguyen et~al\mbox{.}(2024)]%
        {nguyen2024digital}
\bibfield{author}{\bibinfo{person}{Huan~X Nguyen}, \bibinfo{person}{Kexuan Sun}, \bibinfo{person}{Duc To}, \bibinfo{person}{Quoc-Tuan Vien}, {and} \bibinfo{person}{Tuan~Anh Le}.} \bibinfo{year}{2024}\natexlab{}.
\newblock \showarticletitle{Digital twin for o-ran toward 6g}.
\newblock \bibinfo{journal}{\emph{IEEE Communications Magazine}} (\bibinfo{year}{2024}).
\newblock


\bibitem[Nguyen et~al\mbox{.}(2019)]%
        {nguyen2019gee}
\bibfield{author}{\bibinfo{person}{Quoc~Phong Nguyen}, \bibinfo{person}{Kar~Wai Lim}, \bibinfo{person}{Dinil~Mon Divakaran}, \bibinfo{person}{Kian~Hsiang Low}, {and} \bibinfo{person}{Mun~Choon Chan}.} \bibinfo{year}{2019}\natexlab{}.
\newblock \showarticletitle{Gee: A gradient-based explainable variational autoencoder for network anomaly detection}. In \bibinfo{booktitle}{\emph{2019 IEEE Conference on Communications and Network Security (CNS)}}. IEEE, \bibinfo{pages}{91--99}.
\newblock


\bibitem[Ntassah et~al\mbox{.}(2023)]%
        {ntassah2023xapp}
\bibfield{author}{\bibinfo{person}{Rawlings Ntassah}, \bibinfo{person}{Gian~Michele Dell'Aera}, {and} \bibinfo{person}{Fabrizio Granelli}.} \bibinfo{year}{2023}\natexlab{}.
\newblock \showarticletitle{xApp for traffic steering and load balancing in the O-RAN architecture}. In \bibinfo{booktitle}{\emph{ICC 2023-IEEE International Conference on Communications}}. IEEE, \bibinfo{pages}{5259--5264}.
\newblock


\bibitem[Open5GS(2024)]%
        {Open5GS}
\bibfield{author}{\bibinfo{person}{Open5GS}.} \bibinfo{year}{2024}\natexlab{}.
\newblock \bibinfo{booktitle}{}.
\newblock
\newblock
\shownote{\url{"https://open5gs.org"}}.


\bibitem[Permal et~al\mbox{.}(2024)]%
        {permal2024towards}
\bibfield{author}{\bibinfo{person}{Satis~Kumar Permal}, \bibinfo{person}{Yixi Chen}, \bibinfo{person}{Xin~Zhe Khooi}, \bibinfo{person}{Biqing Qiu}, {and} \bibinfo{person}{Mun~Choon Chan}.} \bibinfo{year}{2024}\natexlab{}.
\newblock \showarticletitle{Towards Continuous Latency Monitoring through Open RAN Interfaces}. In \bibinfo{booktitle}{\emph{Proceedings of the 4th ACM Workshop on 5G and Beyond Network Measurements, Modeling, and Use Cases}}. \bibinfo{pages}{14--20}.
\newblock


\bibitem[Perry et~al\mbox{.}(2023)]%
        {perry2023dote}
\bibfield{author}{\bibinfo{person}{Yarin Perry}, \bibinfo{person}{Felipe~Vieira Frujeri}, \bibinfo{person}{Chaim Hoch}, \bibinfo{person}{Srikanth Kandula}, \bibinfo{person}{Ishai Menache}, \bibinfo{person}{Michael Schapira}, {and} \bibinfo{person}{Aviv Tamar}.} \bibinfo{year}{2023}\natexlab{}.
\newblock \showarticletitle{$\{$DOTE$\}$: Rethinking (predictive)$\{$WAN$\}$ traffic engineering}. In \bibinfo{booktitle}{\emph{20th USENIX Symposium on Networked Systems Design and Implementation (NSDI 23)}}. \bibinfo{pages}{1557--1581}.
\newblock


\bibitem[Polese et~al\mbox{.}(2024)]%
        {polese2024colosseum}
\bibfield{author}{\bibinfo{person}{Michele Polese}, \bibinfo{person}{Leonardo Bonati}, \bibinfo{person}{Salvatore D'Oro}, \bibinfo{person}{Pedram Johari}, \bibinfo{person}{Davide Villa}, \bibinfo{person}{Sakthivel Velumani}, \bibinfo{person}{Rajeev Gangula}, \bibinfo{person}{Maria Tsampazi}, \bibinfo{person}{Clifton~Paul Robinson}, \bibinfo{person}{Gabriele Gemmi}, {et~al\mbox{.}}} \bibinfo{year}{2024}\natexlab{}.
\newblock \showarticletitle{Colosseum: The open RAN digital twin}.
\newblock \bibinfo{journal}{\emph{IEEE Open Journal of the Communications Society}}  \bibinfo{volume}{5} (\bibinfo{year}{2024}), \bibinfo{pages}{5452--5466}.
\newblock


\bibitem[Polese et~al\mbox{.}(2023)]%
        {polese2023understanding}
\bibfield{author}{\bibinfo{person}{Michele Polese}, \bibinfo{person}{Leonardo Bonati}, \bibinfo{person}{Salvatore D’oro}, \bibinfo{person}{Stefano Basagni}, {and} \bibinfo{person}{Tommaso Melodia}.} \bibinfo{year}{2023}\natexlab{}.
\newblock \showarticletitle{Understanding O-RAN: Architecture, interfaces, algorithms, security, and research challenges}.
\newblock \bibinfo{journal}{\emph{IEEE Communications Surveys \& Tutorials}} \bibinfo{volume}{25}, \bibinfo{number}{2} (\bibinfo{year}{2023}), \bibinfo{pages}{1376--1411}.
\newblock


\bibitem[Project(2025)]%
        {OSCRIC}
\bibfield{author}{\bibinfo{person}{O-RAN Project}.} \bibinfo{year}{2025}\natexlab{}.
\newblock \bibinfo{booktitle}{\emph{O-RAN SC RIC}}.
\newblock
\newblock
\shownote{\url{"https://docs.o-ran-sc.org/en/latest/index.html"}}.


\bibitem[Qureshi et~al\mbox{.}(2017)]%
        {qureshi2017coordinating}
\bibfield{author}{\bibinfo{person}{Mubashir~Adnan Qureshi}, \bibinfo{person}{Ajay Mahimkar}, \bibinfo{person}{Lili Qiu}, \bibinfo{person}{Zihui Ge}, \bibinfo{person}{Max Zhang}, {and} \bibinfo{person}{Ioannis Broustis}.} \bibinfo{year}{2017}\natexlab{}.
\newblock \showarticletitle{Coordinating rolling software upgrades for cellular networks}. In \bibinfo{booktitle}{\emph{2017 IEEE 25th International Conference on Network Protocols (ICNP)}}. IEEE, \bibinfo{pages}{1--10}.
\newblock


\bibitem[Ramanathan et~al\mbox{.}(2022)]%
        {ramanathan2022resiliency}
\bibfield{author}{\bibinfo{person}{Shunmugapriya Ramanathan}, \bibinfo{person}{Koteswararao Kondepu}, {and} \bibinfo{person}{Andrea Fumagalli}.} \bibinfo{year}{2022}\natexlab{}.
\newblock \showarticletitle{Resiliency in Open-Source Solutions for Disaggregated 5G Cloud Radio Access and Transport Networks}. In \bibinfo{booktitle}{\emph{2022 IEEE Conference on Network Function Virtualization and Software Defined Networks (NFV-SDN)}}. IEEE, \bibinfo{pages}{124--129}.
\newblock


\bibitem[Schiavo et~al\mbox{.}(2024)]%
        {schiavo2024cloudric}
\bibfield{author}{\bibinfo{person}{Leonardo~Lo Schiavo}, \bibinfo{person}{Gines Garcia-Aviles}, \bibinfo{person}{Andres Garcia-Saavedra}, \bibinfo{person}{Marco Gramaglia}, \bibinfo{person}{Marco Fiore}, \bibinfo{person}{Albert Banchs}, {and} \bibinfo{person}{Xavier Costa-Perez}.} \bibinfo{year}{2024}\natexlab{}.
\newblock \showarticletitle{Cloudric: Open radio access network (o-ran) virtualization with shared heterogeneous computing}. In \bibinfo{booktitle}{\emph{Proceedings of the 30th Annual International Conference on Mobile Computing and Networking}}. \bibinfo{pages}{558--572}.
\newblock


\bibitem[Shafin et~al\mbox{.}(2020)]%
        {shafin2020artificial}
\bibfield{author}{\bibinfo{person}{Rubayet Shafin}, \bibinfo{person}{Lingjia Liu}, \bibinfo{person}{Vikram Chandrasekhar}, \bibinfo{person}{Hao Chen}, \bibinfo{person}{Jeffrey Reed}, {and} \bibinfo{person}{Jianzhong~Charlie Zhang}.} \bibinfo{year}{2020}\natexlab{}.
\newblock \showarticletitle{Artificial intelligence-enabled cellular networks: A critical path to beyond-5G and 6G}.
\newblock \bibinfo{journal}{\emph{IEEE Wireless Communications}} \bibinfo{volume}{27}, \bibinfo{number}{2} (\bibinfo{year}{2020}), \bibinfo{pages}{212--217}.
\newblock


\bibitem[Shen(2024)]%
        {shen2024outlier}
\bibfield{author}{\bibinfo{person}{Zhongyang Shen}.} \bibinfo{year}{2024}\natexlab{}.
\newblock \showarticletitle{Outlier Detect using Vector Cosine Similarity by Adding a Dimension}. In \bibinfo{booktitle}{\emph{{IEEE} {ICAIIC}}}. \bibinfo{pages}{255--259}.
\newblock


\bibitem[srsRAN: Open source 4g/5g software radio~access network(2023)]%
        {srsRAN}
\bibfield{author}{\bibinfo{person}{srsRAN: Open source 4g/5g software radio~access network}.} \bibinfo{year}{2023}\natexlab{}.
\newblock \bibinfo{booktitle}{}.
\newblock
\newblock
\shownote{\url{"https://www.srsran.com/"}}.


\bibitem[Sun et~al\mbox{.}(2016)]%
        {sun2016return}
\bibfield{author}{\bibinfo{person}{Baochen Sun}, \bibinfo{person}{Jiashi Feng}, {and} \bibinfo{person}{Kate Saenko}.} \bibinfo{year}{2016}\natexlab{}.
\newblock \showarticletitle{Return of frustratingly easy domain adaptation}. In \bibinfo{booktitle}{\emph{Proceedings of the AAAI conference on artificial intelligence}}, Vol.~\bibinfo{volume}{30}.
\newblock


\bibitem[Sun et~al\mbox{.}(2024)]%
        {sun2024spotlight}
\bibfield{author}{\bibinfo{person}{Chuanhao Sun}, \bibinfo{person}{Ujjwal Pawar}, \bibinfo{person}{Molham Khoja}, \bibinfo{person}{Xenofon Foukas}, \bibinfo{person}{Mahesh~K Marina}, {and} \bibinfo{person}{Bozidar Radunovic}.} \bibinfo{year}{2024}\natexlab{}.
\newblock \showarticletitle{SpotLight: Accurate, explainable and efficient anomaly detection for Open RAN}. In \bibinfo{booktitle}{\emph{Proceedings of the 30th Annual International Conference on Mobile Computing and Networking}}. \bibinfo{pages}{923--937}.
\newblock


\bibitem[sysmoISIM SJA2(2021)]%
        {sysmoISIM}
\bibfield{author}{\bibinfo{person}{sysmoISIM SJA2}.} \bibinfo{year}{2021}\natexlab{}.
\newblock \bibinfo{booktitle}{}.
\newblock
\newblock
\shownote{\url{"https://sysmocom.de/products/sim/sysmousim/"}}.


\bibitem[Tong et~al\mbox{.}(2025)]%
        {tong2025continual}
\bibfield{author}{\bibinfo{person}{Haonan Tong}, \bibinfo{person}{Mingzhe Chen}, \bibinfo{person}{Jun Zhao}, \bibinfo{person}{Ye Hu}, \bibinfo{person}{Zhaohui Yang}, \bibinfo{person}{Yuchen Liu}, {and} \bibinfo{person}{Changchuan Yin}.} \bibinfo{year}{2025}\natexlab{}.
\newblock \showarticletitle{Continual Reinforcement Learning for Digital Twin Synchronization Optimization}.
\newblock \bibinfo{journal}{\emph{IEEE Transactions on Mobile Computing}} (\bibinfo{year}{2025}).
\newblock


\bibitem[Van~der Maaten and Hinton(2008)]%
        {van2008visualizing}
\bibfield{author}{\bibinfo{person}{Laurens Van~der Maaten} {and} \bibinfo{person}{Geoffrey Hinton}.} \bibinfo{year}{2008}\natexlab{}.
\newblock \showarticletitle{Visualizing data using t-SNE.}
\newblock \bibinfo{journal}{\emph{Journal of machine learning research}} \bibinfo{volume}{9}, \bibinfo{number}{11} (\bibinfo{year}{2008}).
\newblock


\bibitem[Vaswani et~al\mbox{.}(2017)]%
        {vaswani2017attention}
\bibfield{author}{\bibinfo{person}{Ashish Vaswani}, \bibinfo{person}{Noam Shazeer}, \bibinfo{person}{Niki Parmar}, \bibinfo{person}{Jakob Uszkoreit}, \bibinfo{person}{Llion Jones}, \bibinfo{person}{Aidan~N Gomez}, \bibinfo{person}{{\L}ukasz Kaiser}, {and} \bibinfo{person}{Illia Polosukhin}.} \bibinfo{year}{2017}\natexlab{}.
\newblock \showarticletitle{Attention is all you need}.
\newblock \bibinfo{journal}{\emph{Advances in neural information processing systems}}  \bibinfo{volume}{30} (\bibinfo{year}{2017}).
\newblock


\bibitem[Wang et~al\mbox{.}(2024)]%
        {wang2024comprehensive}
\bibfield{author}{\bibinfo{person}{Liyuan Wang}, \bibinfo{person}{Xingxing Zhang}, \bibinfo{person}{Hang Su}, {and} \bibinfo{person}{Jun Zhu}.} \bibinfo{year}{2024}\natexlab{}.
\newblock \showarticletitle{A comprehensive survey of continual learning: Theory, method and application}.
\newblock \bibinfo{journal}{\emph{IEEE transactions on pattern analysis and machine intelligence}} \bibinfo{volume}{46}, \bibinfo{number}{8} (\bibinfo{year}{2024}), \bibinfo{pages}{5362--5383}.
\newblock


\bibitem[Wang et~al\mbox{.}(2018)]%
        {wang2018neural}
\bibfield{author}{\bibinfo{person}{Mowei Wang}, \bibinfo{person}{Yong Cui}, \bibinfo{person}{Shihan Xiao}, \bibinfo{person}{Xin Wang}, \bibinfo{person}{Dan Yang}, \bibinfo{person}{Kai Chen}, {and} \bibinfo{person}{Jun Zhu}.} \bibinfo{year}{2018}\natexlab{}.
\newblock \showarticletitle{Neural network meets DCN: Traffic-driven topology adaptation with deep learning}.
\newblock \bibinfo{journal}{\emph{Proceedings of the ACM on Measurement and Analysis of Computing Systems}} \bibinfo{volume}{2}, \bibinfo{number}{2} (\bibinfo{year}{2018}), \bibinfo{pages}{1--25}.
\newblock


\bibitem[Weiss et~al\mbox{.}(2016)]%
        {weiss2016survey}
\bibfield{author}{\bibinfo{person}{Karl Weiss}, \bibinfo{person}{Taghi~M Khoshgoftaar}, {and} \bibinfo{person}{DingDing Wang}.} \bibinfo{year}{2016}\natexlab{}.
\newblock \showarticletitle{A survey of transfer learning}.
\newblock \bibinfo{journal}{\emph{Journal of Big data}} \bibinfo{volume}{3}, \bibinfo{number}{1} (\bibinfo{year}{2016}), \bibinfo{pages}{9}.
\newblock


\bibitem[Wen et~al\mbox{.}(2024a)]%
        {wen20245g}
\bibfield{author}{\bibinfo{person}{Haohuang Wen}, \bibinfo{person}{Phillip Porras}, \bibinfo{person}{Vinod Yegneswaran}, \bibinfo{person}{Ashish Gehani}, {and} \bibinfo{person}{Zhiqiang Lin}.} \bibinfo{year}{2024}\natexlab{a}.
\newblock \showarticletitle{5G-spector: An O-RAN compliant layer-3 cellular attack detection service}. In \bibinfo{booktitle}{\emph{Proceedings of the 31st Annual Network and Distributed System Security Symposium, NDSS}}, Vol.~\bibinfo{volume}{24}.
\newblock


\bibitem[Wen et~al\mbox{.}(2024b)]%
        {wen20246g}
\bibfield{author}{\bibinfo{person}{Haohuang Wen}, \bibinfo{person}{Prakhar Sharma}, \bibinfo{person}{Vinod Yegneswaran}, \bibinfo{person}{Phillip Porras}, \bibinfo{person}{Ashish Gehani}, {and} \bibinfo{person}{Zhiqiang Lin}.} \bibinfo{year}{2024}\natexlab{b}.
\newblock \showarticletitle{6G-XSec: Explainable Edge Security for Emerging OpenRAN Architectures}. In \bibinfo{booktitle}{\emph{Proceedings of the 23rd ACM Workshop on Hot Topics in Networks}}. \bibinfo{pages}{77--85}.
\newblock


\bibitem[Xing et~al\mbox{.}(2023)]%
        {xing2023enabling}
\bibfield{author}{\bibinfo{person}{Jiarong Xing}, \bibinfo{person}{Junzhi Gong}, \bibinfo{person}{Xenofon Foukas}, \bibinfo{person}{Anuj Kalia}, \bibinfo{person}{Daehyeok Kim}, {and} \bibinfo{person}{Manikanta Kotaru}.} \bibinfo{year}{2023}\natexlab{}.
\newblock \showarticletitle{Enabling resilience in virtualized rans with atlas}. In \bibinfo{booktitle}{\emph{Proceedings of the 29th Annual International Conference on Mobile Computing and Networking}}. \bibinfo{pages}{1--15}.
\newblock


\bibitem[Xu et~al\mbox{.}(2015)]%
        {xu2015magus}
\bibfield{author}{\bibinfo{person}{Xing Xu}, \bibinfo{person}{Ioannis Broustis}, \bibinfo{person}{Zihui Ge}, \bibinfo{person}{Ramesh Govindan}, \bibinfo{person}{Ajay Mahimkar}, \bibinfo{person}{NK Shankaranarayanan}, {and} \bibinfo{person}{Jia Wang}.} \bibinfo{year}{2015}\natexlab{}.
\newblock \showarticletitle{Magus: Minimizing cellular service disruption during network upgrades}. In \bibinfo{booktitle}{\emph{Proceedings of the 11th ACM Conference on Emerging Networking Experiments and Technologies}}. \bibinfo{pages}{1--13}.
\newblock


\bibitem[Yan et~al\mbox{.}(2012)]%
        {yan2012argus}
\bibfield{author}{\bibinfo{person}{He Yan}, \bibinfo{person}{Ashley Flavel}, \bibinfo{person}{Zihui Ge}, \bibinfo{person}{Alexandre Gerber}, \bibinfo{person}{Dan Massey}, \bibinfo{person}{Christos Papadopoulos}, \bibinfo{person}{Hiren Shah}, {and} \bibinfo{person}{Jennifer Yates}.} \bibinfo{year}{2012}\natexlab{}.
\newblock \showarticletitle{Argus: End-to-end service anomaly detection and localization from an ISP's point of view}. In \bibinfo{booktitle}{\emph{2012 Proceedings IEEE INFOCOM}}. IEEE, \bibinfo{pages}{2756--2760}.
\newblock


\bibitem[Ye et~al\mbox{.}(2024)]%
        {ye2024dissecting}
\bibfield{author}{\bibinfo{person}{Wei Ye}, \bibinfo{person}{Xinyue Hu}, \bibinfo{person}{Steven Sleder}, \bibinfo{person}{Anlan Zhang}, \bibinfo{person}{Udhaya~Kumar Dayalan}, \bibinfo{person}{Ahmad Hassan}, \bibinfo{person}{Rostand~AK Fezeu}, \bibinfo{person}{Akshay Jajoo}, \bibinfo{person}{Myungjin Lee}, \bibinfo{person}{Eman Ramadan}, {et~al\mbox{.}}} \bibinfo{year}{2024}\natexlab{}.
\newblock \showarticletitle{Dissecting carrier aggregation in 5G networks: Measurement, QoE implications and prediction}. In \bibinfo{booktitle}{\emph{Proceedings of the ACM SIGCOMM 2024 Conference}}. \bibinfo{pages}{340--357}.
\newblock


\bibitem[Zumegen et~al\mbox{.}(2024)]%
        {zumegen2024beamarmor}
\bibfield{author}{\bibinfo{person}{Frederik~Jonathan Zumegen}, \bibinfo{person}{Ish~Kumar Jain}, {and} \bibinfo{person}{Dinesh Bharadia}.} \bibinfo{year}{2024}\natexlab{}.
\newblock \showarticletitle{Beamarmor: Seamless anti-jamming in 5g cellular networks with mimo null-steering}. In \bibinfo{booktitle}{\emph{Proceedings of the 25th International Workshop on Mobile Computing Systems and Applications}}. \bibinfo{pages}{121--126}.
\newblock


\end{thebibliography}

\appendix


\section{Appendices}
\label{sec:appendix}

Appendices are supporting material that has not been peer-reviewed.


\vspace{-2pt}
\subsection{Descriptions of KPMs:}
\label{subsec:KPMlist}

{\small
\begin{itemize}[leftmargin=0.3cm]
    \item \textit{DRB.UEThpDl} – Average downlink UE throughput.  
    \item \textit{DRB.UEThpUl} – Average uplink UE throughput.  
    \item \textit{DRB.AirIfDelayUl} – Average delay of the downlink air-interface.  
    \item \textit{DRB.RlcDelayUl} – Average RLC packet delay in the uplink.
    \item \textit{DRB.RlcPacketDropRateDl} – Downlink packet drop rate in DU.  
    \item \textit{DRB.RlcSduDelayDl} – Average delay of downlink SDUs in DU.  
    \item \textit{DRB.RlcSduTransmittedVolumeDL} – Volume of transmitted downlink data.  
    \item \textit{DRB.RlcSduTransmittedVolumeUL} – Volume of transmitted uplink data.  
    \item \textit{RRU.PrbTotDl} – Downlink total PRB usage.  
    \item \textit{RRU.PrbTotUl} – Uplink total PRB usage.  
    \item \textit{RRU.PrbUsedDl} – Mean downlink PRBs used for data traffic.  
    \item \textit{RRU.PrbAvailDl} – Total available PRBs in downlink.  
    \item \textit{RRU.PrbUsedUl} – Mean uplink PRBs used for data traffic.  
    \item \textit{RRU.PrbAvailUl} – Total available PRBs in uplink.  
    \item \textit{slot\_decision\_latency} – Interval between scheduling request and decision.  
    \item \textit{slot\_k1} – Minimum scheduling time interval in slots.  
    \item \textit{pucch\_sinr\_fm2} – SINR for PUCCH format 2.  
    \item \textit{pucch\_timeali\_us\_fm2} – Synchronization time between UE and gNB for PUCCH format 2.  
    \item \textit{pucch\_metric\_fm1} – Performance metric for PUCCH format 1.  
    \item \textit{pucch\_sinr\_fm1} – SINR for PUCCH format 1.  
    \item \textit{pucch\_timeali\_us\_fm1} – Synchronization time between UE and gNB for PUCCH format 1.  
    \item \textit{pusch\_tbs} – Transport block size for PUSCH.  
    \item \textit{pusch\_sch\_sinr} – SINR for scheduled PUSCH.  
    \item \textit{pusch\_time\_us} – PUSCH transmission time.  
    \item \textit{pdcch\_time\_us} – PDCCH transmission time.  
    \item \textit{pdsch\_tbs} – Transport block size for PDSCH.  
    \item \textit{pdsch\_time\_us} – PDSCH transmission time.  
    \item \textit{du\_ue1\_cqi}, \textit{du\_ue2\_cqi} – Channel Quality Indicator reported by UE1/UE2.  
    \item \textit{du\_ue1\_dl\_brate\_kbps}, \textit{du\_ue2\_dl\_brate\_kbps} – Downlink bitrate for UE1/UE2.  
    \item \textit{du\_ue1\_dl\_mcs}, \textit{du\_ue2\_dl\_mcs} – Downlink MCS indicator for UE1/UE2.  
    \item \textit{du\_ue1\_dl\_nof\_ok}, \textit{du\_ue2\_dl\_nof\_ok} – Successful DL packets for UE1/UE2.  
    \item \textit{du\_ue1\_dl\_nof\_nok}, \textit{du\_ue2\_dl\_nof\_nok} – Failed DL packets for UE.  
    \item \textit{du\_ue1\_dl\_failrate}, \textit{du\_ue2\_dl\_failrate} – DL failure rate for UE.  
    \item \textit{du\_ue1\_dl\_bs}, \textit{du\_ue2\_dl\_bs} – Downlink buffer size for UE.  
    \item \textit{du\_ue1\_pusch\_snr\_db}, \textit{du\_ue2\_pusch\_snr\_db} – Uplink SNR for UE PUSCH.  
    \item \textit{du\_ue1\_pusch\_rsrp\_db}, \textit{du\_ue2\_pusch\_rsrp\_db} – RSRP measurements for UE PUSCH.  
    \item \textit{du\_ue1\_ul\_mcs}, \textit{du\_ue2\_ul\_mcs} – Uplink MCS indicator for UE.  
    \item \textit{du\_ue1\_ul\_brate\_kbps}, \textit{du\_ue2\_ul\_brate\_kbps} – Uplink bitrate for UE1.  
    \item \textit{du\_ue1\_ul\_nof\_ok}, \textit{du\_ue2\_ul\_nof\_ok} – Successful UL packets for UE.  
    \item \textit{du\_ue1\_ul\_nof\_nok}, \textit{du\_ue2\_ul\_nof\_nok} – Failed UL packets for UE.  
    \item \textit{du\_ue1\_ul\_failrate}, \textit{du\_ue2\_ul\_failrate} – UL failure rate for UE.  
    \item \textit{du\_ue1\_bsr}, \textit{du\_ue2\_bsr} – Buffer status reports for UE uplink.  
    \item \textit{du\_ue1\_ta\_ns}, \textit{du\_ue2\_ta\_ns} – Timing advance for UE uplink.  
    \item \textit{du\_ue1\_last\_phr}, \textit{du\_ue2\_last\_phr} – Power headroom reports for UE uplink.  
\end{itemize}
}

\begin{table}[h]
\centering
\tiny
\begin{tabular}{lll}
\toprule
\textbf{KPM Name} & \textbf{Category} & \textbf{Source} \\
\midrule
DRB.UEThpDl & Wireless Channel Quality & Subscription \\
DRB.UEThpUl & Wireless Channel Quality & Subscription \\
DRB.AirIfDelayUl & QoS & Subscription \\
DRB.RlcDelayUl & QoS & Subscription \\
DRB.RlcSduTransmittedVolumeDL & QoS & Subscription \\
DRB.RlcSduTransmittedVolumeUL & QoS & Subscription \\
DRB.RlcPacketDropRateDl & QoS & Subscription \\
DRB.RlcSduDelayDl & QoS & Subscription \\
RRU.PrbTotDl & Resource Scheduling & Subscription \\
RRU.PrbTotUl & Resource Scheduling & Subscription \\
RRU.PrbUsedDl & Resource Scheduling & Subscription \\
RRU.PrbAvailDl & Resource Scheduling & Subscription \\
RRU.PrbUsedUl & Resource Scheduling & Subscription \\
RRU.PrbAvailUl & Resource Scheduling & Subscription \\
slot\_decision\_latency & Resource Scheduling & Hooks \\
slot\_k1 & Resource Scheduling & Hooks \\
pucch\_sinr\_fm2 & Wireless Channel Quality & Hooks \\
pucch\_timeali\_fm2 & Resource Scheduling & Hooks \\
pucch\_metric\_fm1 & QoS & Hooks \\
pucch\_sinr\_fm1 & Wireless Channel Quality & Hooks \\
pucch\_timeali\_fm1 & Resource Scheduling & Hooks \\
pusch\_tbs & Wireless Channel Quality & Hooks \\
pusch\_sch\_sinr & Wireless Channel Quality & Hooks \\
pusch\_time & Resource Scheduling & Hooks \\
pdcch\_time & Resource Scheduling & Hooks \\
pdsch\_tbs & Wireless Channel Quality & Hooks \\
pdsch\_time & Resource Scheduling & Hooks \\
du\_ue1\_cqi & Wireless Channel Quality & Hooks \\
du\_ue2\_cqi & Wireless Channel Quality & Hooks \\
du\_ue1\_dl\_brate & Wireless Channel Quality & Hooks \\
du\_ue2\_dl\_brate & Wireless Channel Quality & Hooks \\
du\_ue1\_dl\_mcs & Wireless Channel Quality & Hooks \\
du\_ue2\_dl\_mcs & Wireless Channel Quality & Hooks \\
du\_ue1\_dl\_nof\_ok & QoS & Hooks \\
du\_ue2\_dl\_nof\_ok & QoS & Hooks \\
du\_ue1\_dl\_nof\_nok & QoS & Hooks \\
du\_ue2\_dl\_nof\_nok & QoS & Hooks \\
du\_ue1\_dl\_failrate & QoS & Hooks \\
du\_ue2\_dl\_failrate & QoS & Hooks \\
du\_ue1\_dl\_bs & QoS & Hooks \\
du\_ue2\_dl\_bs & QoS & Hooks \\
du\_ue1\_pusch\_snr & Wireless Channel Quality & Hooks \\
du\_ue2\_pusch\_snr & Wireless Channel Quality & Hooks \\
du\_ue1\_pusch\_rsrp & Wireless Channel Quality & Hooks \\
du\_ue2\_pusch\_rsrp & Wireless Channel Quality & Hooks \\
du\_ue1\_ul\_mcs & Wireless Channel Quality & Hooks \\
du\_ue2\_ul\_mcs & Wireless Channel Quality & Hooks \\
du\_ue1\_ul\_brate & Wireless Channel Quality & Hooks \\
du\_ue2\_ul\_brate & Wireless Channel Quality & Hooks \\
du\_ue1\_ul\_nof\_ok & QoS & Hooks \\
du\_ue2\_ul\_nof\_ok & QoS & Hooks \\
du\_ue1\_ul\_nof\_nok & QoS & Hooks \\
du\_ue2\_ul\_nof\_nok & QoS & Hooks \\
du\_ue1\_ul\_failrate & QoS & Hooks \\
du\_ue2\_ul\_failrate & QoS & Hooks \\
du\_ue1\_bsr & QoS & Hooks \\
du\_ue2\_bsr & QoS & Hooks \\
du\_ue1\_ta & QoS & Hooks \\
du\_ue2\_ta & QoS & Hooks \\
du\_ue1\_phr & QoS & Hooks \\
du\_ue2\_phr & QoS & Hooks \\
\bottomrule
\end{tabular}
\caption{KPM database collected from srsRAN}
\vspace{-15pt}
\label{tab:KPM_database}
\end{table}

An increase in the number of UEs can also increase the dimensionality of the KPM samples. To handle such changes, during the virtual O-RAN trace-driven emulation phase, KPMs are structurally separated into UE-specific and non-UE metrics. The dimensionality of non-UE metrics does not grow with the number of UEs. For UE metrics, the virtual O-RAN instantiates an independent, parallel operational emulator of fixed size for each active UE profile. During the meta-augmentation phase, we handle the input dimensionality by reshaping the trace matrices. Specifically, non-UE metrics are temporally aggregated into fixed-size summary tracking profiles, while UE metrics are mapped as independent, standalone sequence arrays into the batch dimension, \ie, number of UEs times per‑UE dimensionality.

\begin{figure*}[t]
\setlength{\abovecaptionskip}{0pt}
        \subfigtopskip=-2pt
        \subfigcapskip=-2pt
\centering
\begin{minipage}[b]{0.235\textwidth}
\subfigure[KLD threshold $\tau$]{
\centering
\includegraphics[height = 3cm]{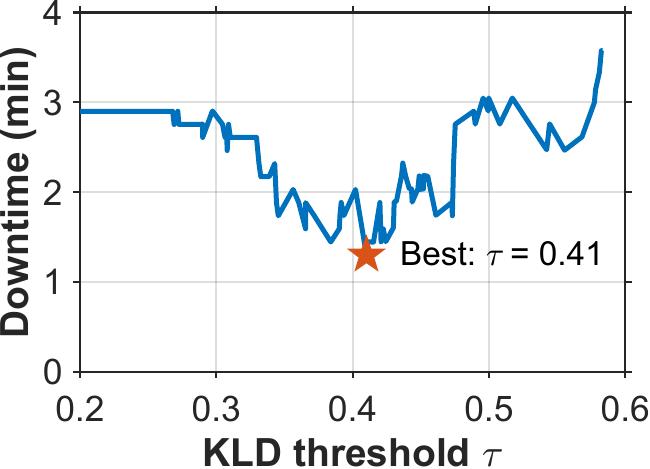}
\label{fig:ru_contention_tau_a}
}
\end{minipage}
\begin{minipage}[b]{0.235\textwidth}
\subfigure[Collection window $t$]{
\centering
\includegraphics[height = 3cm]{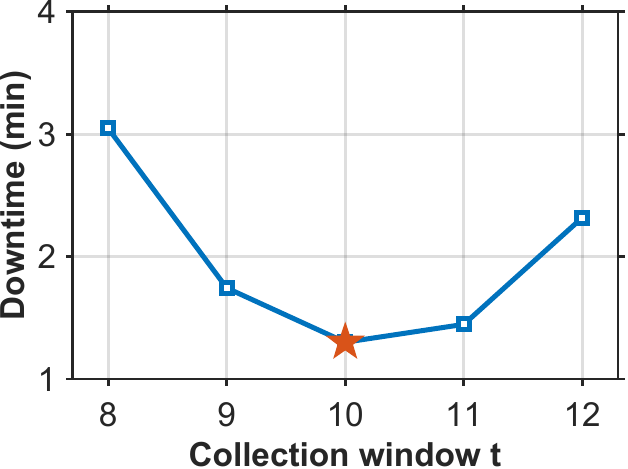}
\label{fig:ru_contention_tau_b}
}
\end{minipage}
\begin{minipage}[b]{0.235\textwidth}
\subfigure[Replay buffer $b$]{
\centering
\includegraphics[height = 3cm]{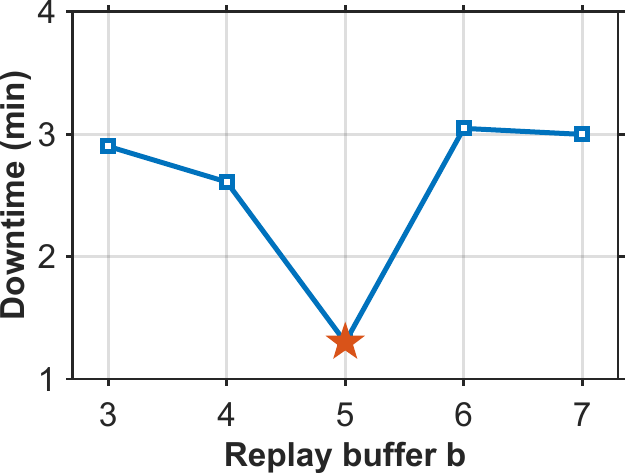}
\label{fig:ru_contention_tau_c}
}
\end{minipage}
\begin{minipage}[b]{0.235\textwidth}
\subfigure[Window length $l$]{
\centering
\includegraphics[height = 3cm]{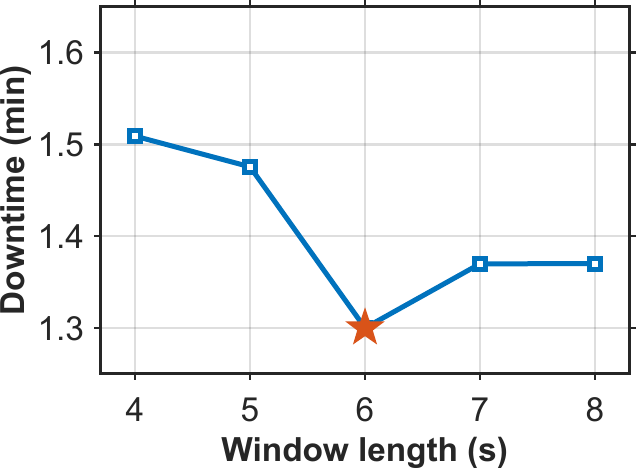}
\label{fig:ru_contention_tau_d}
}
\end{minipage}
\caption{Sensitivity study of hyperparameters in incremental learning: downtime of anomaly detection model under different (a) KLD threshold $\tau$; (b) collection window $t$; (c) replay buffer $b$; and window length $l$.}
\vspace{-5pt}
\label{fig:ru_contention_tau}
\end{figure*}

\vspace{-2pt}
\subsection{Input schema of virtual O-RAN}
\label{subsec:virtualORANInput}

Table~\ref{tab:control_knobs} shows the detailed control knobs of the virtual O-RAN.

\begin{table}[h]
\centering
\small
\begin{tabular}{l|l}
\hline
\textbf{APIs} & \textbf{Input Schema} \\
\hline
Topology      & Graph architecture by adjacency matrix \\\hline
Cell Config   & \makecell[l]{Cell identities (\texttt{cell\_id}, \texttt{pci})\\Frequency bands (\texttt{band})\\Adjacent cell (\texttt{neighborCells})} \\ \hline
RAN Unit Config & \makecell[l]{RU (Tx/Rx gain: [0-30])\\DU (PRB scheduling: [default/xApp])\\CU (Handover: [default/xApp])}\\\hline
xApp            & A containerized xApp with control logic\\
\hline
\end{tabular}
\caption{Control knobs of Virtual O-RAN}
\vspace{-5pt}
\label{tab:control_knobs}
\end{table}

\vspace{-2pt}
\subsection{Detailed algorithm of Meta-Augmentation}

Algorithm~\ref{alg:continuous_meta_augmentation} shows the detailed learning process of continuous meta-augmentation.

\begin{algorithm}[t]
\footnotesize
    \SetAlgoLined
    \SetKwInOut{Input}{Input}\SetKwInOut{Output}{Output}
    \Input{The KPM set $\mathcal{K}$ after data synthesis (\S\ref{SubSec:DataSynthesis})}
    \Output{The augmented KPM dataset $\mathcal{D}$}

    // \ding{182} Autoregressive Pretraining\\
    Initialize $\mathcal{M}$\;
    $\mathcal{K}\xrightarrow[]{\text{slice}}\mathcal{K}_s=\{k_1,k_2,\cdots\}\quad\quad$  // KPM slicing
    
    \For{\textbf{each} $k_i\in\mathcal{K}_s$}{
        $k_i^\prime=\mathcal{M}(k_i);\quad\quad\quad\quad\quad\quad\mathcal{L}_\text{pretrain}=\text{sim}(k_i^\prime,\hat{k_i^\prime})$\;
        $\text{Update}(\mathcal{M}, \mathcal{L}_\text{pretrain});\quad\quad \text{Append }k_i^\prime \text{ to } k_i \rightarrow\hat{k_i}$
    }
    // \ding{183} Meta-Learning\\
    \For{\textbf{each} $\mathcal{K}_i\in\mathcal{D}$}{
        Split $\mathcal{K}_i \rightarrow (\mathcal{S}_i, \mathcal{Q}_i);\quad\quad\quad\quad\quad\quad\mathcal{M}_i=\text{Copy}(\mathcal{M})$\;
        $\mathcal{L}_{\mathcal{S}_i}=\sum_{k_j,\hat{k_j^\prime}\in\mathcal{S}_i}{\text{sim}(\hat{k_j^\prime}, \mathcal{M}_i(k_j)};\quad\text{Update}(\mathcal{M}_i, \mathcal{L}_{\mathcal{S}_i})$\;
        $\mathcal{L}_{\mathcal{Q}_i}=\sum_{k_j,\hat{k_j^\prime}\in\mathcal{Q}_i}\text{sim}(\hat{k_j^\prime}, \mathcal{M}_i(k_j)$\;
    }
    $\mathcal{L}_{\text{sum}}=\sum_{\mathcal{Q}_i \in \mathcal{Q}}{\mathcal{L}_{\mathcal{Q}_i}};\quad\quad\text{Update}(\mathcal{M}, \mathcal{L}_{\text{sum}})\rightarrow\hat{\mathcal{M}}$\;
    // \ding{184} KPM Augmentation\\
    \For{\textbf{each} $k_i\in\mathcal{K}_s$}{
        $\mathcal{D}_i^a=\bigcup({\mathcal{D}_i=\hat{\mathcal{M}}_(k_i)})$\;
    }
    // \ding{185} Continual Tuning\\
    Initialize $\mathcal{D}=\mathcal{K}$ with dynamic maintenance\;
    Repeat step \ding{182} -- \ding{184} with linear decay and append $\mathcal{D}_i^a$ into $\mathcal{D}$\;
    Update O-RAN AI models using $\mathcal{D}$\;
    \caption{\textbf{Continuous meta-augmentation}}
    \label{alg:continuous_meta_augmentation}
\end{algorithm}

\subsection{Wireless channel model}
\label{subsec:wirelessChannelDetails}

$P_{\text{tx}}$ denotes the RU transmit power, and the path loss is defined as $PL(d(t)) = PL_0 + 10n \log_{10}(d(t)/d_0)$, with $PL_0$ representing the reference path loss at distance $d_0$ and $n$ the path loss exponent. $X_\sigma$ models log-normal shadowing as a Gaussian random variable with standard deviation $\sigma$, while $F(t)$ captures fast fading effects, modeled using Rayleigh/Rician distributions depending on NLoS/LoS dominant signal path. The noise floor is denoted by $N_0$. 
Importantly, \projectName enables flexible modeling of UE mobility and deployment environments through customizable parameters: (1) \textit{Mobility Modeling}: UE mobility is incorporated by modeling $d(t)$ as a time-dependent function (\eg, $d(t) = v \cdot t$ for linear motion), allowing emulation of diverse mobility profiles. (2) \textit{Environmental Setting}: Parameters $n$ and $\sigma$ are configurable to reflect different propagation conditions (detailed in \tableautorefname~\ref{tab:pathloss_shadowing}). These parameters are initially estimated by \projectName to align with the KPM traces from the RAN recorder and subsequently adjusted to evaluate performance under varying mobility and environmental scenarios.
Based on the computed SINR, the UE derives CQI and RSRP \cite{3GPP} for KPM reporting. This modeling approach supports sufficient fidelity of wireless channel dynamics required by AI model adaptation, while maintaining minimal computational overhead.

\begin{table}[h]
\centering
\small
\begin{tabular}{l|c|c}
\hline
\textbf{Environment} & \textbf{Path Loss $n$} & \textbf{Shadowing $\sigma$} \\
\hline
Free Space (Ideal)       & 2.0   & 0--2   \\
Indoor (LoS)   & 1.6--2.0 & 2--4   \\
Indoor (NLoS) & 3.0--5.0 & 4--8   \\
Urban (Dense)            & 2.7--4.0 & 6--10  \\
Suburban                 & 2.0--3.0 & 4--6   \\
Rural (Open)             & 2.0--2.5 & 2--4   \\
\hline
\end{tabular}
\caption{Configuration of wireless channel parameters}
\vspace{-5pt}
\label{tab:pathloss_shadowing}
\end{table}

\vspace{-2pt}
\subsection{Sensitivity study of selective replay}
\label{subsec:incrementalLearningSensitivity}

To evaluate the sensitivity of \projectName's incremental learning module, we extend the anomaly detection experiment in Table~\ref{tab:anomaly_RUcontention} by varying the hyperparameters introduced in \S~\ref{sec:hyperparameters}, including KLD threshold  $\tau$, collection window $t$, replay buffer $b$, and window length $l$. We use the AI downtime required for the AI models to match Oracle performance as the metric for benchmarking incremental learning performance. By default, we set $\tau=0.41$, $t=10$, $b=5$, $l=6$.

\begin{figure}[t]
\setlength{\abovecaptionskip}{0pt}
        \subfigtopskip=-2pt
        \subfigcapskip=-2pt
\centering
\begin{minipage}[b]{0.235\textwidth}
\subfigure[Amount of seed data]{
\centering
\includegraphics[height = 3cm]{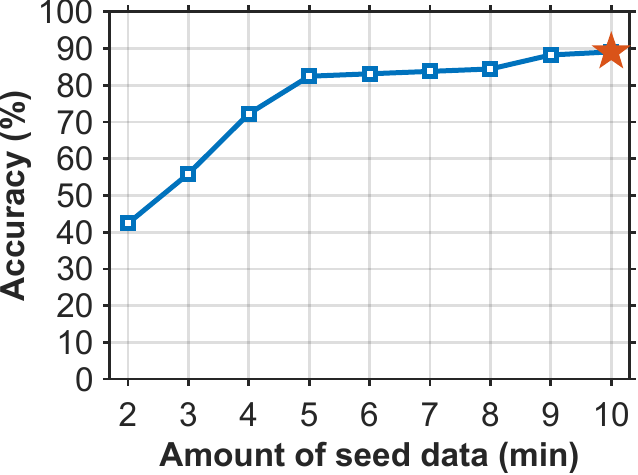}
\label{fig:ru_contention_seed_a}
}
\end{minipage}
\begin{minipage}[b]{0.235\textwidth}
\subfigure[Amount of synthetic data]{
\centering
\includegraphics[height = 3cm]{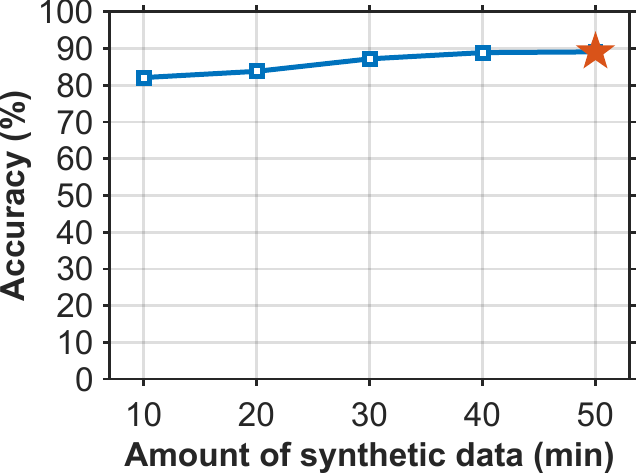}
\label{fig:ru_contention_seed_b}
}
\end{minipage}
\caption{Impact of seed/synthetic data volume on the trained model accuracy.}
\vspace{-5pt}
\label{fig:ru_contention_seed}
\end{figure}

\noindent
\textbf{Results.} The impact of KLD threshold $\tau$ is illustrated in \figureautorefname{}~\ref{fig:ru_contention_tau}(a), when the threshold is too low (\eg, $\tau=0.2$), the selection criteria become overly permissive, causing almost all post-reconfiguration traces to be ingested. This collapses the pipeline into conventional continual learning replay, which uniformly uses all collected data, diluting vital anomaly features and increasing AI downtime. Conversely, when the threshold is too high (\eg, $\tau=0.6$), the filter becomes overly restrictive, erroneously discarding representative abnormal data blocks. This deprives the incremental training set of critical post-reconfiguration dynamics and retards model convergence. The empirical results demonstrate a clear operational sweet spot at $\tau = 0.41$, minimizing AI downtime to approximately $1.3$~minutes. At this optimal threshold, \projectName maximizes the selection of highly representative anomaly traces while precisely filtering out redundant normal data that does not contribute to model adaptation. 
In addition, the collection window $t$ and the replay buffer size $b$ collectively govern the density of useful patterns extracted from the real KPM trace. As shown in \figureautorefname{}~\ref{fig:ru_contention_tau}(b,c), decreasing $t$ or increasing $b$ expands the data inclusion range, which dilutes the proportion of truly useful patterns and degrades downstream training efficiency. Conversely, increasing $t$ or decreasing $b$ narrows the selection range, potentially overlooking useful data patterns. 
Finally, \figureautorefname{}~\ref{fig:ru_contention_tau}(d) shows the impact of the KPM segment sliding window length $l$. A short window fails to capture long-term KPM sequence dependencies, whereas a long window subsumes a large volume of normal data, ultimately flooding the underlying anomaly patterns and suppressing the KLD score.

\vspace{-2pt}
\subsection{Impact of seed/synthetic data volume}

This experiment evaluates how the volumes of both seed and synthetic data affect downstream AI model adaptation performance. Extending the anomaly detection scenario from Table~\ref{tab:anomaly_RUcontention}, we measure the plug-and-play detection accuracy of the models immediately upon deployment to benchmark AI adaptation performance. To assess the impact of seed data volume, we vary the duration of the baseline KPM traces used to drive the Virtual O-RAN data synthesis pipeline, while utilizing Meta-Augmentation to expand the total training set to a fixed 1-hour KPM trace (aligning with the previous experiment). To evaluate the impact of synthetic data volume, we fix the seed data to a 10-minute KPM trace and vary the duration of the synthetic data generated by the Meta-Augmentation module.

\noindent
\textbf{Results.}
As illustrated in \figureautorefname{}~\ref{fig:ru_contention_seed}(a), the plug-and-play accuracy increases sharply as the volume of the initial seed data scales upward. This trend occurs because brief seed windows fail to capture a representative distribution of baseline network traffic characteristics and scheduling dynamics. However, scaling the historical seed trace to 9 minutes yields a significant performance knee, elevating the accuracy to an optimal plateau of around 89\%. Beyond this threshold, further expanding the seed data volume yields diminishing returns. The results indicate that a proper seed KPM trace containing diverse and comprehensive environmental features is essential for effective data synthesis. \figureautorefname{}~\ref{fig:ru_contention_seed}(b) depicts the impact of the generated synthetic data volume. Similar to the seed data dynamics, the plug-and-play accuracy climbs rapidly as the training set is populated with synthetic traces, crossing the threshold at 30 minutes of expansion. Beyond this 30-minute threshold, the accuracy increases only marginally. It demonstrates that further generative expansion provides diminishing performance gains once the target distribution space has been robustly synthesized.

\begin{figure}[h]
\vspace{-2pt}
\setlength{\abovecaptionskip}{4pt}
\centerline{\includegraphics[height = 4.5cm]{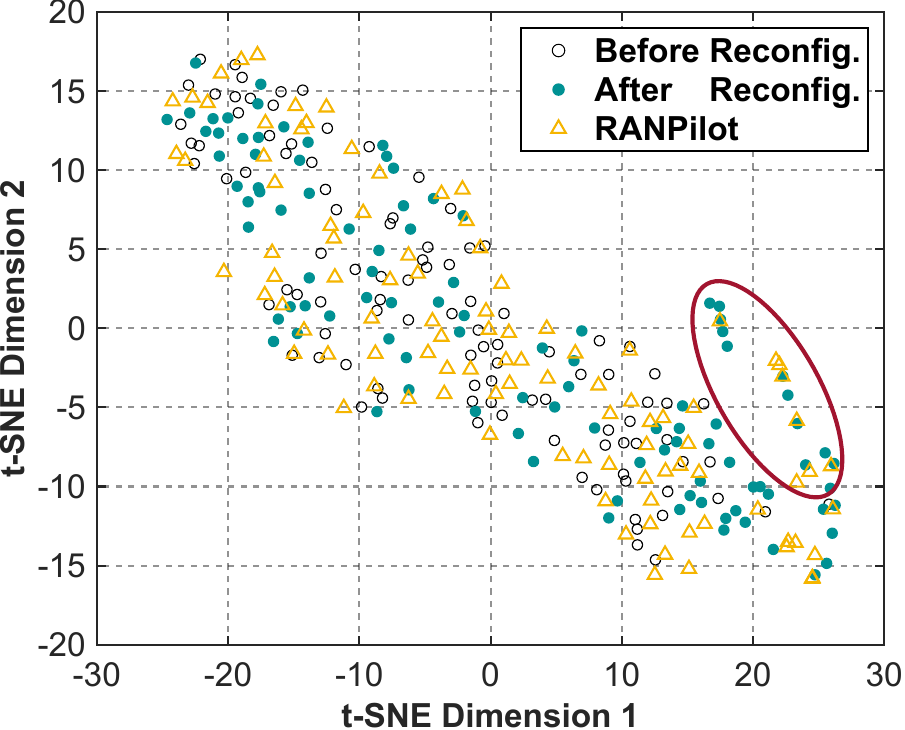}}
\caption{Visualization of KPM distribution.}
\label{fig:visualization}
\vspace{-5pt}
\end{figure}

\vspace{-2pt}
\subsection{Visualization of KPM distribution}
To illustrate how \projectName mitigates data drift and recovers the KPM distribution, we use t-SNE \cite{van2008visualizing} to visualize the KPM trace's data distribution. We collect three sets of segmented KPM traces in the indoor cell addition experiment: (1) live traces before the reconfiguration; (2) live traces after the reconfiguration; and (3) synthetic traces generated by \projectName.

\noindent
\textbf{Results.} As illustrated in \figureautorefname{}~\ref{fig:visualization}, after cell addition, the distribution of most post-reconfiguration KPM traces aligns with the pre-reconfiguration traces, while a small portion of data exhibits severe drift (marked by the red circle). The pre-reconfiguration seed traces are entirely absent from this region, meaning that a model trained purely on raw seed data cannot handle these network variations. \projectName effectively mitigates distribution drift by emulating RAN behavior and synthesizing KPM traces, thereby proactively adapting downstream AI models to upcoming out-of-distribution KPM variations.


\end{document}